\documentclass[prb,twocolumn,notitlepage,superscriptaddress,showkeys,showpacs]{revtex4-1}
\usepackage{amsmath,amsfonts,amssymb,amsthm,epsfig,epstopdf,url,array}
\usepackage{dsfont}
\usepackage{bm}
\usepackage{dcolumn}
\usepackage{mathrsfs}
\usepackage{latexsym}
\usepackage{graphics}
\usepackage{textcomp}
\usepackage{mathrsfs}
\usepackage{indentfirst}
\usepackage{fancyhdr}
\usepackage{titletoc}
\usepackage{titlesec}
\usepackage{float}
\usepackage{verbatim}
\usepackage{multirow}
\usepackage{mathtools}
\usepackage{bbold}
\usepackage{ragged2e}
\usepackage[hidelinks]{hyperref}
\usepackage{makecell}
\usepackage{color}
\usepackage[dvipsnames]{xcolor}
\usepackage{feynmf}
\usepackage[titletoc,title]{appendix}
\usepackage[normalem]{ulem}

\newcommand{\bsl}[1]{\boldsymbol{#1}}

\newcommand{\citen}[1]{\onlinecite{#1}}

\begin{document}
\title{Spin Susceptibility, Upper Critical Field and Disorder Effect in $j=\frac{3}{2}$ Superconductors with Singlet-Quintet Mixing}
\author{Jiabin Yu}
\affiliation{Department of Physics, the Pennsylvania State University, University Park, PA, 16802}
\author{Chao-Xing Liu}
\email{cxl56@psu.edu}
\affiliation{Department of Physics, the Pennsylvania State University, University Park, PA, 16802}
\begin{abstract}
Recently, a new pairing state with the mixing between s-wave singlet channel and isotropic d-wave quintet channel induced by centrosymmetric spin-orbit coupling has been theoretically proposed in the superconducting materials with $j=\frac{3}{2}$ electrons.\cite{yu2017Singlet-Quintetj=3/2SC} 
In this work, we derive the expressions of the zero-temperature spin susceptibility, the upper critical field close to the zero-field critical temperature $T_c$ and the critical temperature with weak random non-magnetic disorders for the singlet-quintet mixed state based on the Luttinger model. Our study revealed the following features of the singlet-quintet mixing. 
(1) The zero-temperature spin susceptibility remains zero for the singlet-quintet mixed state if only the centrosymmetric spin-orbit coupling is taken into account, and will deviate from zero when the non-centrosymmetric spin-orbit coupling is introduced. 
(2) The singlet-quintet mixing can help enhance the upper critical field roughly because it can increase $T_c$.
(3) Although the quintet channel is generally suppressed by the non-magnetic disorder scattering, we find the strong mixing between singlet and quintet channels can help to stabilize the quintet channel. 
As a result, we still find a sizable quintet component mixed into the singlet channel in the presence of weak random non-magnetic disorders. 
Our work provides the guidance for future experiments on spin susceptibility and upper critical field of the singlet-quintet mixed superconducting states, and illustrates the stability of the singlet-quintet mixing against the weak random non-magnetic disorder.
\end{abstract}
\maketitle

\section{Introduction}

Increasing research interests have recently been focused on the superconductivity in half-Heusler materials,
including RPtBi(R=La, Y and Lu) and RPdBi(R = Er, Lu,
Ho, Y, Sm, Tb, Dy and Tm) due to their possible unconventional mechanism  indicated by the low carrier density($10^{18}\sim 10^{19} cm^{-3}$) compared with the critical temperature ($0.5\sim 1.9 K$), the power-law temperature dependence of London penetration depth implying nodal superconductivity (YPtBi) and the large upper critical field.\cite{Goll2008LaBiPtSC,Butch2011SCYPtBi,
Bay2012SCYPtBi,Tafti2013LuPtBiSC,
Pan2013ErPdBiSC,Nakajima2015RPdBiSC,Xu2014LuPdBiSC,
Pavlosiuk2015LuPdBiSC,Nikitin2015HoPdBiSC,
Meinert2015UnconverntialSCYPtBi,TbPdBi2018SC,kim2018YPtBiSCj=3/2}
In these half-Heusler compounds, the low energy excitations have total angular momentum $j=\frac{3}{2}$ given by the addition of $\frac{1}{2}$ spin and angular momentum of p atomic orbitals ($l=1$). Therefore, half-Heusler SCs provide an intriguing platform to study superconductivity with $j=\frac{3}{2}$ fermions\cite{kim2018YPtBiSCj=3/2,Brydon2016j=3/2SC}. Such $j=\frac{3}{2}$ fermions also exist in Anti-perovskite materials\cite{Kawakami2018j=3/2electrons} and the cold atom system\cite{Wu2006spin3/2CAS,Kuzmenko2018F=3/2CFG}.
The effective spin $j=\frac{3}{2}$ of electrons allows the spin of Cooper pairs to take four values,  $S=0$ (singlet), 1 (triplet), 2 (quintet) and 3 (septet), instead of only singlet and triplet for spin-$\frac{1}{2}$ electrons. A variety of pairing states have been studied in such system, including mixed singlet-septet pairing\cite{Brydon2016j=3/2SC,kim2018YPtBiSCj=3/2,
Yang2017Majoranaj=3/2SC,Timm2017nodalj=3/2SC},
mixed singlet-quintet pairing\cite{yu2017Singlet-Quintetj=3/2SC,
Wang2018j=3/2SCSurface}, s-wave quintet pairing
\cite{Brydon2016j=3/2SC,Roy2017j=3/2SC,Timm2017nodalj=3/2SC,
Boettcher2018j=3/2SC}
, d-wave quintet pairing\cite{Yang2016j=3/2Fermions,Venderbos2018j=3/2SC}
, odd-parity (triplet and septet) parings\cite{Yang2016j=3/2Fermions,Venderbos2018j=3/2SC,
Savary2017j=3/2SC,Ghorashi2017j=3/2SCdisorder}, {\it et al}\cite{Venderbos2018j=3/2SC,Brydon2018BFS}.
In particular, the mixing between the s-wave singlet and isotropic d-wave quintet channels proposed in Ref.[\citen{yu2017Singlet-Quintetj=3/2SC}] is the first realistic proposal of the mixing between different spin channels that preserves the inversion symmetry in solid state systems.
The mixing is promising because it is induced by the strong inversion-invariant ``spin orbital coupling (SOC)''(the coupling between the ``$\frac{3}{2}$-spin'' and the orbit) and the resulted topological nodal-line superconductivity(TNLS) is protected by the non-trivial topological invariant.\cite{yu2017Singlet-Quintetj=3/2SC}
In this work, we studied the spin susceptibility, the upper critical field and the non-magnetic disorder effect of such pairing mixing state. We found that the spin susceptibility is isotropic and approaches to a non-zero (zero) value as the temperature decreases in the presence (absence) of the inversion-breaking SOC.
We also found that the upper critical field near the zero-field critical temperature $T_c$ can be isotropic and enhanced by the mixing, and its slope at $T_c$ varies significantly with the band structure.
In presence of the non-magnetic random disorder, it is found that the critical temperature and the portion of the quintet channel of the paring-mixed state are suppressed, while the latter cannot be entirely suppressed due to the singlet-quintet mixing.
Our results show several properties of the singlet-quintet mixed state that can be experimentally measured.

The rest of the paper is organized as the following.
We will describe the model for the mixing between the s-wave singlet and isotropic d-wave quintet channels in Sec.\ref{sec:model_H}, addresses spin susceptibility in Sec.\ref{sec:spin_sus}, study the upper critical field in Sec.\ref{sec:mag_protes}, discuss the disorder effect in Sec.\ref{disorder}, and eventually conclude our work with the discussion about experiments in Sec.\ref{sum}.

\section{Model Hamiltonian}
\label{sec:model_H}
\begin{figure}[t]
\includegraphics[width=\columnwidth]{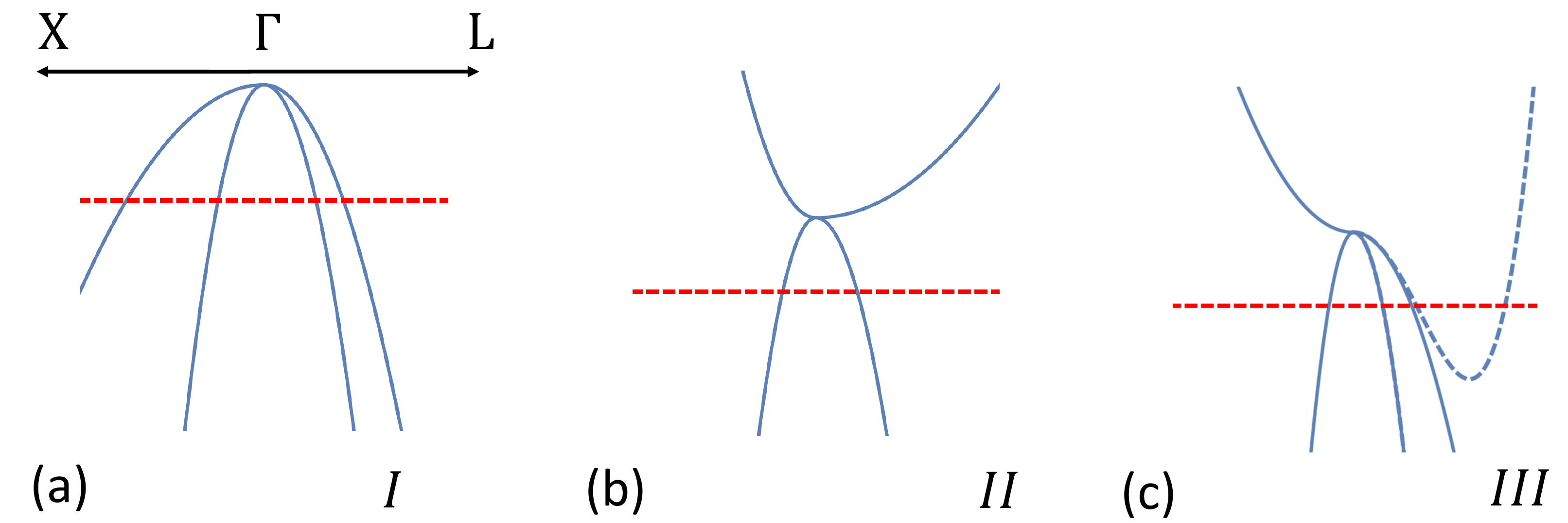}
\caption{
\label{fig:band_struc}
The solid lines in (a), (b) and (c) plot the typical band structure in regimes I, II and III, respectively.
The red dashed lines stand for the position of the chemical potential $\mu$.
The dashed purple line in (c) shows the realistic band structure beyond the Luttinger model.
}
\end{figure}

In this section, we will first review the model without magnetic fields proposed in Ref.[\citen{yu2017Singlet-Quintetj=3/2SC}] and then introduce the modification due to the external magnetic field. 
The effective non-interacting Hamiltonian that describes the low-energy $j=\frac{3}{2}$ fermionic excitations with long wavelengths is the Luttinger model\cite{Luttinger1956LuttingerModel,chadov2010tunable,
Winkler2003SOC, yu2017Singlet-Quintetj=3/2SC}, which reads
\begin{eqnarray}\label{Eqn:h}
h(\bsl{k})=\xi_{\bsl{k}}\Gamma^0+ h_{SSOC}(\bsl{k})+h_{ASOC}(\bsl{k})\ ,
\end{eqnarray}
where
\begin{equation}
h_{SSOC}(\bsl{k})=c_1 \sum_{i=1}^{3}g_{\bsl{k},i}\Gamma^i+c_2 \sum_{i=4}^{5}g_{\bsl{k},i}\Gamma^i
\end{equation}
is the symmetric SOC(SSOC) which is invariant under inversion,
\begin{equation}
h_{ASOC}=\frac{2 C}{\sqrt{3}}(k_x V_x+ k_y V_y+k_z V_z)
\end{equation}
is the anti-symmetric SOC(ASOC) which changes sign under inversion.
Here the bases have total angular momentum $\frac{3}{2}$ as mentioned in the last section and can be labeled as $|j,j_z\rangle$ with $j=\frac{3}{2}$ and  $j_z=3/2, 1/2, -1/2, -3/2$, $\xi_{\bsl{k}}=\frac{1}{2m}k^2-\mu$ with $\mu$ the chemical potential, and
the expressions of five d-orbital cubic harmonics $g_i$'s, six
$4\times 4$ matrices $\Gamma^i$ ($i=0,\dots,5$)
and $V_{x,y,z}$ are shown in Appendix.\ref{app:conv_expn}.
We want to emphasize that both SSOC and ASOC refer to the coupling between the ``3/2-spin'' and the orbit degrees of freedom of $j=3/2$ fermions.
$h(\bsl{k})$ has $O(3)$ point group symmetry for $c_1=c_2$.
$c_1\neq c_2$ reduces $O(3)$ to $O_h$ and $C\neq 0$ further reduces it to $T_d$.
$h(\bsl{k})$ also has time-reversal(TR) symmetry: $\gamma h^*(-\bsl{k})\gamma^{\dagger}=h(\bsl{k})$, where $\gamma=-\Gamma_1\Gamma_3$ is the TR matrix.
If $C=0$, $h(\bsl{k})$ has two doubly degenerate bands $\xi_{\pm}(\mathbf{k})=k^2/(2m_{\pm})-\mu$, where $m_{\pm}=m \widetilde{m}_{\pm}$, $\widetilde{m}_{\pm}=1/(1\pm 2mQ_c)$,  $Q_c=\sqrt{c_1^2 Q_1^2+c_2^2 Q_2^2}$, $Q_1=\sqrt{\hat{g}^2_{1}+\hat{g}^2_{2}+\hat{g}^2_{3}}$, $Q_2=\sqrt{\hat{g}^2_{4}+\hat{g}^2_{5}}$ and $\hat{g}_i=g_i/k^2$.
We also assume $\mu<0$ for p-type carriers\cite{Brydon2016j=3/2SC, Savary2017j=3/2SC}, $m<0$\cite{Yang2017HHTP} and $c_1 c_2>0$ for simplicity.
In this case, we have three regimes(Fig.\ref{fig:band_struc}): (I) $m_+<0$ (normal band structure), (II) $m_+>0$ (inverted band structure),
and (III) the sign of $m_+$ being angular dependent, while $m_-$ is always negative.\cite{yu2017Singlet-Quintetj=3/2SC}
At last, since the Luttinger model is only valid around the $\Gamma$ point, we introduce a momentum cut-off $\Lambda$ and only care about the Fermi surface inside $\Lambda$.
The momentum cut-off $\Lambda$ is not essential in regimes I and II since the Fermi surfaces are closed and finite, and thus we drop it in those regimes.\cite{yu2017Singlet-Quintetj=3/2SC}
In regime III, the $\xi_+$ band would form a saddle point and its corresponding Fermi surface is unbounded, which is just an artifact of Luttinger model(Fig.\ref{fig:band_struc}c) and requires the momentum cut-off $\Lambda$.\cite{yu2017Singlet-Quintetj=3/2SC}

As described in Ref.[\citen{yu2017Singlet-Quintetj=3/2SC}], we focus on a minimal $O(3)$-invariant attractive interaction
\begin{equation}\label{Eqn:HI}
H_I=\frac{1}{2\mathcal{V}}
\sum_{\bsl{q}}\left[
V_0 P_0(\bsl{q}) P_0^{\dagger}(\bsl{q})
+
V_1 P_1(\bsl{q}) P_1^{\dagger}(\bsl{q})
\right]
\end{equation}
in the s-wave singlet and isotropic d-wave quintet channels,
where $P_0(\bsl{q})=\sum_{\bsl{k}}c^{\dagger}_{\bsl{k}+\frac{\bsl{q}}{2}}(\Gamma^0 \gamma/2)(c^{\dagger}_{-\bsl{k}+\frac{\bsl{q}}{2}})^T$,
$P_1(\bsl{q})=\sum_{\bsl{k}}c^{\dagger}_{\bsl{k}+\frac{\bsl{q}}{2}}(a^2 \bsl{g}_{\bsl{k}}\cdot\bsl{\Gamma}
\gamma/2)(c^{\dagger}_{-\bsl{k}+\frac{\bsl{q}}{2}})^T$,  $V_0<0$ and $V_1<0$
stand for the attractive interaction in singlet and quintet channels, respectively.
The above attractive interaction only applies to the electrons near the Fermi energy within the energy cut-off $\epsilon_c$. Here $c^{\dagger}_{\bsl{k}}=(c^{\dagger}_{\bsl{k},\frac{3}{2}},c^{\dagger}_{\bsl{k},\frac{1}{2}},c^{\dagger}_{\bsl{k},-\frac{1}{2}},c^{\dagger}_{\bsl{k},-\frac{3}{2}})$ creates a $j=\frac{3}{2}$ fermion with wavevector $\bsl{k}$,
$\mathcal{V}$ is the volume and $a$ is the lattice constant. Compared to Ref.[\citen{yu2017Singlet-Quintetj=3/2SC}], we include a non-zero $\bsl{q}$ in the interaction term (\ref{Eqn:HI}), which is essential for the study of upper critical field. In this case, the mean-field gap function derived from Eq.\ref{Eqn:HI} reads
\begin{equation}
\label{eq:pairing_gen}
\Delta(\bsl{k},\bsl{q})=\Delta_0(\bsl{q})\frac{\Gamma^0\gamma}{2}+\Delta_1(\bsl{q})\frac{a^2 \bsl{g}_{\bsl{k}}\cdot\bsl{\Gamma}\gamma}{2}\ ,
\end{equation}
where $\Delta_0(\bsl{q})$ and $\Delta_1(\bsl{q})$ are order parameters in the singlet and quintet channels, respectively.

To study spin susceptibility and upper critical field, a uniform magnetic field $\bsl{B}$ is required to couple to the electrons in the above model.
We assume the magnetic field is small enough so that only the first order of $B=|\bsl{B}|$ is kept.
Such assumption is suitable for the calculation of spin susceptibility but restricts the study of the upper critical field to be at the temperature close to the zero-field critical temperature.
The magnetic field has two effects: the Zeeman effect and the orbital effect.\cite{peierls1933theorie,
luttinger1951magnetic,
Kohn1959magnetic,
Wannier1962electricmagnetic,
Blount1962magnetic,
roth1962magnetic,WHHeq1966,Samokhin2004magnetic}
The Zeeman effect is described by the Hamiltonian
\begin{equation}
\label{eq:H_Z_G8}
h_Z^{\Gamma_8}=\frac{2\mu_B}{3} \bsl{B}\cdot \bsl{J}\ ,
\end{equation}
in the basis of the $\Gamma_8$ bands (Appendix.\ref{app:conv_expn}), where
$\bsl{J}=(J_x,J_y,J_z)$ are angular momentum matrices for $j=\frac{3}{2}$ (Appendix.\ref{app:conv_expn}), $\mu_B=\frac{e\hbar}{2m_e}$ is the Bohr magneton, $e$ is the elementary charge and $m_e$ is the rest mass of the electron.
Before including the orbital effect, we first project $h(\bsl{k})+h_Z^{\Gamma_8}$ onto $\xi_{\pm}$ bands and get the effective Hamiltonian
\begin{equation}
\label{eq:XI_k}
\Xi^{\pm}(\bsl{k},\bsl{B})=\xi_{\pm}(\bsl{k})+C k \bsl{p}^{\pm}(\hat{\bsl{k}})\cdot \bsl{\sigma}+\bsl{B}\cdot\bsl{M}^{\pm}(\hat{\bsl{k}})\ ,
\end{equation}
where $\bsl{p}^{\pm}(\hat{\bsl{k}})\cdot \bsl{\sigma}$ and  $\bsl{M}^{\pm}(\hat{\bsl{k}})$ are the corresponding $2\times 2$ blocks of the projected
$\frac{2}{\sqrt{3}}\hat{\bsl{k}}\cdot \bsl{V}$ and $\frac{2\mu_B}{3}\bsl{J}$ on $\xi_{\pm}$ bands, respectively, $\hat{\bsl{k}}=\bsl{k}/k$ , $\bsl{p}^{\pm}(-\hat{\bsl{k}})=-\bsl{p}^{\pm}(\hat{\bsl{k}})$, $\bsl{\sigma}=(\sigma_x,\sigma_y,\sigma_z)$ are Pauli matrices for the double degeneracy of each band, $\bsl{M}^{\pm}(-\hat{\bsl{k}})=\bsl{M}^{\pm}(\hat{\bsl{k}})$ and $\text{Tr}[\bsl{M}^{\pm}(\hat{\bsl{k}})]=0$.
In Eq.\ref{eq:XI_k}, we neglect the terms of order $\frac{C k}{2 Q_c k^2}$.
The reason is that the energy scale of SSOC near the Fermi surface is typically much larger than that of ASOC, e.g. $2 Q_c k_F^2\sim 20 meV$ and $C k_F\sim 4 meV$ for YPtBi\cite{kim2018YPtBiSCj=3/2,Brydon2016j=3/2SC, Savary2017j=3/2SC} with $k_F$ being the magnitude of the Fermi momentum.
For the orbital effect, we can choose the symmetric gauge for the vector potential as $\bsl{A}(\bsl{r})=\frac{\bsl{B}\times \bsl{r}}{2}$ and the vector potential can be included into the Hamiltonian with the Peierls substitution\cite{peierls1933theorie,
luttinger1951magnetic,
Kohn1959magnetic,
Wannier1962electricmagnetic,
Blount1962magnetic,
roth1962magnetic,WHHeq1966,Samokhin2004magnetic}.
As a result, the effective Hamiltonian (\ref{eq:XI_k}) becomes
\begin{equation}
\label{eq:XIK_1B}
 \Xi^{\pm}(\bsl{K},\bsl{B})=h_{\pm}(\bsl{k})+\bsl{B}\cdot\bsl{M}^{\pm}(\hat{\bsl{k}})+\frac{e}{\hbar}\bsl{\nabla}_{\bsl{k}}h_{\pm}(\bsl{k})\cdot \bsl{A}(i\bsl{\nabla}_{\bsl{k}})
\end{equation}
with $\bsl{K}=\bsl{k}+\frac{e}{\hbar}\bsl{A}(i\bsl{\nabla}_{\bf k})$ and $h_{\pm}(\bsl{k})=\xi_{\pm}(\bsl{k})+C k \bsl{p}^{\pm}(\hat{\bsl{k}})\cdot \bsl{\sigma}$.

\section{Spin Susceptibility}
\label{sec:spin_sus}
The spin susceptibility $\chi_{ij}$ can be defined as
\begin{equation}
\label{eq:def_spin_sus}
\chi_{ij}=\left. \frac{\partial M_i^{spin}}{\partial B_j}\right|_{\bsl{B}\rightarrow 0}\ ,
\end{equation}
where $M_{i}^{spin}$ is the $i$th component of the magnetic moment generated by the spins of conduction electrons.\cite{Abrikosov1961KnightShift}
The spin susceptibility of a material in the superconducting phase $\chi^{S}_{ij}$ is typically different from that in the normal metal phase $\chi^N_{ij}$ due to the formation of Cooper pairs.
Such difference cause Knight shifts\cite{KnightShift1949,Reif1957KnightShift} in nuclear-magnetic-resonance(NMR) experiments, which serves as an important experimental tool to identify the pairing form.
In this section, we will study the spin susceptibility of the singlet-quintet mixed superconducting state.

We first analyze the symmetry properties of $\chi^S$ and $\chi^N$.
According to the definition of $\chi_{ij}$ (\ref{eq:def_spin_sus}),
the spin susceptibility satisfies $\chi_{ij}= \sum_{i',j'} R_{ii'} R_{j j'} \chi_{i'j'}$ for any operation $\hat{R}$ in the point group of the material, where $R_{ii'}$ represents the transformation of a pseudo-vector under $\hat{R}$.
The model considered here (\ref{Eqn:h}, \ref{Eqn:HI}) has $T_d$ symmetry, meaning that $\chi^N$ satisfies $T_d$ symmetry.
In the zero magnetic field limit\cite{Abrikosov1961KnightShift,Frigeri2004SpinSus,
Samokhin2005SpinSus}, we consider uniform order parameters in the superconducting phase, i.e. Eq.\ref{eq:pairing_gen} is zero for $\bsl{q}\neq 0$.
Such pairing has $O(3)$ symmetry\cite{yu2017Singlet-Quintetj=3/2SC}, implying that $\chi^S$ is also $T_d$ invariant.
For $T_d$ group,  $R_{ii'}$ belongs to $T_1$ irreducible representation.
As a result, $\chi^S_{ij}=\chi^S\delta_{ij}$ and $\chi^N_{ij}=\chi^N\delta_{ij}$ can be derived from Schur's lemma.\cite{tung1985group}
Thus, $\chi^S_{ij}$ and $\chi^N_{ij}$ are isotropic, which simplifies our calculations.

\begin{figure*}[t]
\includegraphics[width=2\columnwidth]{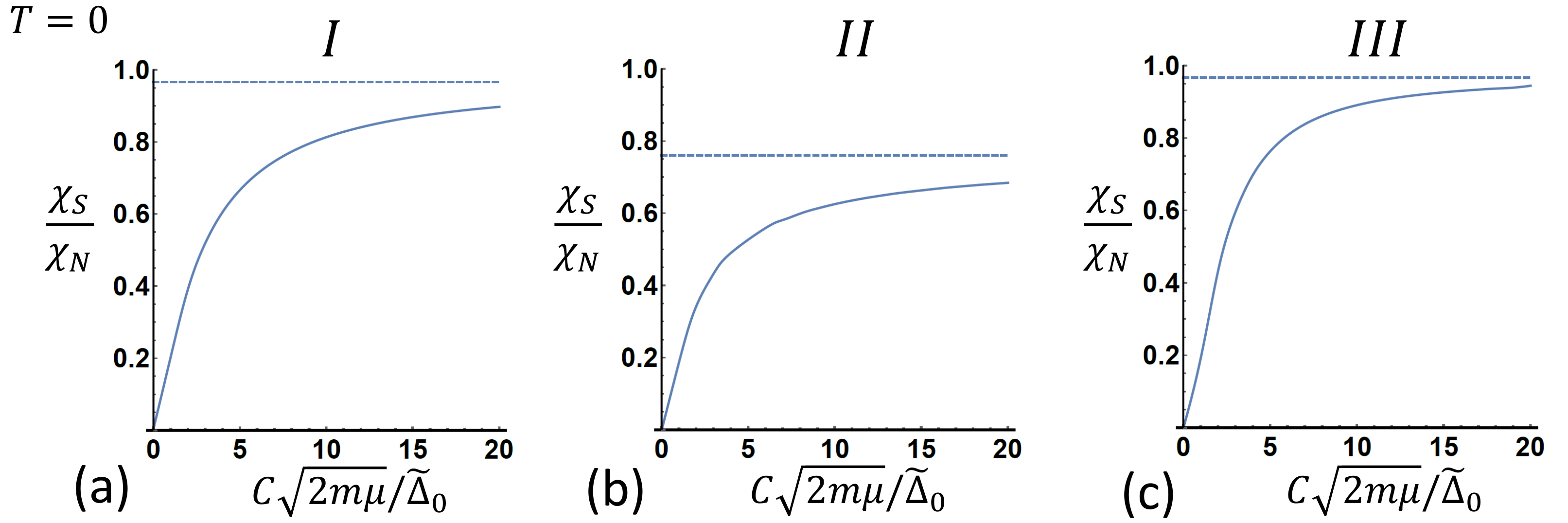}
\caption{
\label{fig:0Tspinsusp}
The solid lines in (a), (b) and (c) plot the zero-temperature $(T=0)$ spin susceptibility $\chi_S/\chi_N$  as a function of the ratio between ASOC and pairing $C\sqrt{2m\mu}/\tilde{\Delta}_0$ in regime I, II and III, respectively.
The dashed lines show the zero-temperature spin susceptibility at large ASOC limit($C\rightarrow \infty$), which is given by Eq.\ref{eq:Spinsus_largeASOC}.
$\widetilde{\Delta}_1/\widetilde{\Delta}_0=1.6$ and $c_2=2c_1$ are chosen for every graph, where $\widetilde{\Delta}_0=\text{sgn}(c_1)\Delta_0$ and $\widetilde{\Delta}_1=2m\mu a^2\Delta_1$.
$|2m|c_1=0.4$,$|2m|c_1=1.2$ and $|2m|c_1=0.6$ are chosen for (a),(b) and (c), respectively, and a finite momentum cut-off $\Lambda/\sqrt{2m\mu}=3$ is set for (c).
}
\end{figure*}

Following Ref.[\citen{Frigeri2004SpinSus}], the spin susceptibilities in superconducting phase and normal metal phase read
\begin{widetext}
\begin{eqnarray}
\label{eq:chi_S_sq}
&&\frac{\chi^S}{\chi^N}=1-\frac{\mathcal{V}N_0 }{\beta \chi^N} \sum_{\lambda,\omega_n}\int\frac{d\Omega}{4\pi} \theta(\widetilde{m}_{\lambda})\widetilde{m}_{\lambda}^{3/2}
\left\{\frac{m_{\lambda}^z(\hat{\bsl{k}})+\bar{m}_{\lambda}^z(\hat{\bsl{k}})}{2} \frac{\pi |d_{\lambda}(\bsl{k}_{F,\lambda})|^2}{(|d_{\lambda}(\bsl{k}_{F,\lambda})|^2+\omega_n^2)^{3/2}}\right.\nonumber\\
&&\left.+\frac{m_{\lambda}^z(\hat{\bsl{k}})-\bar{m}_{\lambda}^z(\hat{\bsl{k}})}{2} \frac{ \pi |d_{\lambda}(\bsl{k}_{F,\lambda})|^2}{ (|d_{\lambda}(\bsl{k}_{F,\lambda})|^2+\omega_n^2)^{1/2}(|d_{\lambda}(\bsl{k}_{F,\lambda})|^2+\alpha_{\lambda}^2(\hat{\bsl{k}})+\omega_n^2)}\right\}
\end{eqnarray}
\end{widetext}
and
\begin{eqnarray}
\label{eq:chi_N}
\chi^N=\mathcal{V}N_0 \sum_{\lambda}\int\frac{d\Omega}{4\pi} \theta(\widetilde{m}_{\lambda})\widetilde{m}_{\lambda}^{3/2} m^z_{\lambda}(\hat{\bsl{k}})\ ,
\end{eqnarray}
respectively.
Here $\beta=\frac{1}{k_B T}$, $k_B$ is the Bolzmann constant, $\omega_n=(2n+1)\pi/\beta$ is the fermionic Matsubara frequency, $N_0=\frac{4\pi}{(2\pi)^3}|m|\sqrt{2m\mu}$, $m^{\lambda}_{z}(\hat{\bsl{k}})=Tr[M_z^{\lambda}(\hat{\bsl{k}})M_z^{\lambda}(\hat{\bsl{k}})]$, $\bar{m}^{\lambda}_{z}(\hat{\bsl{k}})=Tr[M_z^{\lambda}(\hat{\bsl{k}})\hat{\bsl{p}}^{\lambda}(\hat{\bsl{k}})\cdot\bsl{\sigma}M_z^{\lambda}(\hat{\bsl{k}})\hat{\bsl{p}}^{\lambda}(\hat{\bsl{k}})\cdot\bsl{\sigma}]$, $\hat{\bsl{p}}^{\lambda}(\hat{\bsl{k}})=\bsl{p}^{\lambda}(\hat{\bsl{k}})/p^{\lambda}(\hat{\bsl{k}})$,
$p^{\lambda}(\hat{\bsl{k}})=|\bsl{p}^{\lambda}(\hat{\bsl{k}})|$, $d_{\lambda}(\bsl{k})= \frac{\Delta_0}{2}+\lambda\frac{\Delta_1 }{2} a^2 k^2 \text{sgn}(c_1) f_Q$,
$\Delta_{0,1}$ are uniform order parameters in singlet and quintet channels, respectively, $f_Q=(|c_1|Q_1^2+|c_2| Q_2^2)/Q_c$,
$\alpha_{\lambda}(\hat{\bsl{k}})=|C k_{F,\lambda}| p_{\lambda}(\hat{\bsl{k}})$, and the terms of order $1/(\beta \epsilon_c)$, $\alpha_{\lambda}  /\epsilon_c $, $|d_{\lambda}|  /\epsilon_c $ and $\epsilon_c/|\mu|$ are neglected.
(See Appendix.\ref{app:der_spin_sus} for more details.)
In the case where only one of the $\lambda=\pm$ bands is cut by the Fermi energy, $\bsl{M}^{\lambda}(\hat{\bsl{k}})=\mu_B\bsl{\sigma}$ and the system is isotropic,
Eq.\ref{eq:chi_S_sq} and Eq.\ref{eq:chi_N} would match the results in Ref.\citen{Frigeri2004SpinSus}.

In particular, we focus on the zero-temperature limit of  Eq.\ref{eq:chi_S_sq}.
One should be careful that $T\rightarrow 0$ limit and $d_{\lambda}\rightarrow 0$ limit are not exchangeable, and $d_{\lambda}\rightarrow 0$ limit, if needed, should be performed before $T\rightarrow 0$ limit since the later is not physically achievable.
Although TNLS indicates $d_{\lambda}$ can be zero along some lines on the Fermi surface, such lines can be neglected in Eq.\ref{eq:chi_S_sq} since they do not cause any divergence and have zero measure in the surface integration.
After summing over $\omega_n$, Eq.\ref{eq:chi_S_sq} at zero temperature reads
\begin{eqnarray}
\label{eq:chi_S_T0}
&&\left. \frac{\chi^S}{\chi^N}\right|_{T\rightarrow 0}=1-\frac{\mathcal{V}N_0 }{\chi^N}\sum_{\lambda}\int\frac{d\Omega}{4\pi} \theta(\widetilde{m}_{\lambda})\widetilde{m}_{\lambda}^{3/2}\\
&&\left[\frac{m_{\lambda}^z(\hat{\bsl{k}})+\bar{m}_{\lambda}^z(\hat{\bsl{k}})}{2}+\frac{m_{\lambda}^z(\hat{\bsl{k}})-\bar{m}_{\lambda}^z(\hat{\bsl{k}})}{2}\mathcal{J}(\frac{|d_{\lambda}(\bsl{k}_{F,\lambda})|}{\alpha_{\lambda}(\hat{\bsl{k}})})\right]\nonumber\ ,
\end{eqnarray}
where $\mathcal{J}(x)=\frac{x^2}{\sqrt{1+x^2}}\ln(\frac{1+\sqrt{1+x^2}}{x})$.
According to Eq.\ref{eq:chi_S_T0}, a non-vanishing $\left. \frac{\chi^S}{\chi^N}\right|_{T\rightarrow 0}$ comes from the ASOC term\cite{Frigeri2004SpinSus}.
In the limit of zero ASOC, i.e. $C\rightarrow 0$ or equivalently $\alpha_{\lambda}\rightarrow 0$, we find $\left. \frac{\chi^S}{\chi^N}\right|_{T\rightarrow 0}=0$ using $\mathcal{J}(x\rightarrow +\infty)=1$.
On the other hand, if ASOC is much larger than the superconducting gap on the Fermi surface $\alpha_{\lambda}\gg d_{\lambda}$, Eq.\ref{eq:chi_S_T0} is simplified as
\begin{equation}
\label{eq:Spinsus_largeASOC}
\left. \frac{\chi^S}{\chi^N}\right|_{T\rightarrow 0}=1-\frac{\mathcal{V}N_0 }{\chi^N}\sum_{\lambda}\int\frac{d\Omega}{4\pi} \theta(\widetilde{m}_{\lambda})\widetilde{m}_{\lambda}^{3/2}\frac{m_{\lambda}^z(\hat{\bsl{k}})+\bar{m}_{\lambda}^z(\hat{\bsl{k}})}{2}
\end{equation}
using $\mathcal{J}(x\rightarrow 0)=0$. This expression is generally non-zero.
Fig.\ref{fig:0Tspinsusp} a, b, and c show the behavior of $\left. \frac{\chi^S}{\chi^N}\right|_{T\rightarrow 0}$ as a function of the ratio between ASOC and pairing amplitude in regime I, II and III, respectively.
We find the $\left. \frac{\chi^S}{\chi^N}\right|_{T\rightarrow 0}$ drops to zero for zero ASOC and approaches to the limit set by Eq.\ref{eq:Spinsus_largeASOC} (dashed lines in Fig.\ref{fig:0Tspinsusp}) when ASOC increases.
We also can see that $\left. \frac{\chi^S}{\chi^N}\right|_{T\rightarrow 0}$ is not sensitive to SSOC, and Eq.\ref{eq:Spinsus_largeASOC} gives a slightly smaller value in regime II than those in regime I and III.

Based on this calculation, we arrive at the following conclusions. (1) Unlike the singlet-triplet mixing with a non-zero $\left. \frac{\chi^S}{\chi^N}\right|_{T\rightarrow 0}$, zero-temperature spin susceptibility can be zero for singlet-quintet mixing. This is because the singlet-triplet mixing is from ASOC and the singlet-quintet mixing comes from SSOC, while $\left. \frac{\chi^S}{\chi^N}\right|_{T\rightarrow 0}$ is only sensitive to ASOC. This indicates that in some centrosymemtric SCs with $j=3/2$ (e.g. anti-perovskite materials\cite{Kawakami2018j=3/2electrons}), even if one measures a vanishing zero-temperature spin susceptibility, the possibility of singlet-quintet mixing {\it cannot} be excluded. (2) In half-Heusler SCs, such as YPtBi, since the energy scale of ASOC near the Fermi surface $(\sim 4meV)$ is much larger than the gap function of the similar order as $k_B T_c\sim 0.06 meV$, a non-zero $\left. \frac{\chi^S}{\chi^N}\right|_{T\rightarrow 0}$ is expected for the singlet-quintet mixed pairing. We notice that the situation here is similar to the case of other non-centrosymmetric SCs with ``spin-1/2" electrons\cite{Frigeri2004SpinSus}.



\section{Upper Critical Field}
\label{sec:mag_protes}


In this section, we will study the upper critical field $B_{c,2}$, at which the superconductivity is destroyed by the external magnetic field \cite{tinkham1996introductionSC}, in our model.
The upper critical field can be obtained by solving linearized gap equation with a non-zero magnetic field.
The effect of magnetic field is taken into account through the orbital term and the Zeeman term, as discussed in Eq.(\ref{eq:XIK_1B}). \cite{Samokhin2004magnetic}
Although the magnetic field is not infinitesimal, the projection of Zeeman term onto $\Gamma_8$ bands in Eq.\ref{eq:H_Z_G8} and $\xi_{\pm}$ bands in Eq.\ref{eq:XI_k} can still be justified.
The reason is that in half Heusler materials, the energy scale of Zeeman term ($\mu_B B\sim 0.1 meV$ for $B\sim 2 T$ as the typical zero-temperature upper critical field\cite{Nakajima2015RPdBiSC,Butch2011SCYPtBi,Pan2013ErPdBiSC,
Xu2014LuPdBiSC,Nikitin2015HoPdBiSC,Tafti2013LuPtBiSC,
Bay2012SCYPtBi}) is much smaller than the energy gap between $\Gamma_8$ and $\Gamma_7$ bands ($|E_{\Gamma_8}-E_{\Gamma_7}|\sim 1 eV$\cite{lin2010half,kim2018YPtBiSCj=3/2,Xu2014LuPdBiSC,
Yang2017HHTP}) and the energy scale of SSOC near the Fermi surface ($2 Q_c k_F^2\sim 20 meV$ for YPtBi\cite{kim2018YPtBiSCj=3/2,Brydon2016j=3/2SC, Savary2017j=3/2SC}).
For only keeping the leading order of $B$ in Eq.\ref{eq:XIK_1B}, we need to focus on the temperature $T$ close to the zero-field critical temperature $T_c$ for which $B_{c,2}$ is small enough.
In addition, we neglect the ASOC, i.e. $C=0$, for simplicity.
In this case, we have the effective Hamiltonian for each band
\begin{equation}
\label{eq:EK_1B}
 E^{\pm}(\bsl{K},\bsl{B})=\xi_{\pm}(\bsl{k})+\bsl{B}\cdot\bsl{M}^{\pm}(\hat{\bsl{k}})+\frac{e}{\hbar}\bsl{\nabla}_{\bsl{k}}\xi_{\pm}(\bsl{k})\cdot \bsl{A}(i\bsl{\nabla}_{\bsl{k}})
\end{equation}
which is just Eq.\ref{eq:XIK_1B} with $C=0$.
As a result, the corresponding Green function $G^{\pm}(\bsl{r}_1,\bsl{r}_2,\omega_n)$ for each band reads
\begin{equation}
\label{def:green_func}
G^{\pm}(\bsl{r}_1,\bsl{r}_2,\omega_n)=e^{-i\frac{e}{\hbar}\bsl{r}_1\cdot\bsl{A}(\bsl{r}_2)}\widetilde{G}^{\lambda}(\bsl{r}_1-\bsl{r}_2,\omega_n)\ ,
\end{equation}
where $\widetilde{G}^{\lambda}(\bsl{r},\omega_n)=\frac{1}{\mathcal{V}}\sum_{\bsl{k}}e^{i\bsl{k}\cdot\bsl{r}}\widetilde{G}^{\lambda}(\bsl{k},\omega_n)$ and
\begin{equation}
\label{eq:tG_Form}
\widetilde{G}^{\pm}(\bsl{k},\omega_n)=\frac{1}{i\omega_n-\xi_{\pm}(\bsl{k})}+\frac{\bsl{B}\cdot\bsl{M}^{\pm}(\bsl{k})}{(i\omega_n-\xi_{\pm}(\bsl{k}))^2}\ .
\end{equation}
It is clear that, the orbital effect only appears in the phase factor of $G^{\pm}(\bsl{r}_1,\bsl{r}_2,\omega)$ to the first order of $B$.
(See Appendix.\ref{app.G_B} for details.)

As mentioned above, the upper critical field is solved via linearized gap equation\cite{tinkham1996introductionSC}, which is derived from the superconducting Ginzburg-Landau free energy $F_{SC}$ to the second order of the order parameter:
\begin{eqnarray}
\label{eq:F_r_final}
F_{SC}&=&-\sum_{a_1}\int d^3 r \frac{1}{2 \widetilde{V}_{a_1}}|\widetilde{\Delta}_{a_1}(\bsl{r})|^2-\frac{1}{2}\sum_{a_1,a_2}\int d^3 r\nonumber\\
&&\widetilde{\Delta}_{a_1}^*(\bsl{r})(\widetilde{K}_0^{a_1 a_2}+\widetilde{K}_1^{a_1 a_2}\bsl{D}^2)\widetilde{\Delta}_{a_2}(\bsl{r})\ ,
\end{eqnarray}
where $a_1,a_2=0,1$, $\widetilde{\Delta}_0=\text{sgn}(c_1)\Delta_0$, $\widetilde{\Delta}_1=2m\mu a^2\Delta_1$, $\widetilde{V}_0=V_0$, $\widetilde{V}_1=(2m\mu a^2)^2 V_1$ ,
\begin{equation}
\label{eq:K0t}
\widetilde{K}_0=\frac{x N_0}{2}
\left(
\begin{array}{cc}
y_1 & y_2\\
y_2 & y_3\\
\end{array}
\right) \ ,
\end{equation}
\begin{equation}
\label{eq:K1t}
\widetilde{K}_1=-\frac{N_0}{2}\frac{\beta^2\mu}{2m}
\left(
\begin{array}{cc}
z_1 & z_2\\
z_2 & z_3\\
\end{array}
\right)   \ ,
\end{equation}
$x=\ln (2 e^{\bar{\gamma}}\beta\epsilon_c/\pi)$ with $\bar{\gamma}=0.577...$ the Euler's constant,
$\bsl{D}=-i\bsl{\nabla}_{\bsl{r}}+\frac{e}{\hbar}(\bsl{B}\times\bsl{r})$ and the expressions of $y_{1,2,3}$ and $z_{1,2,3}$ are shown in Appendix.\ref{app:conv_expn}.
Here $1/(\beta\epsilon_c)\ll 1$, $\epsilon_c/2Q_c k_F^2\ll 1$ and $\epsilon_c/|\mu|\ll 1$ are assumed.
$F_{SC}$ only contains the orbital effect because the Zeeman term only appears as the second order of $B$ and thus is neglected in $F_{SC}$ .
In addition, $F_{SC}$ does not depend on the direction of $\bsl{B}$ since $\bsl{D}^2$ is isotropic, meaning that the upper critical field is isotropic near the critical temperature.
(See Appendix.\ref{app:der_Fr} for more details.)
The resulted linearized gap equation reads
\begin{equation}
\label{eq:lge_final_mat}
\frac{1}{x}\left(
\begin{array}{c}
\widetilde{\Delta}_0 \\
\widetilde{\Delta}_1 \\
\end{array}
\right)
=
\frac{1}{2}\left(
\begin{array}{cc}
\lambda_0 \widetilde{y}_1 & \lambda_0 \widetilde{y}_{2}\\
\lambda_1 \widetilde{y}_{2} & \lambda_1 \widetilde{y}_3\\
\end{array}
\right)
\left(
\begin{array}{c}
\widetilde{\Delta}_0 \\
\widetilde{\Delta}_1 \\
\end{array}
\right)\ ,
\end{equation}
where $\lambda_{0,1}=-N_0\widetilde{V}_{0,1}$, $\widetilde{y}_1=y_1-\frac{\beta^2\mu^2}{2m\mu x} z_1 l^2$, $\widetilde{y}_{2}=y_2-\frac{\beta^2\mu^2}{2m\mu x} z_2 l^2$, $\widetilde{y}_3=y_3-\frac{\beta^2\mu^2}{2m\mu x} z_3 l^2$ and $l^2=\frac{4eB}{\hbar}(n+\frac{1}{2})+k_3^2$ is the eigenvalue of $\bsl{D}^2$ with $n\geq 0$ and $k_3$ being the component of the momentum in the direction of magnetic field.
The upper critical field $B_{c,2}$ can be obtained by solving the above equation with a fixing temperature $T$ below $T_c$ and the solution gives
\begin{equation}
\label{eq:Bc2_T}
\frac{B_{c,2}}{B_0}=\frac{\frac{T}{T_{c}}-1}{\alpha(\beta_{c}\epsilon_c)^2  x_{c}}\ ,
\end{equation}
where $B_0=\frac{8\hbar m\mu \epsilon_c^2}{2 e \mu^2 }$, $T_{c}$ is given by\cite{yu2017Singlet-Quintetj=3/2SC}:
\begin{equation}
\label{Expression_Tc}
\ln\left(\frac{T_c}{T_0}\right)=\frac{-4}{\sqrt{(\lambda_0 y_{1}-\lambda_1 y_3)^2+4 \lambda_0 \lambda_1 y_{2}^2}+\lambda_0 y_{1}+\lambda_1 y_{3}} ,
\end{equation}
$T_0=\frac{2e^{\bar{\gamma}}\epsilon_c}{\pi k_B}$,
$\beta_{c}=1/(k_B T_{c})$ , $x_{c}=\ln(\frac{2 e^{\bar{\gamma}} \beta_{c}\epsilon_c}{\pi})=\ln(\frac{T_0}{T_c})$, and
\begin{eqnarray}
&&\alpha =-z_1 \lambda_0-z_3 \lambda_1\nonumber\\
&&+\frac{(-\lambda_0 z_1+\lambda_1 z_3) (y_1 \lambda_0-y_3 \lambda_1)-4 y_2 z_2 \lambda_0 \lambda_1}{\sqrt{(y_1 \lambda_0 - y_3 \lambda_1)^2+4 y_2^2 \lambda_0 \lambda_1}}\ .
\end{eqnarray}
As a result, the slope $-d B_{c,2}/d T$ at the zero-field critical temperature has the form
\begin{equation}
\label{eq:dBdTr}
-\frac{d B_{c,2}/B_0}{d T/T_0}
=\frac{T_0}{(-\alpha) (\beta_c \epsilon_c)^2 x_c T_c}
=\frac{1}{-\alpha}\left(\frac{2e^{\bar{\gamma}}}{\pi}\right)^2\frac{T_c}{T_0}\frac{1}{x_c}
\ .
\end{equation}
Below we label $X(\alpha)=\frac{1}{-\alpha}\left(\frac{2e^{\bar{\gamma}}}{\pi}\right)^2$ for short.
The temperature dependence of $B_{c,2}$ for the pure singlet(quintet) channel can be obtained by setting $\lambda_1=0$($\lambda_0=0$) in Eq. (\ref{eq:Bc2_T}). (See Appendix.\ref{app:Tc_Bc} for more details.)

\begin{figure}[t]
\includegraphics[width=\columnwidth]{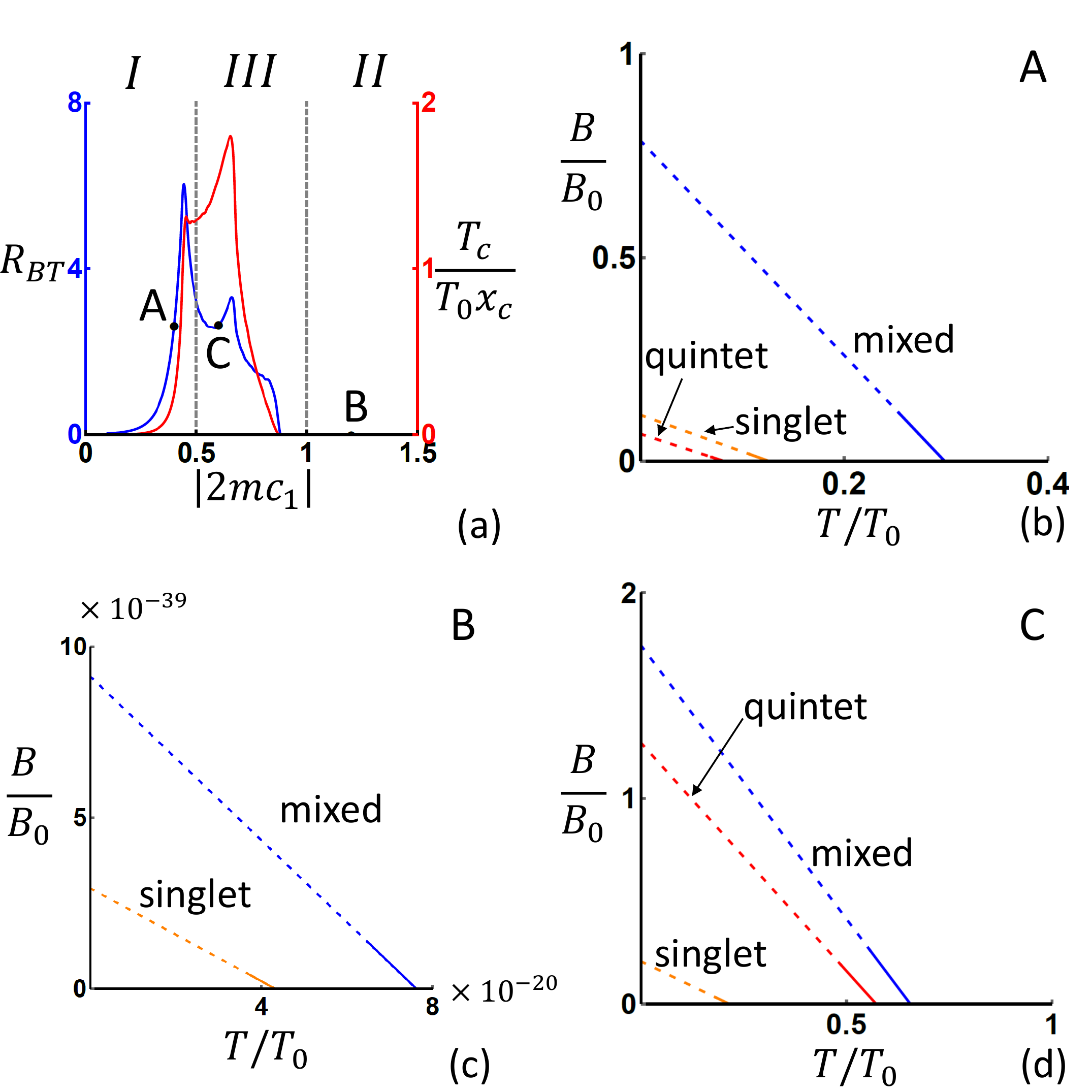}
\caption{
\label{fig:upper_critical_field}
(a) shows the slope(blue) $R_{BT}=-d (B_{c,2}/B_0) / d (T/T_0)$ at the zero-field critical temperature ($T_c$) and the $T_c$ factor(red) of the slope as functions of SSOC $|2 m c_1|$.
 I, II and III stand for the three regimes.
(b), (c) and (d) depict the upper critical field $B_{c,2}$ at various temperature for point A, B and C in (a), receptively.
In (b), (c) and (d),  blue, orange and red lines stand for mixed, singlet and quintet channels, respectively, and the dashed parts are not precise since Eq.\ref{eq:Bc2_T} is only suitable for $T$ close to $T_c$.
The missing quintet channel in (c) is because it is too small.
$c_2=2c_1$ is chosen for all figures,
$|2m c_1|=0.4$ for (b), $|2m c_1|=1.2$ for (c), $|2m c_1|=0.6$ and  $\Lambda=3\sqrt{2m\mu}$ for (d).
The interaction parameter choices are $\lambda_1=0.1\lambda_0=0.02$ for (a) and the mixed channel in (b), (c) and (d).
The singlet(quintet) channel in (b), (c) and (d) is determined by setting $\lambda_0$($\lambda_1$) to be the same as the mixed channel and  $\lambda_1$($\lambda_0$) to be zero.
}
\end{figure}


The blue line of Fig.\ref{fig:upper_critical_field}a shows the slope $R_{BT}=-d B_{c,2}/d T$ at $T=T_c$ as a function of SSOC strength.
The slope in the regime I and III is significantly larger than that in the regime II, which can be attributed to the behavior of the function $\frac{T_c}{T_0x_c}$, as shown by the red line of Fig.\ref{fig:upper_critical_field}a and in Eq.\ref{eq:dBdTr}.
With increasing $|2mc_1|$, we find a peak of the slope $R_{BT}$ appears in the regime I close to the I-III boundary, and then the slope drops rapidly to a dip around the point $C$ in Fig.\ref{fig:upper_critical_field}a. Another small peak is found in the regime III and then the slope drops to almost zero due to the extremely small $T_c$ in the regime II.
The behavior of the slope $R_{BT}$ is mainly determined by the function $\frac{T_c}{T_0x_c}$ (see the red line in Fig.\ref{fig:upper_critical_field}a) except when $|2mc_1|$ is tuned to the boundary between the regime I and III. 
The difference there is attributed to the rapid decrease of the factor $X(\alpha)$ when increasing $|2mc_1|$ towards the I-III boundary, as shown in Fig.\ref{fig:alpha_factor} of Appendix.\ref{app:Tc_Bc}.
Fig.\ref{fig:upper_critical_field}b,c and d show the upper critical field of the pairing mixed state (blue lines) as a function of temperatures (close to $T_c$) for typical parameters in regime I, II, III, respectively.
The upper critical fields for pure singlet pairing (orange lines) and quintet pairing (red lines) are also shown in these figures for comparison. It is clear that the mixing can increase the upper critical fields in Eq.\ref{eq:lge_final_mat}.
Moreover, we can see $B_{c,2}$ of the quintet channel is larger than that of the singlet channel in regime III as shown in Fig.\ref{fig:upper_critical_field}d while the opposite happens in regimes I and II (Fig.\ref{fig:upper_critical_field}b,c),
which coincides with the fact that the quintet channel can be dominant around regime III.

In conclusion, the above results have shown that the upper critical field $B_{c,2}$ close to the zero-field critical temperature $T_c$ is isotropic and can be enhanced by the singlet-quintet mixing.
The slope $d B_{c,2}/d T$ at the zero-field critical temperature is mainly determined by the zero-field critical temperature $T_c$. The slope is much larger in regimes I and III than that in regime II mainly due to its $T_c$ dependence, and reaches its maximum around the boundary between I and III as a result of the interplay between the $T_c$ and $\alpha$ dependence in Eq.\ref{eq:dBdTr}.

\section{Effect of Random Non-magnetic Disorder}
\label{disorder}

\begin{figure}[t]
\includegraphics[width=\columnwidth]{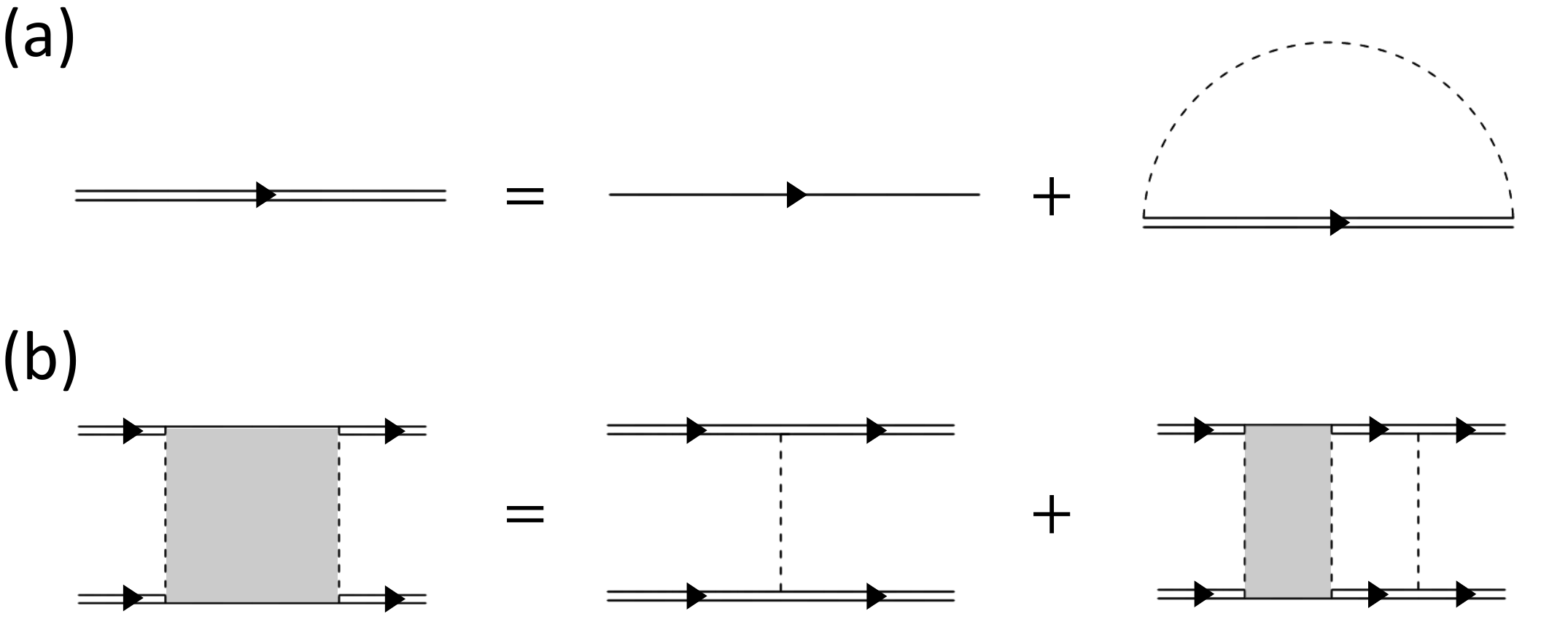}
\caption{
\label{fig:feyndia_dis}
Feynman diagrams (a) and (b) show the equations of the exact propagator and vertex via the replica trick with the Born approximation, respectively.
The dashed line stands for the disorder, and the solid single line and the solid double lines are the bare and exact fermionic propagators, respectively.
}
\end{figure}

\begin{figure*}[t]
\includegraphics[width=2\columnwidth]{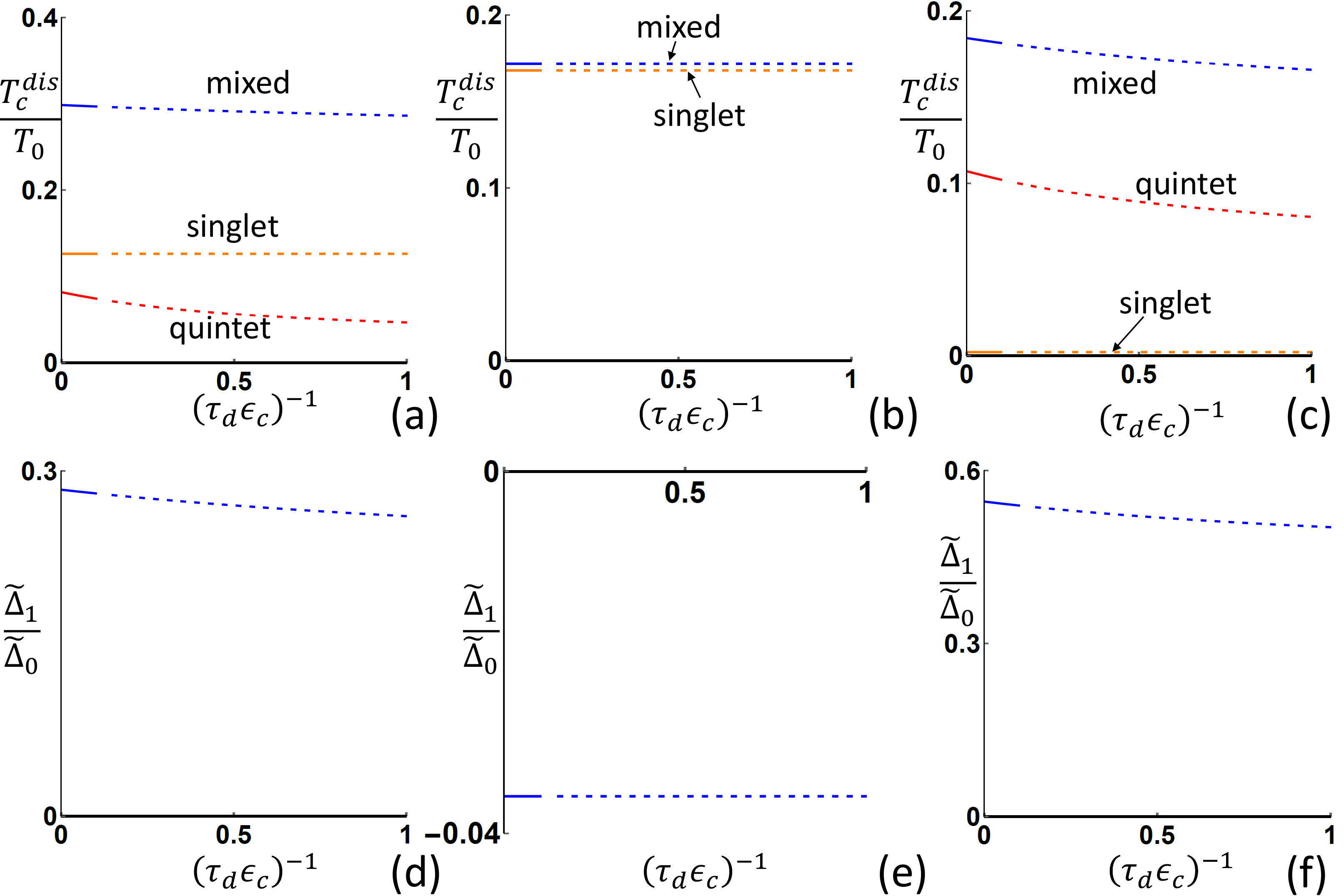}
\caption{\label{fig:T_c_pairing_ratio_dis} The impacts of weak non-magnetic random disorder on the critical temperature $T_c^{dis}$ in singlet(orange), quintet(red) and mixed(blue) channels in regimes I, II and  III are shown in a,b,c, respectively, and the pairing ratios as functions of disorder strength in regime I, II and III are shown in d,e,f, respectively. $(\tau_d\epsilon_c)^{-1}$ measures the strength of disorder with larger $(\tau_d\epsilon_c)^{-1}$ meaning stronger disorder.
Due to the limitation of weak disorder scattering ($(\tau_d\epsilon_c)^{-1}\ll 1$) for Eq.\ref{eq:lge_final_dis}, we use dashed line for a relative large $(\tau_d\epsilon_c)^{-1}$.
$\widetilde{\Delta}_1/\widetilde{\Delta}_0$ stands for the pairing ratio between the quintet and singlet channel.
The band structure parameter choices are  $c_2=2 c_1$ for all figures,
$|2m c_1|=0.4$ for a and d,$|2m c_1|=1.2$ for b and e, $|2m c_1|=0.6$ and  $\Lambda=3\sqrt{2m\mu}$ for c and f.
$\lambda_1=0.1\lambda_0$ is chosen for d,e,f and the mixed channel of a,b,c, with $\lambda_0=0.2$ for d and the mixed channel of a, $\lambda_0=5$ for e and the mixed channel of b and $\lambda_0=0.05$ for f and the mixed channel of c.
The missing quintet channel in b is because it is too small.
The interaction parameters for the pure singlet(quintet) channel in a,b,c  are given by choosing $\lambda_1$($\lambda_0$) to be zero while $\lambda_0$($\lambda_1$) is the same as the corresponding mixed channel.
}
\end{figure*}

In this section, we study the effect of weak random non-magnetic disorder on the singlet-quintet mixed SC. The non-magnetic disorder is included in the Hamiltonian as
\begin{equation}
\label{eq:H_dis}
H_{dis}=\int d^3 r V(\bsl{r}) c^{\dagger}_{\bsl{r}} c_{\bsl{r}}\ ,
\end{equation}
where $c^{\dagger}_{\bsl{r}}$ is the Fourier transformation of $c^{\dagger}_{\bsl{k}}$
and $V(\bsl{r})$ is the random potential describing the disorder scattering.
The probability measure of the disorder configuration is chosen as
\begin{equation}
\label{eq:PV_dis}
P[V]=\exp\left[-\frac{1}{2\gamma^2_d} \int d^3 r V^2 (\bsl{r})\right]\ ,
\end{equation}
and thereby the spatial correlation of $V(\bsl{r})$ is just the delta function $
\left\langle V(\bsl{r})V(\bsl{r}')\right\rangle_{dis}=\gamma^2_d \delta(\bsl{r}-\bsl{r}')$
,
where $\gamma^2_d$ measures the strength of the disorder with larger $\gamma^2_d$ meaning stronger disorder.
In order to carry out the disorder average,
we use the Replica trick\cite{altland2010condensed} which, in our case, is equivalent to eliminating all fermionic loops in Feynman diagrams, as elaborated in Appendix.\ref{app:replica}.
We assume the disorder is weak: $\frac{\gamma^2_d N_F}{|\mu|}\ll 1$ with $\mu$ the chemical potential and $N_F=N_0 y_1$ the density of state at the Fermi energy without spin index.
In this case, we consider the self-energy correction (Fig.\ref{fig:feyndia_dis}a) and vertex correction (Fig.\ref{fig:feyndia_dis}b) with the Born approximation, where all Feymann diagrams with crossed disorder lines are neglected since those terms have higher orders of $\frac{\gamma^2_d N_F}{|\mu|}$.\cite{altland2010condensed} (See Appendix.\ref{app:replica} for the definition of disorder lines.) As a result, the linearized gap equation in presence of the disorder reads (See Appendix.\ref{app:der_lge_dis} for details)
\begin{equation}
\label{eq:lge_final_dis}
\left(
\begin{array}{c}
\tilde{\Delta}_{0}\\
\tilde{\Delta}_{1}
\end{array}
\right)
=
x\left(
\begin{array}{cc}
\frac{\lambda_{0}}{2}y_1 & \frac{\lambda_{0}}{2}y_2\\
\frac{\lambda_{1}}{2}y_2& \frac{\lambda_{1}}{2}y_3 b_1
\end{array}
\right)
\left(
\begin{array}{c}
\tilde{\Delta}_{0}\\
\tilde{\Delta}_{1}
\end{array}
\right)\ ,
\end{equation}
where $\epsilon_c/|\mu|\ll 1$, $\epsilon_c/(2Q_c k_F^2)\ll 1$, $\beta\epsilon_c\gg 1$ and $1/(\epsilon_c \tau_d) \ll 1$ are used. The disorder contribution only appears in the function $b_1$, which is given by
\begin{equation}
b_1=1+\frac{\mathcal{F}(\frac{\beta}{4\pi\tau_d})}{x}(\frac{y_2^2}{y_1 y_3}-1)
\end{equation}
with $\mathcal{F}(\frac{\beta}{4\pi\tau_d})=\Psi^{(0)}(\frac{\beta}{4\pi \tau_d}+\frac{1}{2})-\Psi^{(0)}(\frac{1}{2})$ which is defined in Ref.[\citen{Mineev2007DisSC}], $\Psi^{(0)}(...)$ is the digamma function, and $1/\tau_d=\gamma^2_d \pi N_F$.
The critical temperature $T$ can be solved from Eq.\ref{eq:lge_final_dis} in the presence of disorder for the mixed state as
\begin{equation}
\label{eq:T_mixed_dis}
\text{ln}(\frac{T}{T_0})=\frac{-4}{\sqrt{(\lambda_0 y_{1}-\lambda_1 y_3 b_1)^2+4 \lambda_0 \lambda_1 y_{2}^2}+\lambda_0 y_{1}+\lambda_1 y_{3} b_1}\ .
\end{equation}
The critical temperature expression for the pure singlet(quintet) channel can be given by setting $\lambda_1$($\lambda_0$) to be zero, which gives
\begin{equation}
\label{eq:Ts_dis}
\text{ln}(\frac{T_s}{T_0})=\frac{-2}{\lambda_0 y_{1}}\ ,
\end{equation}
and
\begin{equation}
\label{eq:Tq_dis}
\text{ln}(\frac{T_q}{T_0})=\frac{-2}{\lambda_1 y_{3} b_1}\ .
\end{equation}

Eqs. (\ref{eq:lge_final_dis})-(\ref{eq:Tq_dis}) are the main results of this section and form the basis for our analysis of the disorder effect of singlet-quintet mixing pairing. 
The disorder scattering is controlled by a single function $b_1$ in the linearized gap equation (\ref{eq:lge_final_dis}) and it is found (see Appendix.\ref{app:der_F_lge_final}) that $0<b_1\leq 1$ with $b_1=1$ only occuring either in the clean limit ($1/\tau_d=0$) or for the isotropic case in the regime II. By inspecting Eqs. (\ref{eq:lge_final_dis})-(\ref{eq:Tq_dis}), we can draw the following conclusions. (1) The pure s-wave singlet channel is not influenced by the non-magnetic disorder because we neglect the inter-band scattering. (The s-wave singlet channel can still be influenced by the disorder if one includes the inter-band scattering\cite{Onari2009VioATSC}, typically of higher orders of $\epsilon_c/(2Q_c k_F^2)$ here.) This conclusion is consistent with and required by the Anderson theorem \cite{Anderson1959AT}.
(2) An interesting observation for Eq.\ref{eq:lge_final_dis} is that the disorder scattering ($b_1$ function) only appears in the d-wave quintet channel ($y_3$ term) due to the momentum dependence of d-wave function, but the coupling between singlet and quintet channels is independent of the disorder scattering.
By examining the derivation in Appendix.\ref{app:der_F_lge_final}, we find that such behavior originates from the cancellation between the self-energy correction and the vertex correction for the singlet-quintet coupling term, which is similar to the stable s-wave singlet pairing.
Therefore, our calculation suggests that the cancellation, which appears for the s-wave singlet pairing, also works for the singlet-quintet coupling term, at least at the level of Born approximation.
As shown in the following, such cancellation has a substantial influence on the pairing form in the disordered SCs. 
According to Eq.\ref{eq:lge_final_dis}, if there is no mixing term (the $y_2=0$ case), the quintet pairing is completely controlled by the $y_3b_1$ term and a small value of $b_1$ from disorder effect will greatly suppressed the quintet pairing.
In contrast, for a large $y_2$ term, due to its independence of disorder scattering, a significant quintet pairing channel can still be induced through the mixing effect even if the value of $b_1$ is small. Therefore, the mixing effect stabilizes the quintet pairing channel against the weak non-magnetic disorder scattering.

We further plot the calculated critical temperature with disorder $T_c^{dis}$ as a function of disorder scattering strength in Fig.\ref{fig:T_c_pairing_ratio_dis}a,b,c for singlet-quintet mixed pairing (blue lines), pure singlet pairing (orange lines) and pure quintet pairing (red lines) in the regimes I, II and III, respectively. We find the $T_c^{dis}$ for the pure singlet pairing is always independent of disorder scattering, as expected, while the $T_c^{dis}$ for the pure quintet pairing and mixed pairing decays with increasing the disorder strength $1/\tau_d$. The small decay magnitude is due to the limitation of the approximation for the weak scattering potential $\frac{1}{\tau_d\epsilon_c}\ll 1$ that is used in our theory.


Beside the critical temperatures, the expression of the pairing ratio can also be solved from Eq.\ref{eq:lge_final_dis} and reads
\begin{equation}
\label{Delta_ratio_dis}
\frac{\widetilde{\Delta}_1}{\widetilde{\Delta}_0}=\frac{2}{x \lambda_0 y_2}-\frac{y_1}{y_2}\ .
\end{equation}
Since $x$, which depends on the critical temperature, increases as $1/\tau_d$ increases, the pairing ratio $\widetilde{\Delta}_1/\widetilde{\Delta}_0$ would generally decrease if the disorder strength $1/\tau_d$ increases, with the exception of the isotropic system in regime II.
The decreasing pairing ratio $\widetilde{\Delta}_1/\widetilde{\Delta}_0$ is shown in Fig.\ref{fig:T_c_pairing_ratio_dis}d,e,f for regime I, II and III, respectively.


\section{Conclusion}
\label{sum}
In this work, we studied the zero-temperature spin susceptibility, the upper critical field near the zero-field critical temperature $T_c$ and the non-magnetic disorder scattering of the SCs with $j=\frac{3}{2}$ fermions in the presence of the mixing between s-wave singlet and isotropic d-wave quintet channels.
Our results show that the spin susceptibility is isotropic due to the $T_d$ group  symmetry and zero(non-zero) at zero temperature without(with) ASOC.
As a result,the zero-temperature spin susceptibility given by the singlet-quintet mixing is zero in centrosymmetric SCs, e.g. anti-perovskite materials\cite{Kawakami2018j=3/2electrons},
but non-zero in non-centrosymmetric SC YPtBi due to the large energy scale of ASOC near the Fermi surface $(\sim 4meV)$ compared with the gap function ($k_B T_c\sim 0.06 meV$).
The spin susceptibility can be measured in the NMR-Knight shift experiment.\cite{KnightShift1949,Reif1957KnightShift}
Near $T_c$ and without ASOC, it is found that the upper critical field is isotropic and enhanced by the pairing mixing.
The slope $-d B_{c,2} / d T$ at $T_c$ varies with the SSOC strength, and it is largest in the intermediate region between regime I and III and smallest in regime II.
Finally, our results on the random non-magnetic disorder effect demonstrated that the s-wave singlet channel as well as the singlet-quintet coupling in the linearized gap equation are not influenced by the weak disorder within the Born approximation, if neglecting the interband scattering. 
This suggests that the singlet-quintet mixing, as well as the nodal-line superconductivity, found in Ref.[\onlinecite{yu2017Singlet-Quintetj=3/2SC}] will be stable against the weak non-magnetic disorder scattering in real materials.

\section{Acknowledgment}
J.Y thanks Yang Ge, Rui-Xing Zhang, Jian-Xiao Zhang and Tongzhou Zhao for helpful discussions. C.X.L and J.Y acknowledge the support from the Office
of Naval Research (Grant No. N00014-15-1-2675 and renewal No. N00014-18-1-2793)
and the U.S. Department of Energy (DOE), Office of Science, Basic Energy Sciences (BES) under award No. DE-SC0019064.

\bibliography{Superconductivity}

\begin{appendices}
\section{Convention and Expressions}
\label{app:conv_expn}
The Fourier transformation of creation operators in the continuous limit reads
\begin{equation}
c^{\dagger}_{\bsl{r}}=\frac{1}{\sqrt{\mathcal{V}}}\sum_{\bsl{k}}e^{-i\bsl{k}\cdot\bsl{r}}c^{\dagger}_{\bsl{k}}\ ,
\end{equation}
where $\mathcal{V}$ is the total volume.
The Fourier transformation of corresponding Grassmann field in the continuous limit reads
\begin{equation}
\label{eq:Fourier_field}
\bar{c}_{\tau,\bsl{r}}=\frac{1}{\sqrt{\beta\mathcal{V}}}\sum_{\omega_n,\bsl{k}}e^{i\omega_n \tau-i\bsl{k}\cdot\bsl{r}}\bar{c}_{\bsl{k},\omega_n}\ ,
\end{equation}
where $\beta=1/(k_B T)$.

The five d-orbital cubic harmonics read \cite{Murakami2004SU2}
\begin{equation}
\left\{
\begin{array}{l}
g_{\bsl{k},1}=\sqrt{3} k_y k_z\\
g_{\bsl{k},2}=\sqrt{3} k_z k_x\\
g_{\bsl{k},3}=\sqrt{3} k_x k_y\\
g_{\bsl{k},4}=\frac{\sqrt{3}}{2} (k_x^2-k_y^2)\\
g_{\bsl{k},5}=\frac{1}{2}(2 k_z^2-k_x^2-k_y^2)\\
\end{array}
\right..
\end{equation}

The $j=\frac{3}{2}$ angular momentum matrices are \cite{Winkler2003SOC}
\begin{equation}
J_x=\left(
\begin{array}{cccc}
 0 & \frac{\sqrt{3}}{2} & 0 & 0 \\
 \frac{\sqrt{3}}{2} & 0 & 1 & 0 \\
 0 & 1 & 0 & \frac{\sqrt{3}}{2} \\
 0 & 0 & \frac{\sqrt{3}}{2} & 0 \\
\end{array}
\right)
\end{equation}
\begin{equation}
J_y=\left(
\begin{array}{cccc}
 0 & -\frac{i \sqrt{3}}{2} & 0 & 0 \\
 \frac{i \sqrt{3}}{2} & 0 & -i & 0 \\
 0 & i & 0 & -\frac{i \sqrt{3}}{2}  \\
 0 & 0 & \frac{i \sqrt{3}}{2} & 0 \\
\end{array}
\right)
\end{equation}
\begin{equation}
J_z=\left(
\begin{array}{cccc}
 \frac{3}{2} & 0 & 0 & 0 \\
 0 & \frac{1}{2} & 0 & 0 \\
 0 & 0 & -\frac{1}{2} & 0 \\
 0 & 0 & 0 & -\frac{3}{2} \\
\end{array}
\right).
\end{equation}

The five Gamma matrices are \cite{Murakami2004SU2}
\begin{equation}
\left\{
\begin{array}{l}
\Gamma^1=\frac{1}{\sqrt{3}} (J_y J_z+J_z J_y)\\
\Gamma^2=\frac{1}{\sqrt{3}} (J_z J_x+J_x J_z)\\
\Gamma^3=\frac{1}{\sqrt{3}} (J_x J_y+J_y J_x)\\
\Gamma^4=\frac{1}{\sqrt{3}} (J_x^2-J_y^2)\\
\Gamma^5=\frac{1}{3} (2 J_z^2-J_x^2-J_y^2)\\
\end{array}
\right..
\end{equation}
Clearly, $\{\Gamma^a,\Gamma^b\}=2\delta_{ab}\Gamma^0$ where $\Gamma^0$ is the 4 by 4 identity matrix.
The three $V_i$'s matrices are
 $V_x=\frac{1}{2}\{J_x,J_y^2-J_z^2\}$,
$V_y=\frac{1}{2}\{J_y,J_z^2-J_x^2\}$ and
$V_z=\frac{1}{2}\{J_z,J_x^2-J_y^2\}$.

The Luttinger Hamiltonian $h({\bf k})$ with $C=0$ can be diagonalized by the unitary transformation
\cite{Murakami2004SU2}
$$
U(\hat{\bsl{k}})\equiv \frac{(1+\frac{c_2 g_{\bsl{k},5}}{k^2 Q_c})\Gamma^0 +i\sum_{a=1}^3\frac{c_1 g_{\bsl{k},a}}{k^2 Q_c}\Gamma^{a5}+i\frac{c_2 g_{\bsl{k},4}}{k^2 Q_c}\Gamma^{45}}{\sqrt{2(1+\frac{c_2 g_{\bsl{k},5}}{k^2 Q_c})}}D,
$$
with
$$
D=\left(
\begin{array}{cccc}
 1 & 0 & 0 & 0 \\
 0 & 0 & 1 & 0 \\
 0 & 0 & 0 & 1 \\
 0 & 1 & 0 & 0 \\
\end{array}
\right).
$$
For $C=0$, this leads to
$$
U^{\dagger}(\hat{\bsl{k}})h(\bsl{k})U(\hat{\bsl{k}})=
\left(
\begin{array}{cccc}
 \xi_+ & 0 & 0 & 0 \\
 0 & \xi_+ & 0 & 0 \\
 0 & 0 & \xi_- & 0 \\
 0 & 0 & 0 & \xi_- \\
\end{array}
\right).
$$

The expressions of $|j,j_z\rangle$ with $j=3/2$ and $j_z=\pm 3/2,\pm 1/2$ in terms of the electron spin and atomic $p$ orbitals\cite{Winkler2003SOC} are
\begin{eqnarray}
\nonumber&&\vert 3/2,3/2 \rangle = -\frac{1}{\sqrt{2}}\vert  X + iY \rangle \vert \uparrow \rangle\\
\nonumber&&\vert 3/2,1/2 \rangle =  \frac{1}{\sqrt{6}}(2 \vert Z \rangle \vert \uparrow \rangle - \vert  X + iY \rangle \vert \downarrow \rangle)\\
\nonumber&&\vert 3/2,-1/2 \rangle =  \frac{1}{\sqrt{6}}(2 \vert Z \rangle \vert \downarrow \rangle + \vert  X - iY \rangle \vert \uparrow \rangle)\\
\nonumber&&\vert 3/2,-3/2 \rangle = \frac{1}{\sqrt{2}}\vert  X - iY \rangle \vert \downarrow \rangle,
\label{eqn:basis}
\end{eqnarray} where $\vert X \rangle$, $\vert Y \rangle$ and $\vert Z \rangle$ are atomic $p$ orbitals and real.

The expressions of $y_{1,2,3,4,5}$ and $z_{1,2,3}$ in Eq.\ref{eq:K0t} and Eq.\ref{eq:K1t}:
\begin{eqnarray}
y_1&=&\sum_{\lambda}\int\frac{d\Omega}{4\pi}
\theta(\widetilde{m}_{\lambda})\widetilde{m}_{\lambda}^{3/2}\nonumber\\
y_2&=&\sum_{\lambda} \lambda \int\frac{d\Omega}{4\pi}
\theta(\widetilde{m}_{\lambda})\widetilde{m}_{\lambda}^{5/2}f_Q\nonumber\\
y_3&=&\sum_{\lambda}\int\frac{d\Omega}{4\pi}
\theta(\widetilde{m}_{\lambda})\widetilde{m}_{\lambda}^{7/2}f_Q^2\nonumber\\
y_4&=&\sum_{\lambda}\lambda\int\frac{d\Omega}{4\pi}
\theta(\widetilde{m}_{\lambda}) \widetilde{m}_{\lambda}^{9/2}f_Q^3\nonumber\\
y_5&=&\sum_{\lambda}\int\frac{d\Omega}{4\pi}
\theta(\widetilde{m}_{\lambda})\widetilde{m}_{\lambda}^{11/2}f_Q^4\nonumber\\
z_1&=&\frac{7\zeta(3)}{16\pi^2}\sum_{\lambda}\int\frac{d\Omega}{4\pi}
\theta(\widetilde{m}_{\lambda})\widetilde{m}_{\lambda}^{3/2}(\widetilde{v}_z^{\lambda})^2\nonumber\\
z_2&=&\frac{7\zeta(3)}{16\pi^2}\sum_{\lambda}\lambda\int\frac{d\Omega}{4\pi}
\theta(\widetilde{m}_{\lambda})\widetilde{m}_{\lambda}^{5/2}f_Q(\widetilde{v}_z^{\lambda})^2\nonumber\\
z_3&=&\frac{7\zeta(3)}{16\pi^2}\sum_{\lambda}\int\frac{d\Omega}{4\pi}
\theta(\widetilde{m}_{\lambda})\widetilde{m}_{\lambda}^{7/2}f_Q^2(\widetilde{v}_z^{\lambda})^2\ ,
\end{eqnarray}
where $f_Q=(|c_1|Q_1^2+|c_2| Q_2^2)/Q_c$, $(\widetilde{v}_z^{\lambda})^2=(v^{\lambda}_{z}(\bsl{k}_{F,\lambda}))^2 2m\mu/(\mu^2) $, $v^{\lambda}_{z}(\bsl{k})=\partial_{k_{z}}\xi_{\lambda}(\bsl{k})$ and $\theta(...)$ is the Heaviside step function. To include the momentum cut-off in those $y$'s and $z$'s, just do the following replacement
\begin{equation}
\theta(\widetilde{m}_{\lambda})\rightarrow \theta(\widetilde{m}_{\lambda})\theta(\frac{\Lambda^2}{2m\mu}-\widetilde{m}_{\lambda})=\theta(\frac{1}{\widetilde{m}_{\lambda}}-\frac{2m\mu}{\Lambda^2})\ ,
\end{equation}
where the extra factor in the second expression is given by the fact that the momentum cut-off requires $2m_\lambda \mu\leq \Lambda^2$.
As mentioned in the Sec.\ref{sec:model_H}, we neglect the momentum cut-off $\Lambda$ in regime I and II, which is equivalent to taking $\Lambda\rightarrow\infty$, and choose a finite value for $\Lambda$ only in regime III.

\section{Derivation of Eq.\ref{eq:chi_S_sq} and Eq.\ref{eq:chi_N}}
\label{app:der_spin_sus}
In this section we derive  Eq.\ref{eq:chi_S_sq} and Eq.\ref{eq:chi_N} following Ref.\citen{Frigeri2004SpinSus}.

The magnetic moment generated by the conduction electron spins has the following expression
\begin{equation}
\label{eq:Mspin_gen}
\bsl{M}^{spin}=
\frac{1}{Z}\int D\bar{c} Dc \sum_{\bsl{k},\omega_n} \bar{c}_{\bsl{k},\omega_n}\frac{-2\mu_B}{3\beta}\bsl{J} c_{\bsl{k},\omega_n} e^{-S}\ ,
\end{equation}
where $Z=\int D\bar{c} Dc  e^{-S}$, $\bar{c},c$ are Grassmann fields of the $j=3/2$ fermion.
The action $S$ contains two parts $S=S_{ni}+S_{\Delta}$: the non-interacting part $S_{ni}$ and the pairing part $S_{\Delta}$. Below, we talk about these two parts carefully.

According to Eq.\ref{eq:def_spin_sus}, the derivation of $\chi^{S,N}$ only requires terms up to the first order of the infinitesimal uniform magnetic field.
Therefore, $S_{ni}$ contains three parts $S_{ni}=S_0+S_B^{orb}+S_B^{Z}$, where
\begin{equation}
S_0=\sum_{\bsl{k},\omega_n,\lambda}\bar{\psi}_{\bsl{k},\omega_n,\lambda} [-i\omega_n+h_{\lambda}(\bsl{k})] \psi_{\bsl{k},\omega_n,\lambda}
\end{equation}
is the non-magnetic part,
\begin{equation}
S_B^{orb}=\sum_{\bsl{k},\omega_n,\lambda}\bar{\psi}_{\bsl{k},\omega_n,\lambda} [\frac{e}{\hbar}\bsl{\nabla}_{\bsl{k}}h_{\lambda}(\bsl{k})\cdot \bsl{A}(i\bsl{\nabla}_{\bsl{k}})] \psi_{\bsl{k},\omega_n,\lambda}
\end{equation}
is the orbital part,
\begin{equation}
S_B^{Z}=\sum_{\bsl{k},\omega_n,\lambda}\bar{\psi}_{\bsl{k},\omega_n,\lambda} [\bsl{B}\cdot\bsl{M}^{\lambda}(\hat{\bsl{k}})] \psi_{\bsl{k},\omega_n,\lambda}
\end{equation}
is the Zeeman part, $\lambda=\pm$ and $\bar{\psi}_{\bsl{k},\omega_n,\lambda}, \psi_{\bsl{k},\omega_n,\lambda}$ are Grassmann fields corresponding to eigen-wavefunctions of $\xi_{\lambda}$ band.

Now we discuss the pairing part $S_{\Delta}$.
The reason of using the pairing instead of the interaction is that we consider infinitesimal magnetic field in the superconducting phase where the Cooper pairs are already formed.
And we neglect the change of order parameter due to the magnetic field\cite{Abrikosov1961KnightShift,Frigeri2004SpinSus,
Samokhin2005SpinSus} and only need to consider the uniform order parameters here.
Moreover, since the pairing can only exist within the energy cut-off $\epsilon_c$ of the attractive interaction and $\epsilon_c\ll 2Q_c k_F$, we should also project the pairing onto $\xi_{\pm}$ bands and neglect the inter-band contribution.
Therefore, $S_{\Delta}$ reads
\begin{eqnarray}
&&S_{\Delta}=\frac{1}{2}\sum_{\bsl{k},\omega_n}\bar{\psi}_{\bsl{k},\omega_n,\lambda} \Delta_{\lambda}(\bsl{k}) \bar{\psi}^T_{-\bsl{k},-\omega_n,\lambda}\\
&&+\frac{1}{2}\sum_{\bsl{k},\omega_n}\psi^T_{-\bsl{k},-\omega_n,\lambda} \Delta_{\lambda}^{\dagger}(\bsl{k}) \psi_{\bsl{k},\omega_n,\lambda}\ ,
\end{eqnarray}
where
$
\Delta_{\lambda}(\bsl{k})=\Delta_0 n^0_{\lambda}(\bsl{k})+\Delta_1 n^1_{\lambda}(\bsl{k})
$
, and
\begin{equation}
\label{eq:n0}
n^0_{\pm}(\bsl{k})=\pm \frac{1}{2}i\sigma_y
\end{equation}
and
\begin{equation}
\label{eq:n1}
 n^1_{\pm}(\bsl{k})=\frac{1}{2} k^2 a^2 \text{sgn}(c_1) f_Q i\sigma_y
\end{equation}
are pairing matrices projected to $\xi_{\pm}$ bands.

Now, we have $S=S_0+S_B^{orb}+S_B^{Z}+S_{\Delta}$. Clearly, $S$ have fermion parity symmetry for either of $\lambda=\pm$ subspace since they are decoupled. As a result, Eq.\ref{eq:Mspin_gen} can be re-written as
\begin{equation}
\label{eq:Mspin_proj}
\bsl{M}^{spin}=
-\frac{1}{Z\beta}\int D\bar{\psi} D\psi \sum_{\bsl{k},\omega_n,\lambda} \bar{\psi}_{\bsl{k},\omega_n, \lambda}\bsl{M}^{\lambda}(\hat{\bsl{k}}) \psi_{\bsl{k},\omega_n,\lambda} e^{-S}\ ,
\end{equation}
where we neglect inter-band terms given by $\mu_B \bsl{J}$ because they are odd under the fermion parity for one $\lambda$ subspace.
By defining $\mathcal{S}^M_i=\sum_{\bsl{k},\omega_n,\lambda} \bar{\psi}_{\bsl{k},\omega_n, \lambda}M_i^{\lambda}(\hat{\bsl{k}}) \psi_{\bsl{k},\omega_n,\lambda}$, we have the expression of spin susceptibility
\begin{equation}
\chi_{ij}=\frac{1}{\beta}\langle \mathcal{S}^M_i\frac{\partial(S_B^{orb}+S_B^{Z})}{\partial B_j} \rangle_0-\frac{1}{\beta}\langle \mathcal{S}^M_i\rangle_0\langle\frac{\partial(S_B^{orb}+S_B^{Z})}{\partial B_j} \rangle_0\ ,
\end{equation}
where $\langle X \rangle_{0}=\frac{1}{Z_0}\int D\bar{\psi} D\psi X e^{-S_0-S_{\Delta}}$ with $Z_0=\int D\bar{\psi} D\psi e^{-S_0-S_{\Delta}}$.
In the following, we neglect the orbital contribution to $\chi_{ij}$ as done in Ref.[\citen{Abrikosov1961KnightShift, Samokhin2005SpinSus,
Frigeri2004SpinSus}] and choose the $i,j=z$ since $\chi_{ij}$ is isotropic.
Eventually, the expression of spin susceptibility become
\begin{equation}
\label{eq:spin_sus_final}
\chi=\frac{1}{\beta}(\langle \mathcal{S}^M_z \mathcal{S}^M_z \rangle_0-\langle \mathcal{S}^M_z\rangle_0\langle \mathcal{S}^M_z \rangle_0)\ ,
\end{equation}
where $\frac{\partial S_B^{Z}}{\partial B_j}=\mathcal{S}^M_j$ is used.
Next, we will derive Eq.\ref{eq:chi_S_sq} and Eq.\ref{eq:chi_N} from the expression presented above.

In Nambu representation, $S_0+S_{\Delta}$ is re-written as
\begin{equation}
S_0+S_{\Delta}=\sum_{\bsl{k},\omega_n,\lambda}' \bar{\Psi}_{\bsl{k},\omega_n,\lambda}[-i\omega_n+h^{BdG}_{\lambda}(\bsl{k})]
\Psi_{\bsl{k},\omega_n,\lambda}\ ,
\end{equation}
where
\begin{equation}
h^{BdG}_{\lambda}(\bsl{k})=\left(
\begin{array}{cc}
h_{\lambda}(\bsl{k}) & \Delta_{\lambda}(\bsl{k})\\
\Delta_{\lambda}^{\dagger}(\bsl{k}) & -h_{\lambda}^T(-\bsl{k})\\
\end{array}
\right)\ ,
\end{equation}
$\bar{\Psi}_{\bsl{k},\omega_n,\lambda}=(\bar{\psi}_{\bsl{k},\omega_n,\lambda}, \psi^T_{-\bsl{k},-\omega_n,\lambda})$, $\Psi_{\bsl{k},\omega_n,\lambda}=(\psi^T_{\bsl{k},\omega_n,\lambda}, \bar{\psi}_{-\bsl{k},-\omega_n,\lambda})^T$, the `` $'$ '' on top of $\sum$ means only summing over half the region of $(\bsl{k},\omega_n)$ with the other half obtained by $(\bsl{k},\omega_n)\rightarrow -(\bsl{k},\omega_n)$.
Define $G^{BdG}_{\lambda}(\bsl{k},\omega_n)=[i\omega_n-h^{BdG}_{\lambda}(\bsl{k})]^{-1}$.
The expression of $G^{BdG}_{\lambda}(\bsl{k},\omega_n)$ reads
\begin{equation}
G^{BdG}_{\lambda}(\bsl{k},\omega_n)
=
\left(
\begin{array}{cc}
G_{\lambda}(\bsl{k},\omega_n) & F_{\lambda}(\bsl{k},\omega_n)\\
F_{\lambda}^{\dagger}(\bsl{k},\omega_n) & -G_{\lambda}(-\bsl{k},-\omega_n)\\
\end{array}
\right)\ ,
\end{equation}
where
\begin{equation}
G_{\lambda}(\bsl{k},\omega_n)=\mathcal{G}_{\lambda,+}(\bsl{k},\omega_n)+\mathcal{G}_{\lambda,-}(\bsl{k},\omega_n)\hat{\bsl{p}}^{\lambda}(\hat{\bsl{k}})\cdot\bsl{\sigma}\ ,
\end{equation}
\begin{equation}
F_{\lambda}(\bsl{k},\omega_n)=[\mathcal{F}_{\lambda,+}(\bsl{k},\omega_n)+\mathcal{F}_{\lambda,-}(\bsl{k},\omega_n)\hat{\bsl{p}}^{\lambda}(\hat{\bsl{k}})\cdot\bsl{\sigma}]\Delta_{\lambda}(\bsl{k})\ ,
\end{equation}
\begin{eqnarray}
&&\mathcal{G}_{\lambda,\pm}(\bsl{k},\omega_n)=-\frac{1}{2}\left(
\frac{i\omega_n+E_{\lambda,+}(\bsl{k})}{\omega_n^2+|d_{\lambda}(\bsl{k})|^2+E_{\lambda,+}^2(\bsl{k})}\right.\nonumber\\
&&\left.
\pm
\frac{i\omega_n+E_{\lambda,-}(\bsl{k})}{\omega_n^2+|d_{\lambda}(\bsl{k})|^2+E_{\lambda,-}^2(\bsl{k})}
\right)\ ,
\end{eqnarray}
\begin{eqnarray}
&&\mathcal{F}_{\lambda,\pm}(\bsl{k},\omega_n)=-\frac{1}{2}\left(
\frac{1}{\omega_n^2+|d_{\lambda}(\bsl{k})|^2+E_{\lambda,+}^2(\bsl{k})}\right.\nonumber\\
&&\left.
\pm
\frac{1}{\omega_n^2+|d_{\lambda}(\bsl{k})|^2+E_{\lambda,-}^2(\bsl{k})}
\right)\ ,
\end{eqnarray}\
and
$E_{\lambda,\pm}(\bsl{k})=\xi_{\lambda}(\bsl{k})\pm |C|k p^{\lambda}(\hat{\bsl{k}})$.
Then, $Z_0$ can be expressed as
\begin{equation}
Z_0=\int D\bar{\Psi} D\Psi e^{\sum_{\bsl{k},\omega_n,\lambda}' \bar{\Psi}_{\bsl{k},\omega_n,\lambda}[G^{BdG}_{\lambda}(\bsl{k},\omega_n)]^{-1}
\Psi_{\bsl{k},\omega_n,\lambda}}\ .
\end{equation}
On the other hand, $\mathcal{S}^M_z$ is expressed in Nambu representation as
\begin{equation}
\mathcal{S}^M_z=\sum_{\bsl{k},\omega_n,\lambda}' \bar{\Psi}_{\bsl{k},\omega_n, \lambda}W_z^{\lambda}(\hat{\bsl{k}}) \Psi_{\bsl{k},\omega_n,\lambda}\ ,
\end{equation}
where  $W_z^{\lambda}(\hat{\bsl{k}})=\text{diag}(M^{\lambda}_i(\hat{\bsl{k}}), -[M^{\lambda}_i(-\hat{\bsl{k}})]^T)$.

Now we can work out Eq.\ref{eq:spin_sus_final}.
\begin{equation}
\chi=-\frac{1}{\beta}\sum_{\bsl{k},\omega_n,\lambda}'\text{Tr}[G^{BdG}_{\lambda}(\bsl{k},\omega_n)W_z^{\lambda}(\hat{\bsl{k}})G^{BdG}_{\lambda}(\bsl{k},\omega_n)W_z^{\lambda}(\hat{\bsl{k}})]\ .
\end{equation}
Using $-F^T_{\lambda}(-\bsl{k},-\omega_n)=F_{\lambda}(\bsl{k},\omega_n)$, the equation can be simplified into
\begin{eqnarray}
&&\chi=-\frac{1}{\beta}\sum_{\bsl{k},\omega_n,\lambda}
\left(\text{Tr}[M_z^{\lambda}(\hat{\bsl{k}})G_{\lambda}(\bsl{k},\omega_n)M_z^{\lambda}(\hat{\bsl{k}})G_{\lambda}(\bsl{k},\omega_n)]\right.\nonumber\\
&&\left. -\text{Tr}[M_z^{\lambda}(\hat{\bsl{k}})F_{\lambda}(\bsl{k},\omega_n)[M_z^{\lambda}(-\hat{\bsl{k}})]^T F^{\dagger}_{\lambda}(\bsl{k},\omega_n)]\right)\ .
\end{eqnarray}
Using $Tr[M_z^{\lambda}(\hat{\bsl{k}})M_z^{\lambda}(\hat{\bsl{k}})\hat{\bsl{p}}^{\lambda}(\hat{\bsl{k}})\cdot\bsl{\sigma}]
=0$,
\begin{eqnarray}
\label{eq:chi_GF}
&&\chi=-\frac{1}{\beta}
\sum_{\bsl{k},\omega_n,\lambda}
\left[
m^z_{\lambda}(\hat{\bsl{k}})(\mathcal{G}^2_{\lambda,+}(\bsl{k},\omega_n)+\mathcal{F}^2_{\lambda,+}(\bsl{k},\omega_n)|d_{\lambda}(\bsl{k})|^2)\right.\nonumber\\
&&
\left.
+
\bar{m}^z_{\lambda}(\hat{\bsl{k}})(\mathcal{G}^2_{\lambda,-}(\bsl{k},\omega_n)+\mathcal{F}^2_{\lambda,-}(\bsl{k},\omega_n)|d_{\lambda}(\bsl{k})|^2)
\right]\ .
\end{eqnarray}

The spin susceptibility $\chi^N$ in the normal state can also be obtained from Eq.\ref{eq:chi_GF} by choosing zero value for the order parameters $d_{\lambda}(\bsl{k})=0$. As a result, we can get Eq.\ref{eq:chi_N}  by neglecting terms of order $1/(\beta \epsilon_c)$, $\alpha_{\lambda}  /\epsilon_c $ and $\epsilon_c/|\mu|$.

If the temperature is below $T_c$ and the superconducting order parameters are not zero, the Eq.\ref{eq:chi_GF} gives the superconducting spin susceptibility $\chi^S$. In this case, we can first subtract $\chi^S$ by $\chi^N$ in order to exchange the sum of $\omega_n$ with the energy integration. Then, by neglecting terms of order $1/(\beta \epsilon_c)$, $\alpha_{\lambda}  /\epsilon_c $, $|d_{\lambda}|  /\epsilon_c $ and $\epsilon_c/|\mu|$, we can get Eq.\ref{eq:chi_S_sq}.

\section{Non-interacting Green Function with Magnetic Field}
\label{app.G_B}

In this part, we derive Eq.\ref{eq:tG_Form} following Ref.[\citen{Samokhin2004magnetic}].
In the continuous limit, the corresponding effective Green function for each band satisfies the equation
\begin{equation}
[i\omega_n-E^{\lambda}(\bsl{K}_{\bsl{r}_1},\bsl{B})]G^\lambda(\bsl{r}_1,\bsl{r}_2,\omega_n)
=
\delta(\bsl{r}_1-\bsl{r}_2)\ ,
\end{equation}
where $\bsl{K}_{\bsl{r}}=-i\bsl{\nabla}_{\bsl{r}}+\frac{e}{\hbar}\bsl{A}(\bsl{r})$, $\omega_n=(2n+1)\pi/\beta$ is the fermionic Matusbara frequency, $\lambda=\pm$ and $1/\beta=k_B T$.
Clearly, the Green function $G^{\lambda}(\bsl{r}_1,\bsl{r}_2,\omega_n)$ is not translationally invariant.
Define
\begin{equation}
G^{\lambda}(\bsl{r}_1,\bsl{r}_2,\omega_n)=e^{-i\frac{e}{\hbar}\bsl{r}_1\cdot\bsl{A}(\bsl{r}_2)}\widetilde{G}^{\lambda}(\bsl{r}_1-\bsl{r}_2,\omega_n)\ ,
\end{equation}
resulting that $\widetilde{G}^{\lambda}(\bsl{r}_1-\bsl{r}_2,\omega_n)$ satisfying a translationally invariant equation:
\begin{equation}
[i\omega_n-E^{\lambda}(\bsl{K}_{\bsl{r}},\bsl{B})]\widetilde{G}^{\lambda}(\bsl{r},\omega_n)
=
\delta(\bsl{r})
\end{equation}
or equivalently
\begin{equation}
\label{eq:green_func}
[i\omega_n-E^{\lambda}(\bsl{K},\bsl{B})]\widetilde{G}^{\lambda}(\bsl{k},\omega_n)
=
1
\end{equation}
with $\widetilde{G}^{\lambda}(\bsl{r},\omega_n)=\frac{1}{\mathcal{V}}\sum_{\bsl{k}}e^{i\bsl{k}\cdot\bsl{r}}\widetilde{G}^{\lambda}(\bsl{k},\omega_n)$.
Note that, the derivation shown above uses $\bsl{A}(\bsl{r})=\frac{\bsl{B}\times \bsl{r}}{2}$, $[\bsl{r}\cdot\bsl{k},\ \bsl{r}\cdot\bsl{A}(i\bsl{\nabla}_{\bf k})]=0$ and $[\bsl{r}'\cdot\bsl{\nabla}_{\bf r},\ \bsl{r}'\cdot\bsl{A}(\bsl{r})]=0$.

To solve Eq.\ref{eq:green_func} analytically, we make another assumption that $\bsl{B}$ is sufficiently small so that we can treat the magnetic field dependence in the equation as a perturbation, as mentioned in the main text.
It means $B\mu_B\ll k_B T $ and $\hbar \omega_c\ll k_B T$ for each band, where the latter is for the field dependence in $\bsl{K}$ and $\hbar\omega_c=\frac{2 m_e}{|m_{\lambda}|}B\mu_B$ is the cyclotron frequency of the band under the magnetic field.\cite{Samokhin2004magnetic}
Since the upper critical field approaches to zero as temperature approaches to the zero-field critical temperature $T_c$, that assumption restricts us to consider the temperature near $T_c$ where the upper critical field is small.

Finally, we solve Eq.\ref{eq:green_func} to the first order of $\hbar\omega_c/(k_B T)$ following Ref.[\citen{Samokhin2004magnetic}].
Since $\hbar\omega_c/(k_B T)$ is linear in $B$, we will directly use the order of $B$ to indicate the order of $\hbar\omega_c/(k_B T)$.
If $B=0$, the zero field Green function is easy to solve
\begin{equation}
\widetilde{G}^{\pm}_0(\bsl{k},\omega_n)=\frac{1}{i\omega_n-\xi_{\pm}(\bsl{k})}\ .
\end{equation}
Since the Periels substitution is given by
\begin{equation}
\label{def:EK_conti}
E^{\pm}(\bsl{K},\bsl{B})=\int d^3 r \delta(\bsl{r}) E^{\pm}(i\bsl{\nabla}_{\bsl{r}},\bsl{B})e^{-i\bsl{r}\cdot\bsl{K}}
\end{equation}
with $E^{\pm}(i\bsl{\nabla}_{\bsl{r}},\bsl{B})$ obtained by replacing $\bsl{k}$ in $E^{\pm}(\bsl{k},\bsl{B})$ by $i\bsl{\nabla}_{\bsl{r}}$
, $E^{\pm}(\bsl{K},\bsl{B})$ to the first order of $B$ has the following expression
\begin{widetext}
\begin{eqnarray}
E^{\pm}(\bsl{K},\bsl{B})
&=&
\int d^3 r \delta(\bsl{r})(\xi_{\pm}(i\nabla_{\bsl{r}})+\bsl{B}\cdot\bsl{M}^{\pm}(i\nabla_{\bsl{r}}))e^{-i\bsl{r}\cdot\bsl{K}}\nonumber\\
&=&
\int d^3 r \delta(\bsl{r})(\xi_{\pm}(i\nabla_{\bsl{r}})+\bsl{B}\cdot\bsl{M}^{\pm}(i\nabla_{\bsl{r}}))e^{-i\bsl{r}\cdot\bsl{k}}(1+\frac{e}{2\hbar}\bsl{r}\cdot(\bsl{B}\times \nabla_{\bsl{k}})+O(B^2))\nonumber\\
&=&\xi_{\pm}(\bsl{k})+\bsl{B}\cdot\bsl{M}^{\pm}(\bsl{k})+\int d^3 r \delta(\bsl{r})\xi_{\pm}(i\nabla_{\bsl{r}})e^{-i\bsl{r}\cdot\bsl{k}}\frac{e}{2\hbar}\bsl{r}\cdot(\bsl{B}\times \nabla_{\bsl{k}})+O(B^2)\nonumber\\
&=&\xi_{\pm}(\bsl{k})+\bsl{B}\cdot\bsl{M}^{\pm}(\bsl{k})+\frac{e}{2\hbar}\int d^3 r \delta(\bsl{r}) e^{-i\bsl{r}\cdot\bsl{k}}\xi_{\pm}(\bsl{k}+i\nabla_{\bsl{r}})\bsl{r}\cdot(\bsl{B}\times \nabla_{\bsl{k}})+O(B^2)\nonumber\\
&=&\xi_{\pm}(\bsl{k})+\bsl{B}\cdot\bsl{M}^{\pm}(\bsl{k})+\frac{e}{2\hbar}\int d^3 r \delta(\bsl{r}) e^{-i\bsl{r}\cdot\bsl{k}}[i\nabla_{\bsl{k}}\xi_{\pm}(\bsl{k})\cdot\nabla_{\bsl{r}}]\bsl{r}\cdot(\bsl{B}\times \nabla_{\bsl{k}})+O(B^2)\nonumber\\
&=&\xi_{\pm}(\bsl{k})+\bsl{B}\cdot\bsl{M}^{\pm}(\bsl{k})+i\frac{e}{2\hbar}\bsl{v}^{\pm}\cdot(\bsl{B}\times \nabla_{\bsl{k}})+O(B^2)\ ,
\end{eqnarray}
\end{widetext}
where $\bsl{v}^{\pm}=\nabla_{\bsl{k}}\xi_{\pm}(\bsl{k})$ and the third equality uses the fact that $\bsl{r}\cdot\bsl{k}$ commutes with $\bsl{r}\cdot(\bsl{B}\times \nabla_{\bsl{k}})$.
With the expression shown above, we get the first order correction to the Green function which is
\begin{equation}
\widetilde{G}^{\pm}_0(\bsl{k},\omega_n) (\bsl{B}\cdot\bsl{M}^{\pm}(\bsl{k})+i\frac{e}{2\hbar}\bsl{v}^{\pm}\cdot(\bsl{B}\times \nabla_{\bsl{k}}))\widetilde{G}^{\pm}_0(\bsl{k},\omega_n)\ .
\end{equation}
Note that
\begin{equation}
\bsl{v}^{\pm}\cdot(\bsl{B}\times \nabla_{\bsl{k}})\widetilde{G}^{\pm}_0(\bsl{k},\omega_n)
=
\frac{\bsl{v}^{\pm}\cdot(\bsl{B}\times \bsl{v}^{\pm})}{(i\omega_n-\xi_{\pm}(\bsl{k}))^2}=0\ ,
\end{equation}
we finally get solution to Eq.\ref{eq:green_func} to the first order of $B$
\begin{equation}
\widetilde{G}^{\pm}(\bsl{k},\omega_n)=\frac{1}{i\omega_n-\xi_{\pm}(\bsl{k})}+\frac{\bsl{B}\cdot\bsl{M}^{\pm}(\bsl{k})}{(i\omega_n-\xi_{\pm}(\bsl{k}))^2}\ .
\end{equation}

\section{Derivation of Eq.\ref{eq:F_r_final}}
\label{app:der_Fr}
In this part, we derive Eq.\ref{eq:F_r_final} following Ref.[\citen{Samokhin2004magnetic}].


\subsubsection{General Expression of the Superconducting Free Energy $F_{SC}$}
According to Eq.\ref{Eqn:HI}, the interacting part of the action reads
\begin{equation}
S_I=
\int_{0}^{\beta} d\tau \frac{1}{2\mathcal{V}}\sum_{\bsl{q}}\sum_{a=0,1}\left[
V_a P_a(\bsl{q},\tau)P_a^{\dagger}(\bsl{q},\tau)
\right]\ ,
\end{equation}
where
$\tau$ is the imaginary time,
$\bar{c}_{\bsl{k},\tau}$ is the Grassman field,
\begin{equation}
P_0(\bsl{q},\tau)=\sum_{\bsl{k}}\bar{c}_{\bsl{k}+\frac{\bsl{q}}{2},\tau}(\Gamma^0 \gamma/2)(\bar{c}_{-\bsl{k}+\frac{\bsl{q}}{2},\tau})^T
\end{equation}
and
\begin{equation}
P_1(\bsl{q},\tau)=\sum_{\bsl{k}}\bar{c}_{\bsl{k}+\frac{\bsl{q}}{2},\tau}(a^2 \bsl{g}_{\bsl{k}}\cdot\bsl{\Gamma}
\gamma/2)(\bar{c}_{-\bsl{k}+\frac{\bsl{q}}{2},\tau})^T\ .
\end{equation}

Using Hubbard-Stratonovich transformation, we have
\begin{widetext}
\begin{equation}
\exp(-S_I)= \int D\Delta D\Delta^* \exp[\int_{0}^{\beta} d\tau \sum_{\bsl{q},a}(-\frac{1}{2}P_{a}(\bsl{q},\tau)\Delta_{a}(\bsl{q},\tau)-\frac{1}{2}P_{a}^{\dagger}(\bsl{q},\tau)\Delta_{a}^*(\bsl{q},\tau)+\mathcal{V}\frac{|\Delta_{a}(\bsl{q},\tau)|^2}{2V_a}) ]\ ,
\end{equation}
\end{widetext}
where
\begin{equation}
\int D\Delta D\Delta^*=\prod_{\tau,\bsl{q},a}\frac{\mathcal{V} d\tau }{-2\pi V_a}\int d\Delta_{a}(\bsl{q},\tau) d \Delta_{a}^*(\bsl{q},\tau) \ .
\end{equation}

Assume that $\Delta_i(\bsl{q},\tau)$ is uniform in $\tau$, and thereby it can be re-labeled as $\Delta_i(\bsl{q})$.
Then the partition function becomes
\begin{widetext}
\begin{equation}
Z=\int D c D\bar{c} D\Delta D\Delta^* \exp[-S_0+ \sum_{\bsl{q},\omega_n,a}(-\frac{1}{2}P_{a}(\bsl{q},\omega_n)\Delta_{a}(\bsl{q})-\frac{1}{2}P_{a}^{\dagger}(\bsl{q},\omega_n)\Delta_{a}^*(\bsl{q}))+\sum_{\bsl{q},a}\frac{\beta \mathcal{V}}{2V_a}|\Delta_{a}(\bsl{q})|^2 ]\ ,
\end{equation}
\end{widetext}
where $S_0$ is the non-interacting action
,
\begin{equation}
P_0(\bsl{q},\omega_n)=\sum_{\bsl{k}}\bar{c}_{\bsl{k}+\frac{\bsl{q}}{2},\omega_n}(\Gamma^0 \gamma/2)(\bar{c}_{-\bsl{k}+\frac{\bsl{q}}{2},-\omega_n})^T\ ,
\end{equation}
\begin{equation}
P_1(\bsl{q},\omega_n)=\sum_{\bsl{k}}\bar{c}_{\bsl{k}+\frac{\bsl{q}}{2},\omega_n}(a^2 \bsl{g}_{\bsl{k}}\cdot\bsl{\Gamma}
\gamma/2)(\bar{c}_{-\bsl{k}+\frac{\bsl{q}}{2},-\omega_n})^T
\end{equation}
and the Fourier transformation relation Eq.\ref{eq:Fourier_field} is used.

We can express $P_i$ in eigen-wavefunctions of $\xi_{\pm}$ bands
\begin{equation}
P_a(\bsl{q},\omega_n)=\sum_{\bsl{k}}\bar{\psi}_{\bsl{k}+\frac{\bsl{q}}{2},\omega_n}n^a(\bsl{k},\bsl{q})(\bar{\psi}_{-\bsl{k}+\frac{\bsl{q}}{2},-\omega_n})^T\ ,
\end{equation}
where
\begin{equation}
n^0(\bsl{k},\bsl{q})=U^{\dagger}(\bsl{k}+\frac{\bsl{q}}{2})(\Gamma^0 \gamma/2)U^*(-\bsl{k}+\frac{\bsl{q}}{2})
\end{equation}
and
\begin{equation}
n^1(\bsl{k},\bsl{q})=U^{\dagger}(\bsl{k}+\frac{\bsl{q}}{2})(a^2 \bsl{g}_{\bsl{k}}\cdot\bsl{\Gamma}
\gamma/2)U^*(-\bsl{k}+\frac{\bsl{q}}{2})\ .
\end{equation}

To simplify $n^a(\bsl{k},\bsl{q})$,
we would first neglect the $\bsl{q}$ dependence, i.e. $n^a(\bsl{k},\bsl{q})\approx n^a(\bsl{k},0)$ re-labeled as $n^a(\bsl{k})$.
The reason is given below.
Typically, the order parameter with large $\bsl{q}$ is not the minimum of the Free energy, and thus $|\bsl{q}|$ is small compared with the Fermi momentum $k_F$.
Then, we can expand $U(\bsl{k}+\frac{\bsl{q}}{2})$ in terms of $|\bsl{q}|/k_F$:
\begin{equation}
U(\bsl{k}+\frac{\bsl{q}}{2})=U(\bsl{k})+\frac{\bsl{q}}{2}\cdot\nabla_{\bsl{k}} U(\bsl{k})+...
\end{equation}
Since $U(\bsl{k})$ only depends on the direction of $\bsl{k}$, i.e $U(\bsl{k})=U(\hat{\bsl{k}})$, we thereby have
\begin{equation}
U(\bsl{k}+\frac{\bsl{q}}{2})=U(\hat{\bsl{k}})+\frac{\bsl{q}}{2 k}\cdot\nabla_{\hat{\bsl{k}}} U(\hat{\bsl{k}})+...\ ,
\end{equation}
where $\frac{1}{k}\nabla_{\hat{\bsl{k}}}$ stands for the angular part of $\nabla_{\bsl{k}}$ operator.
Then, we can conclude that,  the $\bsl{q}^n$ term brought by the expression of $U(\bsl{k}+\frac{\bsl{q}}{2})$ on the Fermi surface is of order
$(\frac{|\bsl{q}|}{ k_F})^n$ compared with the original term.
The $\bsl{q}^n$ term in the Free energy can also be given by the Green function.
To estimate that contribution, let us assume an isotropic form of Green function $(i\omega-(\bsl{k}+\bsl{q}/2)^2/(2m^*)+\mu)^{-1}$.
In this case, $|\bsl{q}|^n$ term of the Green function on the Fermi surface is of order $(\frac{|\bsl{q}|}{ k_F}\frac{|\mu|}{k_B T})^n$ compared with the original term.
Here we replace $1/(i\omega)^n$ by $(k_B T)^n$ because the other part of $1/(i\omega)^n$ will just contribute to a convergent dimensionless expression after summing over $\omega$.
Since we assume $\frac{|\mu|}{k_B T}\gg 1$,
the $\bsl{q}$ dependence in the $U(\bsl{k}+\frac{\bsl{q}}{2})$ can be neglected.

We further simplify $n^i(\bsl{k})$ by making the approximation
\begin{equation}
n^a(\bsl{k})\approx
\left(
\begin{matrix}
n^a_+(\bsl{k}) &  \\
 & n^a_-(\bsl{k}) \\
\end{matrix}
\right)\ ,
\end{equation}
where the $n^a_{\pm}(\bsl{k})$ are shown in Eq.\ref{eq:n0} and Eq.\ref{eq:n1}.
This approximation is legitimate since the inter-band contribution is of order $\epsilon_c/(2Q_c k_F^2)\ll 1$, where $\epsilon_c$ is the energy cut-off of the attractive interaction.
Therefore, we have
\begin{equation}
P_a(\bsl{q},\omega_n)=\sum_{\bsl{k}}\sum_{\lambda=\pm}\bar{\psi}_{\bsl{k}+\frac{\bsl{q}}{2},\omega_n,\lambda}n^a_{\lambda}(\bsl{k})(\bar{\psi}_{-\bsl{k}+\frac{\bsl{q}}{2},-\omega_n,\lambda})^T\ .
\end{equation}

Since the Green function does not have translational invariance, it is better to deal with the problem in the position space.
After the Fourier transformation,
we have
\begin{equation}
\sum_{\bsl{q},\omega_n}P_a(\bsl{q},\omega_n)\Delta_{a}(\bsl{q})=\frac{1}{\mathcal{V}}\int d^3 r P_a(\bsl{r}) \Delta_a(\bsl{r})
\end{equation}
with
\begin{equation}
P_a(\bsl{r}_1)=\sum_{\omega_n,\lambda}\int d^3 r_2\bar{\psi}_{\bsl{r}_1+\frac{\bsl{r}_2}{2},\omega_n,\lambda}n^a_{\lambda}(\bsl{r}_2)(\bar{\psi}_{\bsl{r}_1-\frac{\bsl{r}_2}{2},-\omega_n,\lambda})^T
\end{equation}
and
\begin{equation}
\mathcal{V}\sum_{\bsl{q}}|\Delta_{a}(\bsl{q})|^2=
\int d^3r |\Delta_{a}(\bsl{r})|^2\ ,
\end{equation}
where $n^a_{\lambda}(\bsl{r})=\sum_{\bsl{k}}n^a_{\lambda}(\bsl{k})e^{i\bsl{k}\cdot\bsl{r}}$ and $\Delta_a(\bsl{r})=\sum_{\bsl{q}}e^{i\bsl{q}\cdot\bsl{r}}\Delta_a(\bsl{q})$\ .
Then, we have
\begin{widetext}
\begin{equation}
Z=\int D c D\bar{c} D\Delta D\Delta^* \exp[-S_0+ \sum_{a}\int d^3 r(-\frac{1}{2\mathcal{V}}P_{a}(\bsl{r})\Delta_{a}(\bsl{r})-\frac{1}{2\mathcal{V}}P_{a}^{\dagger}(\bsl{r})\Delta_{a}^*(\bsl{r})+\frac{\beta }{2V_a}|\Delta_{a}(\bsl{r})|^2 ) ]\ ,
\end{equation}
\end{widetext}
where
\begin{equation}
-S_0=\sum_{\omega_n,\lambda}\int d^3 r \bar{\psi}_{\bsl{r},\omega_n,\lambda}(i\omega_n-E^{\lambda}(\bsl{K}_{\bsl{r}},\bsl{B}))\psi_{\bsl{r},\omega_n,\lambda}\ .
\end{equation}

Using the expression of Green function Eq.\ref{def:green_func} and Eq.\ref{eq:tG_Form},
we can integrate out the fermionic field and get the effective action $S_{eff}[\Delta]$ with the partition function being
\begin{equation}
Z=\int D\Delta D\Delta^* \exp(-S_{eff}[\Delta])\ .
\end{equation}
Under the mean-field approximation,
we have
\begin{equation}
Z\approx \exp(-S_{eff}[\Delta])
\end{equation}
with $\Delta$ satisfying
\begin{equation}
\frac{\delta S_{eff}[\Delta]}{\delta \Delta_a^*(\mathbf{r})}=0\ .
\end{equation}
Then the mean-field free energy (Ginzburg-Landau free energy) reads
\begin{equation}
F=-\frac{1}{\beta}\ln(Z)=\frac{1}{\beta} S_{eff}\ ,
\end{equation}
which gives the superconducting Free energy
\begin{equation}
F_{SC}=F-F_N=\frac{1}{\beta} (S_{eff}[\Delta]-S_{eff}[0])\ ,
\end{equation}
where $F_N$ means the mean-field free energy with zero $\Delta$.

In order to get the critical temperature of this second-order phase transition, we only need to derive $F_{SC}$ to the second order of $\Delta$, which is
\begin{widetext}
\begin{equation}
\label{eq:F_r}
F_{SC}=-\sum_{a}\int d^3 r \frac{1}{2V_a}|\Delta_{a}(\bsl{r})|^2-\frac{1}{2\mathcal{V}^2}\sum_{a_1,a_2}\int d^3 r_1\int d^3 r_2 \Delta_{a_1}^*(\bsl{r}_1)S^{a_1 a_2}(\bsl{r}_1,\bsl{r}_2)\Delta_{a_2}(\bsl{r}_2)+O(|\Delta|^4)\ .
\end{equation}
Here
\begin{eqnarray}
&& S^{a_1 a_2}(\bsl{r}_1,\bsl{r}_2)
=
\sum_{\omega_n,\lambda}\int d^3 \rho_1 d^3 \rho_2\frac{1}{\beta} \text{Tr}[G_{\lambda}(\bsl{r}_1+\frac{\bsl{\rho}_1}{2},\bsl{r}_2+\frac{\bsl{\rho}_2}{2},\omega_n)n^{a_2}_{\lambda}(\bsl{\rho}_2)G_{\lambda}^T(\bsl{r}_1-\frac{\bsl{\rho}_1}{2},\bsl{r}_2-\frac{\bsl{\rho}_2}{2},-\omega_n)[n^{a_1}_{\lambda}(\bsl{\rho}_1)]^{\dagger}]
\nonumber\\
&=&
\sum_{\omega_n,\lambda}\int d^3 \rho_1 d^3 \rho_2\frac{1}{\beta} \text{Tr}[\widetilde{G}_{\lambda}(\bsl{r}_1+\frac{\bsl{\rho}_1}{2}-\bsl{r}_2-\frac{\bsl{\rho}_2}{2},\omega_n)n^{a_2}_{\lambda}(\bsl{\rho}_2)\widetilde{G}_{\lambda}^T(\bsl{r}_1-\frac{\bsl{\rho}_1}{2}-\bsl{r}_2+\frac{\bsl{\rho}_2}{2},-\omega_n)[n^{a_1}_{\lambda}(\bsl{\rho}_1)]^{\dagger}]e^{-i\frac{e}{\hbar}[2\bsl{r}_1\cdot\bsl{A}(\bsl{r}_2)+\bsl{\rho}_1\cdot\bsl{A}(\frac{\bsl{\rho}_2}{2})]}
\nonumber\\
&=&
e^{-i\frac{2e}{\hbar}\bsl{r}_1\cdot\bsl{A}(\bsl{r}_2)}\sum_{\omega_n,\lambda}\frac{1}{\beta}\sum_{\bsl{k},\bsl{q}} e^{i\bsl{q}\cdot(\bsl{r}_1-\bsl{r}_2)}
[\widetilde{G}_{\lambda}(\bsl{k}+\frac{\bsl{q}}{2},\omega_n)]_{\beta_2\alpha_1}[\widetilde{G}_{\lambda}^T(-\bsl{k}+\frac{\bsl{q}}{2},-\omega_n)]_{\alpha_2\beta_1}
[\Lambda(\bsl{k})]^{a_1a_2}_{\alpha_1\alpha_2\beta_1\beta_2}
\end{eqnarray}
with the summation over $\alpha_1\alpha_2\beta_1\beta_2$ implied,
and
\begin{eqnarray}
&& [\Lambda(\bsl{k})]^{a_1 a_2}_{\alpha_1\alpha_2\beta_1\beta_2}
=
\frac{1}{\mathcal{V}^2}\int d^3 \rho_1 d^3 \rho_2 e^{-i\frac{e}{\hbar}\bsl{\rho}_1\cdot\bsl{A}(\frac{\bsl{\rho}_2}{2})} e^{i\bsl{k}\cdot(\bsl{\rho}_1-\bsl{\rho}_2)} [n^{a_2}_{\lambda}(\bsl{\rho}_2)]_{\alpha_1\alpha_2} [n^{a_1}_{\lambda}(\bsl{\rho}_1)]^{\dagger}_{\beta_1\beta_2}
\nonumber\\
&=&
\frac{1}{\mathcal{V}^2}\int d^3 \rho_1 d^3\rho_2 \sum_{\bsl{k}',\bsl{k}''} e^{-i\frac{e}{\hbar}\bsl{\rho}_1\cdot\bsl{A}(\frac{\bsl{\rho}_2}{2})} e^{i(\bsl{k}-\bsl{k}')\cdot\bsl{\rho}_1}e^{i(\bsl{k}''-\bsl{k})\cdot\bsl{\rho}_2} [n^{a_2}_{\lambda}(\bsl{k}'')]_{\alpha_1\alpha_2} [n^{a_1}_{\lambda}(\bsl{k}')]^{\dagger}_{\beta_1\beta_2}
\nonumber\\
&=&
\left. e^{-i\frac{e}{4\hbar}\bsl{\nabla}_{\bsl{k}'}\cdot(\bsl{B}\times \bsl{\nabla}_{\bsl{k}''})}  [n^{a_2}_{\lambda}(\bsl{k}'')]_{\alpha_1\alpha_2} [n^{a_1}_{\lambda}(\bsl{k}')]^{\dagger}_{\beta_1\beta_2}\right|_{\bsl{k}',\bsl{k}''\rightarrow \bsl{k}}
=
\left. e^{i\frac{e}{4\hbar}\bsl{B}\cdot(\bsl{\nabla}_{\bsl{k}'}\times \bsl{\nabla}_{\bsl{k}''})}  [n^{a_2}_{\lambda}(\bsl{k}'')]_{\alpha_1\alpha_2} [n^{a_1}_{\lambda}(\bsl{k}')]^{\dagger}_{\beta_1\beta_2} \right|_{\bsl{k}',\bsl{k}''\rightarrow \bsl{k}}\ .\nonumber\\
\end{eqnarray}
\end{widetext}

Clearly, as long as $a_1=0$ or $a_2=0$, $[\Lambda(\bsl{k})]^{a_1 a_2}_{\alpha_1\alpha_2\beta_1\beta_2}$ has no magnetic field dependence since $n^0_{\lambda}(\bsl{k})$ is $\bsl{k}$ independent as shown in Eq.$\ref{eq:n0}$.
According to Eq.$\ref{eq:n1}$,
the contribution to $[\Lambda(\bsl{k})]^{11}_{\alpha_1\alpha_2\beta_1\beta_2}$ of first order of $B$ vanishes since it is proportional to the cross product of two same gradients.
Therefore, we have
\begin{equation}
\label{eq:Lambda_B1}
[\Lambda(\bsl{k})]^{a_1 a_2}_{\alpha_1\alpha_2\beta_1\beta_2}
=
[n^i_{\lambda}(\bsl{k})]^{\dagger}_{\beta_1\beta_2}[n^j_{\lambda}(\bsl{k})]_{\alpha_1\alpha_2}
\end{equation}
to the first order of $B$.
Using Eq.\ref{eq:n0}, Eq.\ref{eq:n1} and Eq.\ref{eq:Lambda_B1} and to the first order of $B$,
we have
\begin{widetext}
\begin{eqnarray}
&&[\widetilde{G}_{\lambda}(\bsl{k}+\frac{\bsl{q}}{2},\omega_n)]_{\beta_2\alpha_1}[\widetilde{G}_{\lambda}^T(-\bsl{k}+\frac{\bsl{q}}{2},-\omega_n)]_{\alpha_2\beta_1}[\Lambda(\bsl{k})]^{a_1 a_2}_{\alpha_1\alpha_2\beta_1\beta_2}=\text{Tr}[\widetilde{G}_{\lambda}(\bsl{k}+\frac{\bsl{q}}{2},\omega_n)n^{a_2}_{\lambda}(\bsl{k})\widetilde{G}_{\lambda}^T(-\bsl{k}+\frac{\bsl{q}}{2},-\omega_n)[n^{a_1}_{\lambda}(\bsl{k})]^{\dagger}]\nonumber\\
&=& \frac{\text{Tr}[n^{a_2}_{\lambda}(\bsl{k})[n^{a_1}_{\lambda}(\bsl{k})]^{\dagger}]}{(i\omega_n-\xi_{\lambda}(\bsl{k}+\frac{\bsl{q}}{2}))(-i\omega_n-\xi_{\lambda}(-\bsl{k}+\frac{\bsl{q}}{2}))}+\frac{\text{Tr}[\bsl{B}\cdot\bsl{M}^{\lambda}(\bsl{k}+\frac{\bsl{q}}{2})n^{a_2}_{\lambda}(\bsl{k})[n^{a_1}_{\lambda}(\bsl{k})]^{\dagger}]}{(i\omega_n-\xi_{\lambda}(\bsl{k}+\frac{\bsl{q}}{2}))^2(-i\omega_n-\xi_{\lambda}(-\bsl{k}+\frac{\bsl{q}}{2}))}\nonumber\\
&+&
\frac{\text{Tr}[ n^{a_2}_{\lambda}(\bsl{k})[\bsl{B}\cdot\bsl{M}^{\lambda}(-\bsl{k}+\frac{\bsl{q}}{2})]^T[n^{a_1}_{\lambda}(\bsl{k})]^{\dagger}]}{(i\omega_n-\xi_{\lambda}(\bsl{k}+\frac{\bsl{q}}{2}))(-i\omega_n-\xi_{\lambda}(-\bsl{k}+\frac{\bsl{q}}{2}))^2}
= \frac{\text{Tr}[n^{a_2}_{\lambda}(\bsl{k})[n^{a_1}_{\lambda}(\bsl{k})]^{\dagger}]}{(i\omega_n-\xi_{\lambda}(\bsl{k}+\frac{\bsl{q}}{2}))(-i\omega_n-\xi_{\lambda}(-\bsl{k}+\frac{\bsl{q}}{2}))}\ ,
\end{eqnarray}
\end{widetext}
where the summation over $\alpha_1\alpha_2\beta_1\beta_2$ is implied in the first expression and the last equality uses the fact that $\bsl{B}\cdot \bsl{M}^{\pm}(\bsl{k})$ are traceless. Therefore, the Zeeman coupling does not contribute to the first order term of magnetic fields in the free energy. Eventually, we have
\begin{eqnarray}
&&S^{a_1 a_2}(\bsl{r}_1,\bsl{r}_2)=e^{-i\frac{2e}{\hbar}\bsl{r}_1\cdot\bsl{A}(\bsl{r}_2)}\sum_{\omega_n,\lambda}\frac{1}{\beta}\sum_{\bsl{k},\bsl{q}} e^{i\bsl{q}\cdot(\bsl{r}_1-\bsl{r}_2)}\nonumber\\
&&\frac{\text{Tr}[n^{a_2}_{\lambda}(\bsl{k})[n^{a_1}_{\lambda}(\bsl{k})]^{\dagger}]}{(i\omega_n-\xi_{\lambda}(\bsl{k}+\frac{\bsl{q}}{2}))(-i\omega_n-\xi_{\lambda}(-\bsl{k}+\frac{\bsl{q}}{2}))}\ .
\end{eqnarray}

\subsubsection{Simplification of $S^{a_1 a_2}(\mathbf{r}_1,\mathbf{r}_2)$}
\label{app:dev_eq_K}
In this part, we further simplify $S^{a_1 a_2}(\bsl{r}_1,\bsl{r}_2)$ following Ref.[\citen{Samokhin2004magnetic}].
Einstein summation notation for repeated indexes is used in this part.

First consider the expansion of the following expression to the second order of $|\bsl{q}|$:
\begin{eqnarray}
&&\sum_{\omega_n} \frac{1}{(i\omega_n-\xi_{\lambda}(\bsl{k}+\frac{\bsl{q}}{2}))(-i\omega_n-\xi_{\lambda}(-\bsl{k}+\frac{\bsl{q}}{2}))}\nonumber\\
&&
=S_0(\xi_{\lambda})
+
q_{i}q_{j}[S_2(\xi_{\lambda})v^{\lambda}_{i}v^{\lambda}_{j}
+S_1(\xi_{\lambda})w^{\lambda}_{i j}]\ ,
\end{eqnarray}
where
$\xi_{\lambda}$ is short for $\xi_{\lambda}(\bsl{k})$
,
$\xi_{\lambda}(-\bsl{k})=\xi_{\lambda}(\bsl{k})$ is used
,
$v^{\lambda}_{i}=\partial_{k_{i}}\xi_{\lambda}(\bsl{k})$
,
$w^{\lambda}_{i j}=\partial_{k_{i}}\partial_{k_{j}}\xi_{\lambda}(\bsl{k})$
,
\begin{equation}
S_0(\xi_{\lambda})=\sum_{\omega_n} \frac{1}{\omega_n^2+\xi_{\lambda}^2}=\frac{\beta  \tanh \left(\frac{\beta  \xi_{\lambda} }{2}\right)}{2 \xi_{\lambda} }\ ,
\end{equation}
\begin{equation}
S_1(\xi_{\lambda})=\sum_{\omega_n} \frac{1}{4(i \omega_n-\xi_{\lambda})^2(-i\omega_n-\xi_{\lambda})}=\frac{1}{8}S'_0(\xi_{\lambda})\ ,
\end{equation}
and
\begin{eqnarray}
&&S_2(\xi_{\lambda})=\sum_{\omega_n} \frac{1}{2(i\omega_n-\xi_{\lambda})^3(-i\omega_n-\xi_{\lambda})}-\nonumber\\
&&
\sum_{\omega_n} \frac{1}{4(i\omega_n-\xi_{\lambda})^2(-i\omega_n-\xi_{\lambda})^2}\nonumber\\
&&
=-\frac{\beta^3\cosh^{-3}(\frac{\beta\xi_{\lambda}}{2})\sinh(\frac{\beta\xi_{\lambda}}{2})}{32\xi_{\lambda}}\ .
\end{eqnarray}
Then, we have
\begin{eqnarray}
&&\frac{1}{\mathcal{V}\beta}\sum_{\omega_n,\lambda,\bsl{k}} \frac{\text{Tr}[n^{a_2}_{\lambda}(\bsl{k})[n^{a_1}_{\lambda}(\bsl{k})]^{\dagger}]}{(i\omega_n-\xi_{\lambda}(\bsl{k}+\frac{\bsl{q}}{2}))(-i\omega_n-\xi_{\lambda}(-\bsl{k}+\frac{\bsl{q}}{2}))}\nonumber\\
=
&& K_0^{a_1 a_2}+q_{i} q_{j} K^{a_1 a_2}_{1,i j}+O(|\bsl{q}|^4)\ ,
\end{eqnarray}
where
\begin{equation}
K_0^{a_1 a_2}=\frac{1}{\mathcal{V}\beta}\sum_{\lambda,\bsl{k}} \text{Tr}[n^{a_2}_{\lambda}(\bsl{k})[n^{a_1}_{\lambda}(\bsl{k})]^{\dagger}]S_0(\xi_{\lambda})\ ,
\end{equation}
and
\begin{eqnarray}
&&
K^{a_1 a_2}_{1,i j}=\frac{1}{\mathcal{V}\beta}\sum_{\lambda,\bsl{k}} \text{Tr}[n^{a_2}_{\lambda}(\bsl{k})[n^{a_1}_{\lambda}(\bsl{k})]^{\dagger}]\nonumber\\
&& [S_2(\xi_{\lambda})v^{\lambda}_{i}v^{\lambda}_{j}
+S_1(\xi_{\lambda})w^{\lambda}_{i j}]\ .
\end{eqnarray}

$K_0^{a_1 a_2}$ has been carried out in Ref.[\citen{yu2017Singlet-Quintetj=3/2SC}], which has the expression
\begin{equation}
\label{eq:expan_K0}
K_0=\frac{x N_0}{2}
u\left(
\begin{array}{cc}
y_1 & y_2\\
y_2 & y_3\\
\end{array}
\right) u\ ,
\end{equation}
where $x=\ln (2 e^{\bar{\gamma}}\beta\epsilon_c/\pi)$, $\bar{\gamma}$ is Euler's constant, $u=\text{diag}(\text{sgn}(c_1),2m\mu a^2)$, and expressions of $y_{1,2,3}$ are in Appendix.\ref{app:conv_expn}.

Now we simplify $K_{1,i j}^{a_1 a_2}$.
Firstly we show that $K_{1,i j}^{a_1 a_2}$ is proportional to $\delta_{i j}$.
Since $\xi_{\lambda}(R^{-1}\bsl{k})=\xi_{\lambda}(\bsl{k})$ for any operation $R$ in $O_h$ group with $(R^{-1}\bsl{k})_{i}=R_{i i'}^{-1}k_{i'}$,
we have
\begin{equation}
v_{i}^{\lambda}(\bsl{k})=\frac{\partial \xi_{\lambda}(\bsl{k})}{\partial k_{i}}=\frac{\partial k'_{i'}}{\partial k_{i}}\frac{\partial \xi_{\lambda}(R^{-1}\bsl{k}')}{\partial k'_{i'}}=v_{i'}^{\lambda}(\bsl{k}')R_{i' i}
\end{equation}
and
\begin{equation}
w_{i j}^{\lambda}(\bsl{k})=w_{i' j'}^{\lambda}(\bsl{k}')R_{i' i}R_{j' j}\ ,
\end{equation}
where $\bsl{k}'=R\bsl{k}$.
Due to $n^{a_1}_{\lambda}(R^{-1}\bsl{k})=n^{a_1}_{\lambda}(\bsl{k})$, we can derive that
\begin{equation}
K_{1,i j}^{a_1 a_2}=K_{1,i' j'}^{a_1 a_2} R_{i' i}R_{j' j}
\end{equation}
holds for any operation $R$ in $O_h$ group, which leads to
\begin{equation}
K_{1,i j}^{a_1 a_2}=K_{1,z z}^{a_1 a_2}\delta_{ij}=K_{1}^{a_1 a_2}\delta_{i j}\ .
\end{equation}

Among $K_{1}^{a_1 a_2}$, the term including $S_1$ reads
\begin{equation}
I_1^{a_1 a_2}=\frac{1}{\mathcal{V}\beta}\sum_{\lambda,\bsl{k}}\text{Tr}[n^{a_2}_{\lambda}(\bsl{k})[n^{a_1}_{\lambda}(\bsl{k})]^{\dagger}]w^{\lambda}_{zz}S_1(\xi_{\lambda})
\end{equation}
,the term including $S_2$ reads
\begin{equation}
I_2^{a_1 a_2}=\frac{1}{\mathcal{V}\beta}\sum_{\lambda,\bsl{k}}\text{Tr}[n^{a_2}_{\lambda}(\bsl{k})[n^{a_1}_{\lambda}(\bsl{k})]^{\dagger}]v^{\lambda}_{z}v^{\lambda}_{z}S_2(\xi_{\lambda})\ ,
\end{equation}
and $K_{1}^{a_1 a_2}=I_1^{a_1 a_2}+I_2^{a_1 a_2}$.

For $I_1^{a_1 a_2}$ and $I_2^{a_1 a_2}$, we have
\begin{widetext}
\begin{eqnarray}
I_1^{a_1 a_2}&=&\frac{1}{8\beta}\sum_{\lambda}\int \frac{d\Omega}{4\pi}\theta(\widetilde{m}_{\lambda})N_{\lambda}(0)\int_{-\epsilon_c}^{\epsilon_c} d\xi_{\lambda}\text{Tr}[n^{a_2}_{\lambda}(\bsl{k})[n^{a_1}_{\lambda}(\bsl{k})]^{\dagger}]w^{\lambda}_{zz}S_0'(\xi_{\lambda})\sqrt{\frac{\xi_{\lambda}}{\mu}+1}\nonumber\\
&\approx&-\frac{1}{8\beta}\sum_{\lambda}\int \frac{d\Omega}{4\pi}\theta(\widetilde{m}_{\lambda})N_{\lambda}(0)\int_{-\epsilon_c}^{\epsilon_c} d\xi_{\lambda}\frac{d}{d\xi_{\lambda}}\left(\text{Tr}[n^{a_2}_{\lambda}(\bsl{k})[n^{a_1}_{\lambda}(\bsl{k})]^{\dagger}]w^{\lambda}_{zz}\sqrt{\frac{\xi_{\lambda}}{\mu}+1}\right)S_0(\xi_{\lambda})\nonumber\\
&\approx&-\frac{x}{8}\sum_{\lambda}\int \frac{d\Omega}{4\pi}\theta(\widetilde{m}_{\lambda})N_{\lambda}(0)\left.\frac{d}{d\xi_{\lambda}}\left(\text{Tr}[n^{a_2}_{\lambda}(\bsl{k})[n^{a_1}_{\lambda}(\bsl{k})]^{\dagger}]w^{\lambda}_{zz}\sqrt{\frac{\xi_{\lambda}}{\mu}+1}\right)\right|_{\xi_{\lambda}\rightarrow 0}\ ,
\end{eqnarray}
and
\begin{eqnarray}
I_2^{a_1 a_2}&=&\frac{1}{\beta}\sum_{\lambda}\int \frac{d\Omega}{4\pi}\theta(\widetilde{m}_{\lambda})N_{\lambda}(0)\int_{-\epsilon_c}^{\epsilon_c} d\xi_{\lambda}\text{Tr}[n^{a_2}_{\lambda}(\bsl{k})[n^{a_1}_{\lambda}(\bsl{k})]^{\dagger}](v^{\lambda}_{z})^2 S_2(\xi_{\lambda})\sqrt{\frac{\xi_{\lambda}}{\mu}+1}\nonumber\\
&\approx&\sum_{\lambda}\int \frac{d\Omega}{4\pi} \theta(\widetilde{m}_{\lambda}) N_{\lambda}(0)\text{Tr}[n^{a_2}_{\lambda}(\bsl{k}_{F,\lambda})[n^{a_1}_{\lambda}(\bsl{k}_{F,\lambda})]^{\dagger}](v^{\lambda}_{z}(\bsl{k}_{F,\lambda}))^2 \frac{-7\beta^2}{16\pi^2}\zeta(3)\ ,
\end{eqnarray}
\end{widetext}
where $|\bsl{k}_{F,\lambda}|=\sqrt{2m_\lambda \mu}$, $N_{\lambda}(0)=N_0\widetilde{m}_{\lambda}^{3/2}$, $\zeta(...)$ is the Riemann $\zeta$ function and the result is to the leading order of $1/(\beta\epsilon_c)\ll 1$ and $\epsilon_c/|\mu|\ll 1$.
Before further derivation, let us first estimate the order of those two terms in the isotropic case.
In that case, $m_{\lambda}$, $\xi_{\lambda}$, $n^{a_1}_{\lambda}$ and $N_{\lambda}(0)$ are independent of the angle. Then we can get the magnitude dependence for the following quantities:
$W^{\lambda}_{zz}\propto (m_{\lambda})^{-1}$, $(v^{\lambda}_{z}(\bsl{k}_{F,\lambda}))^2\propto \mu (m_{\lambda})^{-1}$, $\text{Tr}[n^{a_2}_{\lambda}(\bsl{k}_{F,\lambda})[n^{a_1}_{\lambda}(\bsl{k}_{F,\lambda})]^{\dagger}]\propto (m_{\lambda} \mu a^2)^{a_1+a_2}$.
Finally, we have
\begin{equation}
I_1^{a_1 a_2}\propto \frac{x}{\mu}  \sum_{\lambda}\theta(\widetilde{m}_{\lambda}) N_{\lambda}(0)  \frac{(m_{\lambda}\mu a^2)^{a_1+a_2}}{m_{\lambda}}
\end{equation}
compared with
\begin{equation}
I_2^{a_1 a_2}\propto \beta^2 \mu  \sum_{\lambda} \theta(\widetilde{m}_{\lambda}) N_{\lambda}(0)  \frac{(m_{\lambda}\mu a^2)^{a_1+a_2}}{m_{\lambda}}\ .
\end{equation}
Since $(\beta \mu)^2\gg (\beta \epsilon_c)^2 \gg \beta \epsilon_c \gg \ln(\beta \epsilon_c)\sim x$, we have $I_2^{a_1 a_2}\gg I_1^{a_1 a_2}$ and can neglect $I_1^{a_1 a_2}$ to get $K_{1}^{a_1 a_2}\approx I_2^{a_1 a_2}$.
Then, by defining $(\widetilde{v}_z^{\lambda})^2$ and $z_{1,2,3}$ as in Appendix.\ref{app:conv_expn}, we have
\begin{equation}
\label{eq:expan_K1}
K_1=-\frac{N_0}{2}\frac{\beta^2\mu}{2m}
u\left(
\begin{array}{cc}
z_1 & z_2\\
z_2 & z_3\\
\end{array}
\right) u  \ ,
\end{equation}
where the expressions of $z_{1,2,3}$ are shown in Appendix.\ref{app:conv_expn}.


As a result, we have
\begin{eqnarray}
&&S^{a_1 a_2}(\bsl{r}_1,\bsl{r}_2)=\mathcal{V}^2[K_0^{a_1 a_2}\delta(\bsl{r}_1-\bsl{r}_2)\nonumber\\
&&
+
e^{-i\frac{2e}{\hbar}\bsl{r}_1\cdot\bsl{A}(\bsl{r}_2)}K^{a_1 a_2}_{1}(i\bsl{\nabla}_{\bsl{r}_2})^2\delta(\bsl{r}_1-\bsl{r}_2)]\ .
\end{eqnarray}
Substituting the expression shown above into Eq.\ref{eq:F_r}, we can get Eq.\ref{eq:F_r_final}.

\section{Derivation of Eq.\ref{eq:lge_final_mat} and Eq.\ref{eq:Bc2_T}}
\label{app:Tc_Bc}
At first, we derive the eigenvalues of $\bsl{D}^2$.
Suppose $\bsl{B}=B \hat{e}_3$.
Assume that $\hat{e}_1$ and $\hat{e}_2$ are the two orthogonal directions perpendicular to $\hat{e}_3$ and satisfy $\hat{e}_1\times \hat{e}_2=\hat{e}_3$.
Then, $\bsl{D}^2=D_1^2+D_2^2+D_3^2$.
It is the similar to the Landau level problem.
Since $\bsl{D}=-i\bsl{\nabla}_{\bsl{r}}+\frac{e}{\hbar}(\bsl{B}\times\bsl{r})$, we have $D_3=-i\partial_{r_3}$,
\begin{equation}
[D_1,D_2]=-i\frac{2 e}{\hbar}B
\end{equation}
as well as $[D_1,D_3]=[D_2,D_3]=0$.
Define $\hat{a}=\sqrt{\frac{\hbar}{4e B}}(D_1-i D_2)$, we have
\begin{equation}
[\hat{a},\hat{a}^{\dagger}]=\frac{\hbar}{4e B}2i[D_1,D_2]=1\ .
\end{equation}
In this case, $\bsl{D}^2$ can be re-written as
\begin{equation}
\bsl{D}^2=\frac{4eB}{\hbar}(\hat{a}^{\dagger}\hat{a}+\frac{1}{2})+(-i\partial_{r_3})^2\ ,
\end{equation}
of which the eigenvalue is
\begin{equation}
l^2=\frac{4eB}{\hbar}(n+\frac{1}{2})+(k_3)^2
\end{equation}
with $n\geq 0$ and $k_3$ being the component of the momentum along the magnetic field direction.

Next, we solve for the upper critical field.
The linearized gap equation directly given by Eq.\ref{eq:F_r_final} reads
\begin{equation}
\label{eq:lge}
\widetilde{\Delta}_{a_1}(\bsl{r})=-\sum_{a_2}\widetilde{V}_{a_1}(\widetilde{K}_0^{a_1 a_2}+\widetilde{K}_1^{a_1 a_2}\bsl{D}^2)\widetilde{\Delta}_{a_2}(\bsl{r})\ .
\end{equation}
Since Eq.\ref{eq:lge} is linear and $\bsl{D}^2$ is Hermitian, eigenfunctions of $\bsl{D}^2$ with different eigenvalues can not be coupled.
Suppose $\widetilde{\Delta}_{a}(\bsl{r})$'s are the eigenfunctions of $\bsl{D}^2$ with eigenvalue $l^2$, then the linearized gap equation becomes
\begin{equation}
\label{eq:lge_final}
\widetilde{\Delta}_{a_1}(\bsl{r})=-\sum_{a_2}\widetilde{V}_{a_1}(\widetilde{K}_0^{a_1 a_2}+\widetilde{K}_1^{a_1 a_2} l^2)\widetilde{\Delta}_{a_2}(\bsl{r})\ .
\end{equation}
Eq.\ref{eq:lge_final_mat} is just the matrix version of Eq.\ref{eq:lge_final}.

Assume the $l^2$ is of the same order as $eB/\hbar$ meaning that the order of $k_3$ is no larger than the order of $eB/\hbar$ and $n$ is not large.
The resulted expression of the transition temperature $T$ to the first order of $B$ reads
\begin{equation}
\label{eq:T_Bc2}
\frac{T}{T_{c}}=1+ \frac{\beta_{c}^2 \mu^2 x_{c}}{8m \mu}\alpha l^2\ .
\end{equation}
Typically we have $\alpha<0$, meaning that the highest $T$ is given by smallest $l^2$ that is $2 e B/\hbar$.
Replacing $B$ by $B_{c,2}$, we have Eq.\ref{eq:Bc2_T}.

\begin{figure}[t]
\includegraphics[width=\columnwidth]{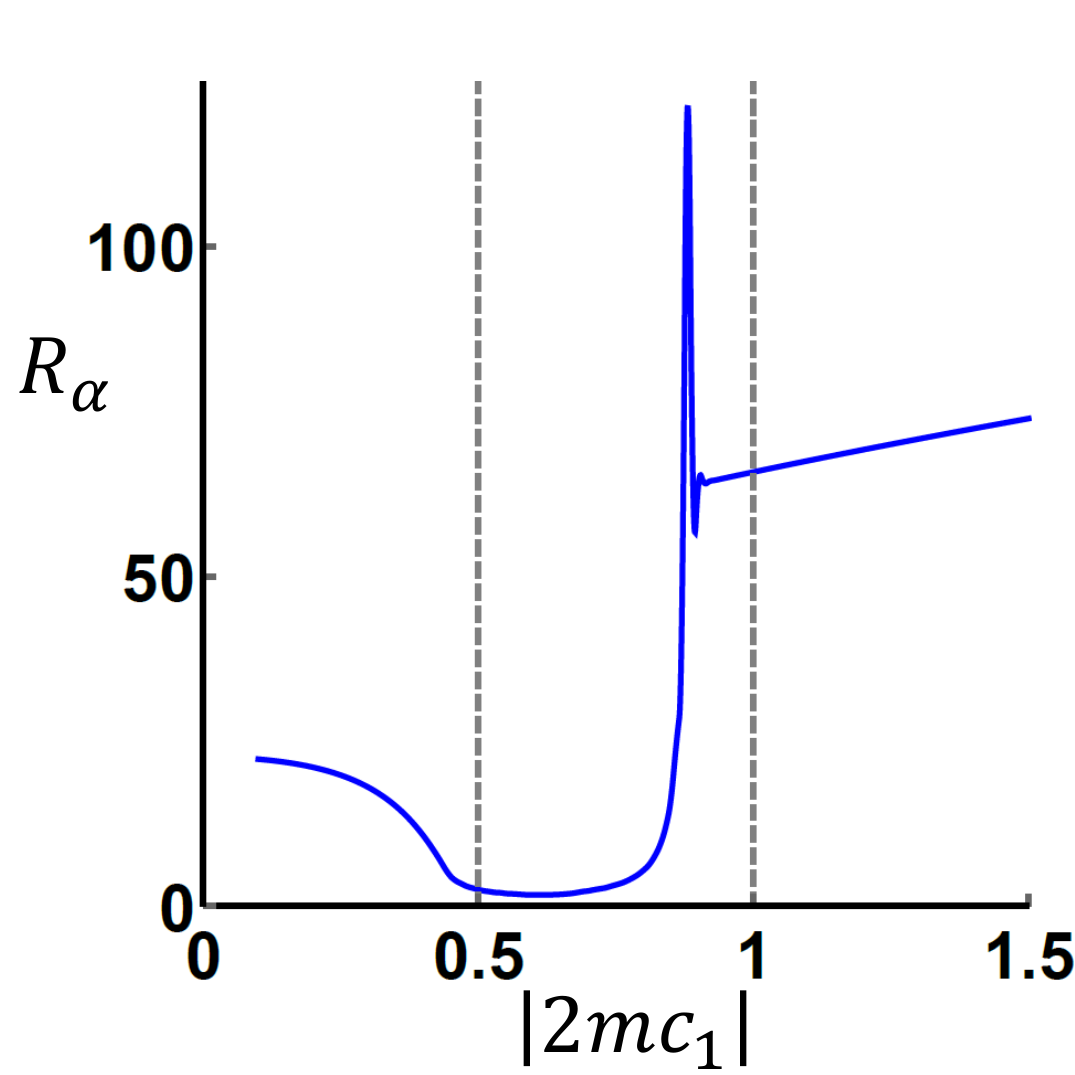}
\caption{
\label{fig:alpha_factor}
This shows the $\alpha$ factor $\frac{1}{-\alpha}\left(\frac{2e^{\bar{\gamma}}}{\pi}\right)^2$ of the slope $-\frac{d B_{c,2}/B_0}{d T/T_0}$ as a function of SSOC $|2m c_1|$. The large change of the $\alpha$ factor in the intermediate region between regime III and II does not have much effects on the slope due to the small $T_c$ factor, as shown in Fig.\ref{fig:upper_critical_field}a.
}
\end{figure}

\section{Disorder Average and Replica Trick}
\label{app:replica}
In this part, we follow Ref.[\citen{altland2010condensed}] to introduce the Replica trick based on our model.
Note that, in this part, we temporarily abandon the previous defined $x=\frac{2 e^{\bar{\gamma}}\beta\epsilon_c}{\pi}$ and define $x=(\tau,\bsl{r})$ instead.

We start from discussing the disorder average of a certain observable for the disorder term Eq.\ref{eq:H_dis} and the probability measure Eq.\ref{eq:PV_dis}.
Given a non-interacting partition function with the random disorder
\begin{equation}
\label{eq:Z_0_dis}
Z_0[V]=\int D\bar{c} Dc \exp(-S[\bar{c},c,V])\ ,
\end{equation}
where
\begin{equation}
\label{eq:S_dis}
S[\bar{c},c,V]=S_0[\bar{c},c]+\int dx V(\bsl{r}) \bar{c}_x c_x\ ,
\end{equation}
$\bar{c},c$ are Grassmann fields if appearing in the action, \begin{equation}
\label{eq:S_0}
-S_0=\sum_{\bsl{k},\omega_n}\bar{c}_{\bsl{k},\omega_n}(i\omega_n-h(\bsl{k}))c_{\bsl{k},\omega_n}\ ,
\end{equation}
$x=(\tau,\bsl{r})$ and $\tau$ is the imaginary time.
Suppose we want to compute thermal average of certain observable $O_i(c^{\dagger},c)$ in the presence of the random disorder:
\begin{eqnarray}
\label{gen_aver_O}
&&\left\langle O_i(\bar{c},c)\right\rangle=\frac{\int D c^{\dagger} Dc O_i(\bar{c},c)\exp(-S[\bar{c},c,V])}{Z_0[V]}\nonumber\\
&&=\left. \frac{\delta}{\delta J_i} \ln(Z[V,J])\right|_{J\rightarrow 0}\ ,
\end{eqnarray}
where
\begin{equation}
Z[V,J]=\int D\bar{c} Dc \exp(-S[\bar{c},c,V]+\int dX \sum_i J_i O_i)\ ,
\end{equation}
and $X$ denotes the imaginary time and position dependence of $O_i$.
Now, one may take the disorder average of $\left\langle O_i(\bar{c},c)\right\rangle$.
However, due to $Z_0[V]$ in denominator of Eq.\ref{gen_aver_O}, the disorder average is hard to carry out directly.
One way to overcome it is the replica trick.
Since $\ln(Z[V,J])=\lim_{R\rightarrow 0} (Z[V,J]^R-1)/R$, we have
\begin{equation}
\left. \frac{\delta}{\delta J_i} \ln(Z[V,J])\right|_{J\rightarrow 0}=\lim_{R\rightarrow 0}\frac{1}{R}\left. \frac{\delta}{\delta J_i}Z[V,J]^R\right|_{J\rightarrow 0}\ .
\end{equation}
If $R$ is integer,
\begin{eqnarray}
&&Z[V,J]^R= \int D\bar{\Psi} D\Psi\exp(-S[\bar{\Psi},\Psi,V]\nonumber\\
&&+\int dX \sum_i J_i O_i(\bar{\Psi},\Psi))\ ,
\end{eqnarray}
where $\Psi=(c_1,...,c_R)^T$, $\bar{\Psi}=(\bar{c}_1,...,\bar{c}_R)$, $O_i(\bar{\Psi},\Psi)=\sum_a O_i(\bar{c}_a,c_a)$, $S[\bar{\Psi},\Psi,V]=\sum_{a=1}^R S[\bar{c}_a,c_a,V]$ and $a=1,...,R$ is the replica index.
Then, we have
\begin{equation}
\left. \frac{\delta}{\delta J_i}Z[V,J]^R\right|_{J\rightarrow 0}= \int D\bar{\Psi} D\Psi O_i(\bar{\Psi},\Psi) \exp(-S[\bar{\Psi},\Psi,V])\ .
\end{equation}
The the disorder average becomes
\begin{eqnarray}
\label{Gen_disaver_O_final}
&&\left\langle\left\langle O_i(\bar{c},c)\right\rangle\right\rangle_{dis}
=\nonumber\\
&&
\lim_{R\rightarrow 0}\frac{1}{R}\int D\bar{\Psi} D\Psi O_i(\bar{\Psi},\Psi) \langle\exp(-S[\bar{\Psi},\Psi,V]) \rangle_{dis} \ ,
\end{eqnarray}
where 
\begin{eqnarray}
\label{Z:int_out_disorder}
&&\langle\exp(-S[\bar{\Psi},\Psi,V]) \rangle_{dis}=\exp(-S_0[\bar{\Psi},\Psi])\times\nonumber\\
&& \langle\exp(-\int dx V(\bsl{r})\bar{\Psi}(x)\Psi(x)) \rangle_{dis}
\end{eqnarray}
with
\begin{eqnarray}
&&\langle \exp(-\int dx V(\bsl{r})\bar{\Psi}(x)\Psi(x)) \rangle_{dis}\nonumber\\
&& =\frac{\int DV P[V]\exp(-\int d^3 r V(\bsl{r})\int d\tau \bar{\Psi}(x)\Psi(x))}{\int DV P[V]}\nonumber\\
&& =\frac{\int DV \exp(-\frac{1}{2\gamma^2_d} \int d^3 r[ V^2 (\bsl{r})+2\gamma_d^2 V(\bsl{r})\int d\tau \bar{\Psi}(x)\Psi(x)])}{\int DV P[V]}\nonumber\\
&& =\exp[\frac{\gamma_d^2}{2}\int d^3 r (\int d\tau \bar{\Psi}(x)\Psi(x))^2]\nonumber\\
&& = \exp[\frac{\gamma^2_d}{2}\int dx dx' \delta(\bsl{r}-\bsl{r}') \bar{\Psi}(x)\Psi(x)\bar{\Psi}(x')\Psi(x')].
\end{eqnarray}
Here the probability measure $P[V]$ is defined in Eq.\ref{eq:PV_dis}, and the limitation of $R$ should be taken as the limitation of the analytic continuity of the function of integer $R$'s.
Although the failure of this trick is possible since the limit of the analytic continuity may not be the real limit, the trick works well for most of the times.

Next, we discuss the Feymann rule.
Recall the non-interacting action,
\begin{equation}
-S_0[\bar{\Psi},\Psi] =\sum_a \sum_{\bsl{k},\omega_n} c^{\dagger}_{a,\bsl{k},\omega_n} G^{-1}_0(\bsl{k},\omega_n) c_{a,\bsl{k},\omega_n}\ ,
\end{equation}
where
\begin{equation}
\label{eq:G_0}
G_0(\bsl{k},\omega_n)=(i\omega_n-h(\bsl{k}))^{-1}\ .
\end{equation}
Based on the expression, if using
\begin{equation}
\frac{\delta \bar{c}_{a,\bsl{k},\omega_n,\alpha_1}}{\delta \bar{c}_{a',\bsl{k}',\omega_n',\alpha_2}}=\delta_{\bsl{k},\bsl{k}'}\delta_{\omega_n,\omega_n'}\delta_{a,a'}\delta_{\alpha_1,\alpha_2}\ ,
\end{equation}
the fermion line corresponds to $-[G_0(\bsl{k},\omega_n)]_{\alpha_1\alpha_2}\delta_{a_1 a_2}$, which conserves the replica index and momentum $(\bsl{k},\omega_n)$.

The Fourier transform of the four fermionic field interaction generated by integrating out the disorder potential in Eq.\ref{Z:int_out_disorder} reads
\begin{eqnarray}
&&\frac{\gamma^2_d}{2}\int dx dx' \delta(\bsl{r}-\bsl{r}') \bar{\Psi}(x)\Psi(x)\bar{\Psi}(x')\Psi(x')\nonumber\\
&&=\frac{\gamma_d^2}{2\mathcal{V}}
\sum_{\omega,\omega'}\sum_{\bsl{k}_1,\bsl{k}_2,\bsl{k}_3,\bsl{k}_4}\sum_{a_1,a_2}\sum_{\alpha_1,\alpha_2}\delta_{\bsl{k}_1+\bsl{k}_3,\bsl{k}_2+\bsl{k}_4}\nonumber\\
&&\bar{c}_{a_1,\bsl{k}_1,\omega,\alpha_1}c_{a_1,\bsl{k}_2,\omega,\alpha_1}\bar{c}_{a_2,\bsl{k}_3,\omega',\alpha_2}c_{a_2,\bsl{k}_4,\omega',\alpha_2}\ .
\end{eqnarray}
This is clear that this effective $(\bar{c}c)^2$ vortex corresponds to $\frac{\gamma^2_d}{\mathcal{V}}$.
The vortex can be noted as a dashed line (disorder line) connected with two fermionic lines at either end.
It conserves replica index $a$, spin index $\alpha$ and frequency $\omega_n$ only for the fermionic lines and conserves spacial momentum $\bsl{k}$ for the entire vortex.

For any disorder average of fermionic operators (must contain equal number of $\bar{c}$ and $c$ due to global $U(1)$ symmetry), say $n$ pairs of $\bar{c}$ and $c$, the graph must contain $n$ fermionic lines without forming fermionic loops.
Each of the $n$ lines would give $\delta_{aa}=1$, which eventually leads to a factor of $R$ after the summation of replica index in the definition of the observable with replica index.
That $R$ cancels the one in the denominator of Eq.\ref{Gen_disaver_O_final}.
If a graph contains at least one fermionic loop, each loop would give a factor of $R$.
After the limitation $R\rightarrow 0$, it is clear that all graphs with fermionic loops would give zero.
Therefore, we only need to consider graphs without fermion loops.
In this case, we can simplify the disorder average of fermionic fields $O_i(\bar{c},c)$ by neglecting the replica index and considering the all graphs without fermionic loops of the following expression:
\begin{eqnarray}
\label{eq:disaver_final}
&&\left\langle\left\langle O_i(\bar{c},c)\right\rangle\right\rangle_{dis}=
\int D\bar{c} Dc O_i(\bar{c},c)\exp\{-S_0[\bar{c},c]\nonumber\\
&&+\frac{\gamma^2_d}{2}\int dx dx' \delta(\bsl{r}-\bsl{r}') \bar{c}_x c_x \bar{c}_{x'}c_{x'}\}\ .
\end{eqnarray}

\section{Derivation of Linearized Gap Equation with Disorder Eq.\ref{eq:lge_final_dis}}
\label{app:der_lge_dis}
In this section, we derive Eq.\ref{eq:lge_final_dis}.
We first derive the disorder-averaged normal state green function, and then derive the superconducting free energy with disorders, and finally get Eq.\ref{eq:lge_final_dis}.

\subsubsection{Disorder-averaged Normal State Green Function}
\label{app:dis_Ge}
The disorder-averaged Green function is defined as
\begin{equation}
\bar{G}_{\alpha\beta}(\bsl{k},\omega_n)=-\left\langle\left\langle c_{\bsl{k},\omega_n,\alpha} \bar{c}_{\bsl{k},\omega_n,\beta}\right\rangle\right\rangle_{dis}\ .
\end{equation}
It can be expressed as
\begin{equation}
[\bar{G}(\bsl{k},\omega)]^{-1}=[G_0(\bsl{k},\omega)]^{-1}-\Sigma(\bsl{k},\omega)\ ,
\end{equation}
where $\Sigma(\bsl{k},\omega)$ is called the self-energy.

Since we adopted the Born approximation, $\Sigma(\bsl{k},\omega)$ only depends on $\omega$ and satisfies the self-consistent Born approximation (SCBA) equation:
\begin{equation}
\label{eq:SCBA}
\Sigma(\omega)=\frac{\gamma^2_d}{\mathcal{V}}\sum_{\bsl{k}'}G_0(\bsl{k}',\omega) + \frac{\gamma^2_d }{\mathcal{V}}\sum_{\bsl{k}'}G_0(\bsl{k}',\omega)\Sigma(\omega)G_0(\bsl{k}',\omega)\ .
\end{equation}

Define
\begin{equation}
\label{eq:proj_P}
P_{\pm}(\bsl{k})=\frac{1}{2}\pm \frac{h(\bsl{k})-\xi_{\bsl{k}}}{2 k^2 Q_c}
\end{equation}
to be the projection operators to $\xi_{\pm}$ bands respectively.
The normal state Green function without disorder is given in Eq.\ref{eq:G_0}, which can be expressed in terms of projection operators
\begin{equation}
G_0(\bsl{k},\omega)=\sum_{\lambda=\pm}\frac{1}{i\omega-\xi_{\lambda}(\bsl{k})}P_{\lambda}(\bsl{k})\ .
\end{equation}
Using the expression of $P_{\lambda}(\bsl{k})$, we have
\begin{eqnarray}
&&\sum_{\bsl{k}}G_0(\bsl{k},\omega_n)
=
\sum_{\bsl{k}}
\frac{1}{2}(\frac{1}{i\omega_n-\xi_{+}(\bsl{k})}+\frac{1}{i\omega_n-\xi_{-}(\bsl{k})})\nonumber\\
&&+\sum_{\bsl{k}}(\frac{1}{i\omega_n-\xi_{+}(\bsl{k})}-\frac{1}{i\omega_n-\xi_{-}(\bsl{k})})\frac{h(\bsl{k})-\xi_{\bsl{k}}}{2 k^2 Q_c}\ .
\end{eqnarray}
Since the second term can be rewritten as
$
\sum_{\bsl{k}}\sum_{i=1}^5 f_i(\bsl{k}) g_{\bsl{k},i} \Gamma^i
$
with $f_i(\bsl{k})$ being $O_h$ invariant,
the second term should be zero.
Then, we have
\begin{equation}
\sum_{\bsl{k}}G_0(\bsl{k},\omega_n)
=
\sum_{\bsl{k}}
\frac{1}{2}(\frac{1}{i\omega_n-\xi_{+}(\bsl{k})}+\frac{1}{i\omega_n-\xi_{-}(\bsl{k})})\ ,
\end{equation}
which is proportional to identity matrix.
Similarly, we have
\begin{eqnarray}
&&\sum_{\bsl{k}}G_0(\bsl{k},\omega_n) G_0(\bsl{k},\omega_n)\nonumber\\
&&=
\sum_{\bsl{k}}\left[\frac{1}{(i\omega_n-\xi_{+}(\bsl{k}))^2}P_+(\bsl{k})+\frac{1}{(i\omega_n-\xi_{-}(\bsl{k}))^2}P_-(\bsl{k})\right]\nonumber\\
&&=
\sum_{\bsl{k}}
\frac{1}{2}\left[\frac{1}{(i\omega_n-\xi_{+}(\bsl{k}))^2}+\frac{1}{(i\omega_n-\xi_{-}(\bsl{k}))^2}\right]\ .
\end{eqnarray}
Then, by induction, we get that $\Sigma(\omega)$ is proportional to the identity matrix.
Thereby, Eq.\ref{eq:SCBA} can be re-written as
\begin{equation}
\Sigma(\omega)=\frac{\gamma^2_d}{\mathcal{V}}\sum_{\bsl{k}'}G_0(\bsl{k}',\omega)[1- \frac{\gamma^2_d }{\mathcal{V}}\sum_{\bsl{k}''}G_0^2(\bsl{k}'',\omega)]^{-1}\ .
\end{equation}

Now estimate the order of $G_0^2(\omega,\bsl{k}')$ term.
The term can be re-written as
\begin{equation}
\frac{\gamma^2_d }{\mathcal{V}}\sum_{\bsl{k}'}G_0^2(\bsl{k}',\omega)
=
(\gamma^2_d) \int d\varepsilon \frac{N(\varepsilon)}{2} \frac{1}{(i\omega-\varepsilon)^2} \ ,
\end{equation}
where
\begin{equation}
N(\varepsilon)=\langle N_+(\epsilon)\rangle_{\Omega}+\langle N_-(\epsilon)\rangle_{\Omega}\ ,
\end{equation}
and $ \langle N_{\pm}(\varepsilon)\rangle_{\Omega}=\frac{1}{\mathcal{V}}\sum_{\bsl{k}'}\delta(\varepsilon-\xi_{\pm}(\bsl{k}'))$ are density of states of $\xi_{\pm}$ bands at $\varepsilon$ without spin index and $\langle ...\rangle_{\Omega}$ is the average over the solid angle.
Then, we have
\begin{equation}
\gamma^2_d\int d\varepsilon \frac{ N(\varepsilon)}{2(i\omega-\varepsilon)^2}\sim  \gamma^2_d N'(0) \sim  \frac{\gamma^2_d N(0)}{\mu}\ ,
\end{equation}
where we assume that $|\omega|$ is no larger than the energy cut-off $\epsilon_c$ which is small compared with chemical potential $\mu$.
This means, when dealing with the disorder problem, we will integrate the energy band first and then sum up the frequency, which is the same as the other way to the leading order of $1/\beta\epsilon_c\ll 1$.
Since  we assume $\frac{\gamma^2_d N_F}{|\mu|}\ll 1$ with $N_F=N(0)=N_0 y_1$,
we can neglect the $G_0^2(\omega,\bsl{k}')$ term since we only keep the leading order of $\frac{\gamma^2_d N_F}{|\mu|}$.
Then, we have
\begin{equation}
\Sigma(\omega)=\frac{\gamma^2_d}{\mathcal{V}}\sum_{\bsl{k}'}G_0(\bsl{k}',\omega)\ .
\end{equation}

Since $\frac{1}{\mathcal{V}}\sum_{\bsl{k}'}G_0(\bsl{k}',\omega)$ is equal to
\begin{equation}
\int d\varepsilon \frac{N(\varepsilon)}{2} \frac{1}{i\omega-\varepsilon}\approx -i\pi\text{sgn}(\omega) \frac{N_F}{2}-\frac{\epsilon_0}{\gamma_d^2}
\end{equation}
with $\epsilon_0=\gamma^2_d \text{P}\left[\int d\varepsilon\frac{N(\epsilon)}{2\varepsilon}\right]$, we have
\begin{equation}
\bar{G}(\bsl{k},\omega)=\frac{1}{i\omega-h(\bsl{k})+i\frac{1}{2\tau_d} \text{sgn}(\omega)+\epsilon_0}
\end{equation}
with $1/\tau_d=\gamma^2_d \pi N_F$.
Moreover, if choosing the isotropic limit, we can estimate the order of $\epsilon_0$ by choosing the range of integration to be $(-|\mu|,|\mu|)$, which is $\epsilon_0 \tau_d\sim 1$.
In terms of the projection operators, the disorder-averaged Green function reads
\begin{equation}
\label{eq:Gbar_proj}
\bar{G}(\bsl{k},\omega)=\sum_{\lambda=\pm}\bar{G}_{\lambda}(\bsl{k},\omega)P_{\lambda}(\bsl{k})
\end{equation}
with
\begin{equation}
\bar{G}_{\lambda}(\bsl{k},\omega)=\frac{1}{i\omega-\xi_{\lambda}(\bsl{k})+i\frac{1}{2\tau_d} \text{sgn}(\omega)+\epsilon_0}\ .
\end{equation}

\subsubsection{Disorder-averaged Superconducting Free Energy}
\label{app:F_2_dis}

The mean-field free energy with disorder reads
\begin{equation}
F=-k_B T \ln\{\int Dc^{\dagger} Dc \exp[-S-S_{\Delta}+\beta f_{\Delta}]\}\ ,
\end{equation}
where
$S$ is shown in Eq.\ref{eq:S_dis},
\begin{equation}
\Delta(\bsl{k},\bsl{q})=\Delta_0(\bsl{q})\frac{\Gamma^0\gamma}{2}+\Delta_1(\bsl{q})\frac{a^2 \bsl{g}_{\bsl{k}}\cdot\bsl{\Gamma}\gamma}{2}\ ,
\end{equation}
\begin{eqnarray}
&&S_{\Delta}= \frac{1}{2}
[
\sum_{\omega,\bsl{k},\bsl{q}}\bar{c}_{\omega,\bsl{k}+\frac{\bsl{q}}{2}}\Delta(\bsl{k},\bsl{q})(\bar{c}_{-\omega,-\bsl{k}+\frac{\bsl{q}}{2}})^T\nonumber\\
&&+
\sum_{\omega,\bsl{k},\bsl{q}}c^T_{\omega,-\bsl{k}+\frac{\bsl{q}}{2}}\Delta^{\dagger}(\bsl{k},\bsl{q})c_{-\omega,\bsl{k}+\frac{\bsl{q}}{2}}]\ ,
\end{eqnarray}
\begin{equation}
f_{\Delta}=\sum_{\bsl{q},a}\frac{ \mathcal{V}}{2V_a}|\Delta_{a}(\bsl{q})|^2 \ ,
\end{equation}
and the order parameter is assumed to be uniform with respect to the imaginary time.
It is clear that $-\Delta^T(-\bsl{k},\bsl{q})=\Delta(\bsl{k},\bsl{q})$.

Then the mean-field superconducting Free energy reads
\begin{equation}
F_{SC}=-f_{\Delta}-k_B T \ln\{\left\langle\exp[-S_{\Delta}]\right\rangle\}
\end{equation}
where
\begin{equation}
\left\langle\exp[-S_{\Delta}]\right\rangle=\frac{\int D\bar{c} Dc \exp[-S-S_{\Delta}]}{Z_0[V]}\ ,
\end{equation}
and $Z_0[V]$ is shown in Eq.\ref{eq:Z_0_dis}.
Since $\left\langle\exp[-S_{\Delta}]\right\rangle=e^W$ with $W$ contains all the connected graphs,
we have the disorder-averaged $F_{SC}$
\begin{equation}
\label{eq:F_SC_dis_app}
\left\langle F_{SC}\right\rangle_{dis}=-f_{\Delta}-k_B T \left\langle W\right\rangle_{dis}\ .
\end{equation}
According to Eq.\ref{eq:disaver_final}, $\left\langle W\right\rangle_{dis}$ to the second order of $\Delta$ reads
\begin{widetext}
\begin{eqnarray}
&&\left\langle W^{(2)}\right\rangle_{dis}=\left\langle\left\langle \frac{(-S_{\Delta})^2}{2!}\right\rangle\right\rangle_{dis}=\int D\bar{c} Dc \frac{(-S_{\Delta})^2}{2!}\exp[-S_{dis}]\\
&&=\frac{1}{4}
\sum_{\omega,\bsl{k},\bsl{q}}\sum_{\omega',\bsl{k}',\bsl{q}'}\Delta_{\alpha_1\alpha_2}(\bsl{k},\bsl{q})\Delta^{*}_{\alpha_4,\alpha_3}(\bsl{k}',\bsl{q}')\int D\bar{c} Dc \bar{c}_{\omega,\bsl{k}+\frac{\bsl{q}}{2},\alpha_1}\bar{c}_{-\omega,-\bsl{k}+\frac{\bsl{q}}{2},\alpha_2}
c_{\omega',-\bsl{k}'+\frac{\bsl{q}'}{2},\alpha_3}c_{-\omega',\bsl{k}'+\frac{\bsl{q}'}{2},\alpha_4}
\exp[-S_{dis}]\nonumber\ ,
\end{eqnarray}
\end{widetext}
where
the summation of $\alpha_{1,2,3,4}$ is implied,
\begin{equation}
-S_{dis}=-S_0+\frac{\gamma^2_d}{2}\int dx dx' \delta(\bsl{r}-\bsl{r}') \bar{c}_x c_x \bar{c}_{x'}c_{x'}\ ,
\end{equation}
the internal fermionic loops are abandoned, and it is not necessary to specifically rule out the disconnected graphs to this order since all non-zero contribution is given by connected graphs.

We adopt Born approximation to abandon all graphs with crossed disorder lines and only include the cooperon modes\cite{altland2010condensed}.
In this case, we have
\begin{eqnarray}
&&\left\langle W^{(2)}\right\rangle_{dis}=\frac{1}{2}
\sum_{\omega,\bsl{k},\bsl{q}}\\
&&\text{Tr}[\bar{G}(\bsl{k}+\frac{\bsl{q}}{2},\omega)D(\bsl{k},\bsl{q},\omega)\bar{G}^T(-\bsl{k}+\frac{\bsl{q}}{2},-\omega)\Delta^{\dagger}(\bsl{k},\bsl{q})]\nonumber\ ,
\end{eqnarray}
where
\begin{eqnarray}
&&D(\bsl{k},\bsl{q},\omega)=
\Delta(\bsl{k},\bsl{q})
+\\
&&\frac{\gamma^2_d}{\mathcal{V}}\sum_{\bsl{k}'}\bar{G}(\bsl{k}'+\frac{\bsl{q}}{2},\omega)D(\bsl{k}',\bsl{q},\omega)\bar{G}^T(-\bsl{k}'+\frac{\bsl{q}}{2},-\omega)\nonumber\ .
\end{eqnarray}

If the order parameter is uniform, then we can choose $\Delta(\bsl{k},\bsl{q})=\Delta(\bsl{k})\delta_{\bsl{q},0}$.
In this case, combining Eq.\ref{eq:F_SC_dis_app} and the two equations shown above, we can get
\begin{eqnarray}
\label{eq:F_dis_uniform}
&&F_{SC}=-\mathcal{V}\sum_a\frac{|\Delta_a|^2}{2V_a}\\
&&-\frac{1}{2\beta}
\sum_{\omega_n,\bsl{k}}\text{Tr}[\bar{G}(
\bsl{k},\omega_n)D(\bsl{k},\omega_n)\bar{G}^T(-\bsl{k},-\omega_n)\Delta^{\dagger}(\bsl{k})]\nonumber\ ,
\end{eqnarray}
where
\begin{equation}
\label{eq:D_delta}
D(\bsl{k},\omega)=
\Delta(\bsl{k})
+\frac{\gamma_d^2}{\mathcal{V}}\sum_{\bsl{k}'}\bar{G}(\bsl{k}',\omega)D(\bsl{k}',\omega)\bar{G}^T(-\bsl{k}',-\omega)\ .
\end{equation}


\subsubsection{Further Simplification of $F_{SC}$ with Disorder}
\label{app:der_eq_F_dis}

First, recall the property of projection operator in our case:
\begin{eqnarray}
\label{eq:P_reln}
&&P_{\lambda}(\bsl{k})=P_{\lambda}^2(\bsl{k})\nonumber\\
&&P_{\lambda}(\bsl{k})=P_{\lambda}^{\dagger}(\bsl{k})\nonumber\\
&&P_{\lambda}(-\bsl{k})=P_{\lambda}(\bsl{k})\nonumber\\
&&P_{\lambda}(\bsl{k})\Gamma^i \gamma P_{\lambda}^T(\bsl{k})\propto P_{\lambda}(\bsl{k}) \gamma P_{\lambda}^T(\bsl{k}) \ ,
\end{eqnarray}
where the first two are general, the third one is due to the inversion symmetry of our model, and the last one is for $i=0,...,5$. In the following, we will use the four relations again and again, and we will not refer to them for convenience.

The trace term in Eq.\ref{eq:F_dis_uniform} can be expressed by the projection operator using Eq.\ref{eq:Gbar_proj} to the leading order of $\epsilon_c/(2 Q_c k_F^2)\ll 1$:
\begin{eqnarray}
\label{eq:F_D_lambda}
&&\sum_{\omega_n,\bsl{k}}
\text{Tr}[\bar{G}(\bsl{k},\omega_n)D(\bsl{k},\omega_n)\bar{G}^T(-\bsl{k},-\omega_n)\Delta^{\dagger}(\bsl{k})]=\nonumber\\
&&\sum_{\omega_n,\bsl{k},\lambda}\bar{G}_{\lambda}(\bsl{k},\omega_n)\bar{G}_{\lambda}(-\bsl{k},-\omega_n)d_{\lambda}^*(\bsl{k})\text{Tr}[D_{\lambda}(\bsl{k},\omega_n)\gamma_{\lambda}^{\dagger}(\bsl{k})]\ ,\nonumber\\
\end{eqnarray}
where $\gamma_{\lambda}(\bsl{k})=P_{\lambda}(\bsl{k})\gamma P_{\lambda}^T(-\bsl{k})$, $P_{\lambda}(\bsl{k})\Delta(\bsl{k})P_{\lambda}^T(-\bsl{k})=d_{\lambda}(\bsl{k})\gamma_{\lambda}(\bsl{k})$, $d_{\lambda}(\bsl{k})= \frac{\Delta_0}{2}+\lambda\frac{\Delta_1 }{2} a^2 k^2 \text{sgn}(c_1) f_Q$, and $D_{\lambda}(\bsl{k},\omega)=P_{\lambda}(\bsl{k})D(\bsl{k},\omega)P_{\lambda}^T(-\bsl{k})$. Using Eq.\ref{eq:D_delta}, the equation satisfied by $D_{\lambda}(\bsl{k},\omega)$ reads
\begin{eqnarray}
\label{eq:D_lambda}
&&D_{\lambda}(\bsl{k},\omega_n)=d_{\lambda}(\bsl{k})\gamma_{\lambda}(\bsl{k})+\frac{\gamma^2_d}{\mathcal{V}}\sum_{\bsl{k}',\lambda'}\\
&&\bar{G}_{\lambda'}(\bsl{k}',\omega_n)\bar{G}_{\lambda'}(-\bsl{k}',-\omega_n)P_{\lambda}(\bsl{k})D_{\lambda'}(\bsl{k}',\omega_n)P_{\lambda}^T(-\bsl{k})\ ,\nonumber
\end{eqnarray}
where we also neglect the interband contribution as before, i.e. only keep terms to the leading order of $\epsilon_c/(2 Q_c k_F^2)\ll 1$.

Since $\gamma_{\lambda}^T(\bsl{k})=\gamma_{\lambda}^T(-\bsl{k})=-\gamma_{\lambda}(\bsl{k})$, we can show that $-D^T_{\lambda}(\bsl{k},\omega_n)$ satisfies the same equation as $D_{\lambda}(\bsl{k},\omega_n)$, meaning that $-D^T_{\lambda}(\bsl{k},\omega_n)=D_{\lambda}(\bsl{k},\omega_n)$.
Thereby $D_{\lambda}(\bsl{k},\omega_n)$ can be expressed in terms of $\Gamma^i \gamma$ with $i=0,...,5$.
Further, we have
\begin{equation}
\label{eq:D=Dbargamma}
D_{\lambda}(\bsl{k},\omega_n)=\bar{D}_{\lambda}(\bsl{k},\omega_n)\gamma_{\lambda}(\bsl{k})\ ,
\end{equation}
where $\bar{D}_{\lambda}(\bsl{k},\omega_n)$ is a scalar function.

Using $\text{Tr}(\gamma_{\lambda}(\bsl{k})\gamma_{\lambda}^{\dagger}(\bsl{k}))=2$, Eq.\ref{eq:D_lambda} and Eq.\ref{eq:D=Dbargamma}, we have
\begin{equation}
\label{Eq:Dlambda_tr}
\text{Tr}[D_{\lambda}(\bsl{k},\omega_n)\gamma_{\lambda}^{\dagger}(\bsl{k})]=2\bar{D}_{\lambda}(\bsl{k},\omega_n)\ ,
\end{equation}
and
\begin{eqnarray}
\label{eq:Dlambda_dlambda}
&&\bar{D}_{\lambda}(\bsl{k},\omega_n)=d_{\lambda}(\bsl{k})+\frac{\gamma^2_d}{2\mathcal{V}}\sum_{\bsl{k}',\lambda'}\\
&&\bar{G}_{\lambda'}(\bsl{k}',\omega_n)\bar{G}_{\lambda'}(-\bsl{k}',-\omega_n)\bar{D}_{\lambda'}(\bsl{k}',\omega_n)\text{Tr}[\gamma_{\lambda'}(\bsl{k}')\gamma^{\dagger}_{\lambda}(\bsl{k})]\ .\nonumber
\end{eqnarray}

Since the non-interacting Hamiltonian is $O_h$ invariant, we have $U_R h(R^{-1}\bsl{k}) U_R^{\dagger}=h(\bsl{k})$ for any $R\in O_h$, where $U_R$ is the unitary representation of $R$ for $j=\frac{3}{2}$ fermions. Using Eq.\ref{eq:proj_P}, we have $U_R P_{\lambda}(R^{-1}\bsl{k})U_R^{\dagger}=P_{\lambda}(\bsl{k})$ and thereby $U_R\gamma_{\lambda}(R^{-1}\bsl{k})U_R^{T}=\gamma_{\lambda}(\bsl{k})$. Using this relation and the fact that $d_{\lambda}(\bsl{k})$ and $\bar{G}_{\lambda}(\bsl{k},\omega_n)$ are $O_h$ invariant, we can get that $\bar{D}_{\lambda}(R^{-1}\bsl{k},\omega_n)$ satisfies the same equation as $\bar{D}_{\lambda}(\bsl{k},\omega_n)$, meaning that $\bar{D}_{\lambda}(\bsl{k},\omega_n)$ is $O_h$ invariant: $\bar{D}_{\lambda}(R^{-1}\bsl{k},\omega_n)=\bar{D}_{\lambda}(\bsl{k},\omega_n)$.
Moreover, according to Schur's lemmas\cite{tung1985group}, we have
$\sum_{\bsl{k}}f(\bsl{k})g_{\bsl{k},i}=0$ and $\sum_{\bsl{k}}f(\bsl{k})g_{\bsl{k},i}g_{\bsl{k},j}=\sum_{\bsl{k}}f(\bsl{k})g_{\bsl{k},i}^2 \delta_{ij}$ if $f(\bsl{k})$ is $O_h$ invariant and $i,j=1,..,5$.
Combining the previous facts,
we can get
\begin{eqnarray}
&&\sum_{\bsl{k}'}\bar{G}_{\lambda'}(\bsl{k}',\omega_n)\bar{G}_{\lambda'}(-\bsl{k}',-\omega_n)\bar{D}_{\lambda'}(\bsl{k}',\omega_n)\gamma_{\lambda'}(\bsl{k}') \nonumber\\
&&=\sum_{\bsl{k}'}\bar{G}_{\lambda'}(\bsl{k}',\omega_n)\bar{G}_{\lambda'}(-\bsl{k}',-\omega_n)\bar{D}_{\lambda'}(\bsl{k}',\omega_n)\frac{\gamma}{2}\ ,
\end{eqnarray}
where $\Gamma_i\gamma\Gamma_i^T=\gamma$ for $i=1,...,5$ is used.
Then Eq.\ref{eq:Dlambda_dlambda} becomes
\begin{eqnarray}
\label{eq:Dlambda_dlambda_final}
&&\bar{D}_{\lambda}(\bsl{k},\omega_n)=d_{\lambda}(\bsl{k})+\frac{\gamma^2_d}{2\mathcal{V}}\sum_{\bsl{k}',\lambda'}\\
&&\bar{G}_{\lambda'}(\bsl{k}',\omega_n)\bar{G}_{\lambda'}(-\bsl{k}',-\omega_n)\bar{D}_{\lambda'}(\bsl{k}',\omega_n)\ .\nonumber
\end{eqnarray}
Define
\begin{equation}
\bar{L}_0(\omega_n)=\sqrt{\frac{\gamma^2_d}{2\mathcal{V}}}\sum_{\bsl{k},\lambda}\bar{G}_{\lambda}(\bsl{k},\omega_n)\bar{G}_{\lambda}(-\bsl{k},-\omega_n)d_{\lambda}(\bsl{k})\ ,
\end{equation}
and
\begin{equation}
\bar{L}(\omega_n)=\sqrt{\frac{\gamma^2_d}{2\mathcal{V}}}\sum_{\bsl{k},\lambda}\bar{G}_{\lambda}(\bsl{k},\omega_n)\bar{G}_{\lambda}(-\bsl{k},-\omega_n)\bar{D}_{\lambda}(\bsl{k},\omega_n)\ .
\end{equation}
Then Eq.\ref{eq:Dlambda_dlambda_final} relates $\bar{L}$ and $\bar{L}_0$:
\begin{equation}
\label{eq:Lbar_Lbar0}
\bar{L}(\omega_n)=\bar{L}_0(\omega_n)+\bar{b}(\omega_n)\bar{L}(\omega_n)\ ,
\end{equation}
which gives
\begin{equation}
\bar{L}(\omega_n)=\frac{\bar{L}_0(\omega_n)}{1-\bar{b}(\omega_n)}
\end{equation}
with
\begin{equation}
\bar{b}(\omega_n)=\frac{\gamma_d^2}{2\mathcal{V}}\sum_{\bsl{k},\lambda}\bar{G}_{\lambda}(\bsl{k},\omega_n)\bar{G}_{\lambda}(-\bsl{k},-\omega_n)\ .
\end{equation}

Eventually, Eq.\ref{eq:F_D_lambda} becomes
\begin{eqnarray}
\label{eq:F_D_lambda_final}
&&\sum_{\omega_n,\bsl{k}}
\text{Tr}[\bar{G}(\bsl{k},\omega_n)D(\bsl{k},\omega_n)\bar{G}^T(-\bsl{k},-\omega_n)\Delta^{\dagger}(\bsl{k})]\nonumber\\
&&=\sum_{\omega_n,\bsl{k},\lambda}\bar{G}_{\lambda}(\bsl{k},\omega_n)\bar{G}_{\lambda}(-\bsl{k},-\omega_n)d_{\lambda}^*(\bsl{k})2\bar{D}_{\lambda}(\bsl{k},\omega_n)\nonumber\\
&&=2\sum_{\omega_n,\bsl{k},\lambda}\bar{G}_{\lambda}(\bsl{k},\omega_n)\bar{G}_{\lambda}(-\bsl{k},-\omega_n)d_{\lambda}^*(\bsl{k})d_{\lambda}(\bsl{k})\nonumber\\
&&+2\sum_{\omega_n}\bar{L}_0^*(\omega_n)\bar{L}(\omega_n)\nonumber\\
&&=2\sum_{\omega_n,\bsl{k},\lambda}\bar{G}_{\lambda}(\bsl{k},\omega_n)\bar{G}_{\lambda}(-\bsl{k},-\omega_n)d_{\lambda}^*(\bsl{k})d_{\lambda}(\bsl{k})\nonumber\\
&&+2\sum_{\omega_n}\frac{|\bar{L}_0(\omega_n)|^2}{1-\bar{b}(\omega_n)}\ ,
\end{eqnarray}
where the second equality uses Eq.\ref{eq:F_D_lambda} and Eq.\ref{Eq:Dlambda_tr}, the third equality uses Eq.\ref{eq:Dlambda_dlambda_final}, $\bar{G}_{\lambda}(\bsl{k},\omega_n)\bar{G}_{\lambda}(-\bsl{k},-\omega_n)$ is real and definitions of $\bar{L}(\omega_n)$ and $\bar{L}_0(\omega_n)$, and the last equality uses Eq.\ref{eq:Lbar_Lbar0}.
Combined with Eq.\ref{eq:F_dis_uniform}, we eventually have
\begin{eqnarray}
\label{eq:F_L_omega}
&&F_{SC}=-\frac{1}{\beta}
\sum_{\omega_n,\bsl{k},\lambda}\bar{G}_{\lambda}(\bsl{k},\omega_n)\bar{G}_{\lambda}(-\bsl{k},-\omega_n)d_{\lambda}^*(\bsl{k})d_{\lambda}(\bsl{k})\nonumber\\
&&-\frac{1}{\beta}\sum_{\omega_n}\frac{|\bar{L}_0(\omega_n)|^2}{1-\bar{b}(\omega_n)}-\mathcal{V}\sum_a\frac{|\widetilde{\Delta}_a|^2}{2\widetilde{V}_a}\ .
\end{eqnarray}

\subsubsection{Derivation of Eq.\ref{eq:lge_final_dis}}
\label{app:der_F_lge_final}

First we derive a general expression that will be used repeatedly later:
\begin{eqnarray}
&&\frac{1}{\mathcal{V}}\sum_{\lambda,\bsl{k}}\bar{G}_{\lambda}(\bsl{k},\omega_n)\bar{G}_{\lambda}(-\bsl{k},-\omega_n)f_{\lambda}(\bsl{k})\approx \nonumber\\
&&\sum_{\lambda}\int \frac{d\Omega}{4\pi} N_0 \widetilde{m}_{\lambda}^{3/2}\theta(\widetilde{m}_{\lambda})f_{\lambda}(\bsl{k}_{F,\lambda})
\int d\xi_{\lambda} \frac{1}{(|\omega_n|+\frac{1}{2\tau_d})^2+(\xi_{\lambda})^2}\nonumber\\
&&=\frac{\pi}{|\omega_n|+\frac{1}{2\tau_d}}\sum_{\lambda}\langle N_0 \widetilde{m}_{\lambda}^{3/2}\theta(\widetilde{m}_{\lambda})f_{\lambda}(\bsl{k}_{F,\lambda})\rangle_{\Omega}\ ,
\end{eqnarray}
where the first equality uses two things:
(i) $\epsilon_0\sim 1/\tau_d$ and we neglect terms of order $1/(\tau_d\mu)$;
(ii) $1/(\xi^2+\epsilon^2)$ has a peak at $\xi=0$ and drops fast away from the peak  when $\epsilon$ is small.
The range of the integration of $\xi_{\lambda}$ is from $-\infty$ to $\infty$ since the energy cut-off $\epsilon_c$ is included in the limit of the summation of $\omega_n$ as $|\omega_n|+\frac{1}{2\tau_d}\leq \epsilon_c$.

Using the formula derived above,
we have
\begin{eqnarray}
\label{eq:F_GGdd_dis}
&&
\sum_{\bsl{k},\lambda}\bar{G}_{\lambda}(\bsl{k},\omega_n)\bar{G}_{\lambda}(-\bsl{k},-\omega_n)d_{\lambda}^*(\bsl{k})d_{\lambda}(\bsl{k})\nonumber\\
&&=\frac{N_0\mathcal{V}}{4}\frac{\pi}{|\omega_n|}\frac{|\omega_n|}{|\omega_n|+\frac{1}{2\tau_d}}\widetilde{\Delta}^{\dagger}
\left(
\begin{matrix}
y_1 & y_2 \\
y_2 & y_3 \\
\end{matrix}
\right)
\widetilde{\Delta}
\end{eqnarray}
with $\widetilde{\Delta}=(\widetilde{\Delta}_0,\widetilde{\Delta}_1)^T$ ,
\begin{equation}
1-\bar{b}(\omega_n)=\frac{|\omega_n|}{|\omega_n|+\frac{1}{2\tau_d}}\ ,
\end{equation}
\begin{equation}
\bar{L}_0(\omega_n)=\mathcal{V}\sqrt{\frac{\gamma^2_d}{2\mathcal{V}}}\frac{\pi}{|\omega_n|+\frac{1}{2\tau_d}}N_0 \text{sgn}(c_1)(\frac{\widetilde{\Delta}_0}{2}y_1+\frac{\widetilde{\Delta}_1}{2}y_2)\ ,
\end{equation}
and thereby
\begin{eqnarray}
\label{eq:F_L0sq_dis}
&&\frac{|\bar{L}_0(\omega_n)|^2}{1-\bar{b}(\omega_n)}=\frac{N_0\mathcal{V}}{4}\frac{\pi}{|\omega_n|}\frac{\frac{1}{2\tau_d}}{(|\omega_n|+\frac{1}{2\tau_d})}\frac{1}{y_1}| \widetilde{\Delta}_0 y_1+ \widetilde{\Delta}_1 y_2|^2\nonumber\\
&&=\frac{N_0\mathcal{V}}{4}\frac{\pi}{|\omega_n|}\frac{\frac{1}{2\tau_d}}{(|\omega_n|+\frac{1}{2\tau_d})}\widetilde{\Delta}^{\dagger}
\left(
\begin{matrix}
y_1 & y_2 \\
y_2 & y_2^2/y_1 \\
\end{matrix}
\right)
\widetilde{\Delta}
\ .
\end{eqnarray}
Here $1/\tau_d=\gamma^2_d \pi N_F$ is used.

Substituting Eq.\ref{eq:F_GGdd_dis} and Eq.\ref{eq:F_L0sq_dis} into Eq.\ref{eq:F_L_omega}, we have
\begin{eqnarray}
&&F_{SC}=-\mathcal{V}\sum_a\frac{|\widetilde{\Delta}_a|^2}{2\widetilde{V}_a}\\
&&-\frac{N_0\mathcal{V}}{4\beta}
\sum_{\omega_n}\frac{\pi}{|\omega_n|}
\widetilde{\Delta}^{\dagger}
\left(
\begin{matrix}
y_1 & y_2 \\
y_2 & \frac{|\omega_n|}{|\omega_n|+\frac{1}{2\tau_d}}y_3+\frac{\frac{1}{2\tau_d}}{(|\omega_n|+\frac{1}{2\tau_d})}y_2^2/y_1 \\
\end{matrix}
\right)
\widetilde{\Delta}\nonumber\ .
\end{eqnarray}

Assuming $\epsilon_c \tau_d \gg 1$ and $\beta\epsilon_c\gg 1$, we have
\begin{equation}
\frac{\pi}{\beta}\sum_{\omega_n}\frac{1}{|\omega_n|}=\ln(\frac{2e^{\bar{\gamma}}\beta\epsilon_c}{\pi})+O(\frac{1}{\beta\epsilon_c},\frac{1}{\epsilon_c \tau_d})\ ,
\end{equation}
\begin{equation}
\frac{\pi}{\beta}\sum_{\omega_n}\frac{1}{|\omega_n|+\frac{1}{2\tau_d}}= \ln(\frac{2e^{\bar{\gamma}}\beta\epsilon_c}{\pi})-\mathcal{F}(\frac{\beta}{4\pi\tau_d})+O(\frac{1}{\beta\epsilon_c})\ ,
\end{equation}
and
\begin{equation}
\frac{\pi}{\beta}\sum_{\omega_n}\frac{\frac{1}{2\tau_d}}{|\omega_n|(|\omega_n|+\frac{1}{2\tau_d})}=\mathcal{F}(\frac{\beta}{4\pi\tau_d})+O(\frac{1}{\beta\epsilon_c},\frac{1}{\epsilon_c \tau_d})\ ,
\end{equation}
where $\mathcal{F}(\frac{\beta}{4\pi\tau_d})=\Psi^{(0)}(\frac{\beta}{4\pi \tau_d}+\frac{1}{2})-\Psi^{(0)}(\frac{1}{2})$ ,$\Psi^{(0)}(x)$ is the digamma function and the range of the sum is $|\omega_n|+\frac{1}{2\tau_d}\leq \epsilon_c$.
Then, the free energy becomes
\begin{eqnarray}
&&F_{SC}=-\frac{\mathcal{V}}{2}\widetilde{\Delta}^{\dagger}
\left(
\begin{matrix}
\frac{1}{\widetilde{V}_0} &  \\
  & \frac{1}{\widetilde{V}_1} \\
\end{matrix}
\right)
\widetilde{\Delta}\nonumber\\
&&-\frac{N_0\mathcal{V}}{4} x
\widetilde{\Delta}^{\dagger}
\left(
\begin{matrix}
y_1 & y_2 \\
y_2 & y_3+\frac{\mathcal{F}}{x}(y_2^2/y_1-y_3) \\
\end{matrix}
\right)
\widetilde{\Delta}\ ,
\end{eqnarray}
where $\widetilde{\Delta}=(\widetilde{\Delta}_0,\widetilde{\Delta}_1)^T$.
Then linearized gap equation reads
\begin{equation}
\left(
\begin{array}{c}
\tilde{\Delta}_{0}\\
\tilde{\Delta}_{1}
\end{array}
\right)
=
x\left(
\begin{array}{cc}
\frac{\lambda_{0}}{2}y_1 & \frac{\lambda_{0}}{2}y_2\\
\frac{\lambda_{1}}{2}y_2& \frac{\lambda_{1}}{2}y_3 b_1
\end{array}
\right)
\left(
\begin{array}{c}
\tilde{\Delta}_{0}\\
\tilde{\Delta}_{1}
\end{array}
\right)\ ,
\end{equation}
where
\begin{equation}
b_1=1+\frac{\mathcal{F}(\frac{\beta}{4\pi\tau_d})}{x}(\frac{y_2^2}{y_1 y_3}-1)\ .
\end{equation}

Next we will show $0<b_1\leq 1$. Let us define $t_{\lambda}=\widetilde{m}_{\lambda}\theta(\widetilde{m}_{\lambda})$, and thereby $t_{\lambda}\geq 0$.
In this case,  $y_1=\langle (t_+^{3/2}+t_-^{3/2})\rangle_{\Omega}$,  $y_2=\langle f_Q (t_+^{5/2}-t_-^{5/2})\rangle_{\Omega}$ and $y_3=\langle f_Q^2 (t_+^{7/2}+t_-^{7/2})\rangle_{\Omega}$.
According to Cauchy-Schwarz inequality, $t_{\pm}\geq 0$ and $f_Q\geq 0$, we have
\begin{eqnarray}
&&y_1y_3=\langle f_Q^2 (t_+^{7/2}+t_-^{7/2})\rangle_{\Omega}
\langle (t_+^{3/2}+t_-^{3/2})\rangle_{\Omega}\nonumber\\
&&
\geq \langle \sqrt{f_Q^2 (t_+^{7/2}+t_-^{7/2})  (t_+^{3/2}+t_-^{3/2})}\rangle_{\Omega}^2\nonumber\\
&&=\langle f_Q\sqrt{(t_+^{5}+t_-^{5}+t_+^{3/2}t_-^{7/2}+t_+^{7/2}t_-^{3/2})}\rangle_{\Omega}^2\nonumber\\
&&\geq
\langle f_Q\sqrt{(t_+^{5}+t_-^{5}+2t_+^{5/2}t_-^{5/2})}\rangle^2_{\Omega}\nonumber\\
&&\geq
\langle f_Q\sqrt{(t_+^{5}+t_-^{5}-2t_+^{5/2}t_-^{5/2})}\rangle^2_{\Omega}\nonumber\\
&&=
\langle f_Q|t_+^{5/2}-t_-^{5/2}|\rangle^2_{\Omega}\nonumber\\
&&\geq
|\langle f_Q(t_+^{5/2}-t_-^{5/2})\rangle_{\Omega}|^2=|y_2|^2=y_2^2\ .
\end{eqnarray}
It gives $0\leq y_2^2/(y_1 y_3)\leq 1$ since $y_{1,3}>0$.
Combined with $\mathcal{F}(\frac{\beta}{4\pi\tau_d})<x$, we have
$0<b_1\leq 1$.
To have $b_1=1$, we either need $\frac{1}{\tau_d}=0$ meaning that there is no disorder or need the system to be in regime II($t_+ t_-=0$ in any direction) and isotropic $(c_1=c_2)$.
\end{appendices}

\end{document}